\documentclass{aa}
\usepackage[utf8]{inputenc} 

\usepackage{graphicx} 
\usepackage{txfonts} 

\usepackage{xcolor} 
\usepackage{natbib}

\pdfpageattr {/Group << /S /Transparency /I true /CS /DeviceRGB>>}

\bibpunct{(}{)}{;}{a}{}{,} 

\newcommand{\ie}{{i.e.}} 
\newcommand{\eg}{{e.g.}}

\newcommand{\emm}[1]{\ensuremath{#1}}
\newcommand{\emr}[1]{\emm{\mathrm{#1}}}
\newcommand{\chem}[1]{\emr{\,#1}} 
\newcommand{\unit}[1]{\emr{\,#1}}

\newcommand{\pc}{\unit{pc}} 
\newcommand{\mum}{\unit{\mu m}}
\newcommand{\pccm}{\unit{cm^{-3}}} 
\newcommand{\pscm}{\unit{cm^{-2}}} 
\newcommand{\Kkms}{\unit{K\,km\,s^{-1}}} 
\newcommand{\kms}{\unit{km\,s^{-1}}} 
\newcommand{\kHz}{\unit{kHz}}
 
\newcommand{\K}{\unit{K}}
\newcommand{\GHz}{\unit{GHz}} 

\newcommand{\thCO}{\chem{^{13}CO}} 
\newcommand{\twCO}{\chem{^{12}CO}}
\newcommand{\CeiO}{\chem{C^{18}O}} 
\newcommand{\Jone}{(1--0)}

\definecolor{ochre}{rgb}{0.8, 0.47, 0.13}

\newcommand{\TabRadexMedians}{%
  \begin{table*}
    \caption{Median gas volume density, gas temperature, C$^{18}$O, HCO$^+$ and CN 
      integrated intensities and associated column densities derived from
      RADEX models for each CN group.}
    \label{tab:Radex_medians}
    \resizebox{\textwidth}{!}{%
      \begin{tabular} {c|cccccc|ccc}
        \hline 
         & median $n_\mathrm{H}$ & median $T$ & median $N_H$ & median $I($C$^{18}$O$)$ & median $I($HCO$^+)$ & median $I($CN$)$ & $N($C$^{18}$O$)$ & $N($HCO$^+)$ & $N($CN$)$ \\
         & $[\mathrm{cm}^{-3}]$ & $[\mathrm{K}]$ & $[\mathrm{cm}^{-2}]$ & $[\mathrm{K.km/s}]$ & $[\mathrm{K.km/s}]$ & $[\mathrm{K.km/s}]$ & $[\mathrm{cm}^{-2}]$ & $[\mathrm{cm}^{-2}]$ & $[\mathrm{cm}^{-2}]$ \\
         \hline
         1 & $2.2\times 10^2$ & $24$ & $3.7\times10^{21}$ & $4.9\times 10^{-2}$ & $3.3\times 10^{-1}$ & $7.9\times 10^{-2}$ & $1.3\times 10^{14}$ & $2.1\times 10^{13}{}^{(b)}$ & $5.0\times 10^{14}{}^{(b)}$ \\
         2 & $8.6\times 10^2$ & $28$ & $9.3\times10^{21}$ & $4.7\times 10^{-1}$ & $1.2$               & $4.4\times 10^{-1}$ & $5.1\times 10^{14}$ & $2.8\times 10^{13}{}^{(b)} - 1.5\times 10^{14}{}^{(a)}$ & $8.9\times 10^{14}{}^{(b)} - 4.4\times 10^{15}{}^{(a)}$ \\
         3 & $3.8\times 10^3$ & $29$ & $1.9\times10^{22}$ & $2.0$               & $2.3$               & $7.6\times 10^{-1}$ & $1.8\times 10^{15}$ & $7.7\times 10^{13}{}^{(a)}$ & $2.0\times 10^{15}{}^{(a)}$ \\
         4 & $2.5\times 10^3$ & $42$ & $1.6\times10^{22}$ & $8.3\times 10^{-1}$ & $3.0$               & $1.3$               & $6.7\times 10^{14}$ & $3.9\times 10^{13}{}^{(b)} - 1.4\times 10^{14}{}^{(a)}$ & $1.3\times 10^{15}{}^{(b)} - 4.6\times 10^{15}{}^{(a)}$ \\
         5 & $4.1\times 10^4$ & $39$ & $6.6\times10^{22}$ & $5.8$               & $5.4$               & $1.9$               & $1.1\times 10^{16}$ & $2.2\times 10^{13}{}^{(a)}$ & $5.7\times 10^{14}{}^{(a)}$ \\
        \hline 
      \end{tabular}}
    \tablefoot{%
      \tablefoottext{a}{RADEX result for $x_e=0$.}
      \tablefoottext{b}{RADEX result for $x_e=1.4\times10^{-4}$}.}
  \end{table*}
}

\newcommand{\FigDustPDFs}{
  \begin{figure}
    \includegraphics[width=1\linewidth]{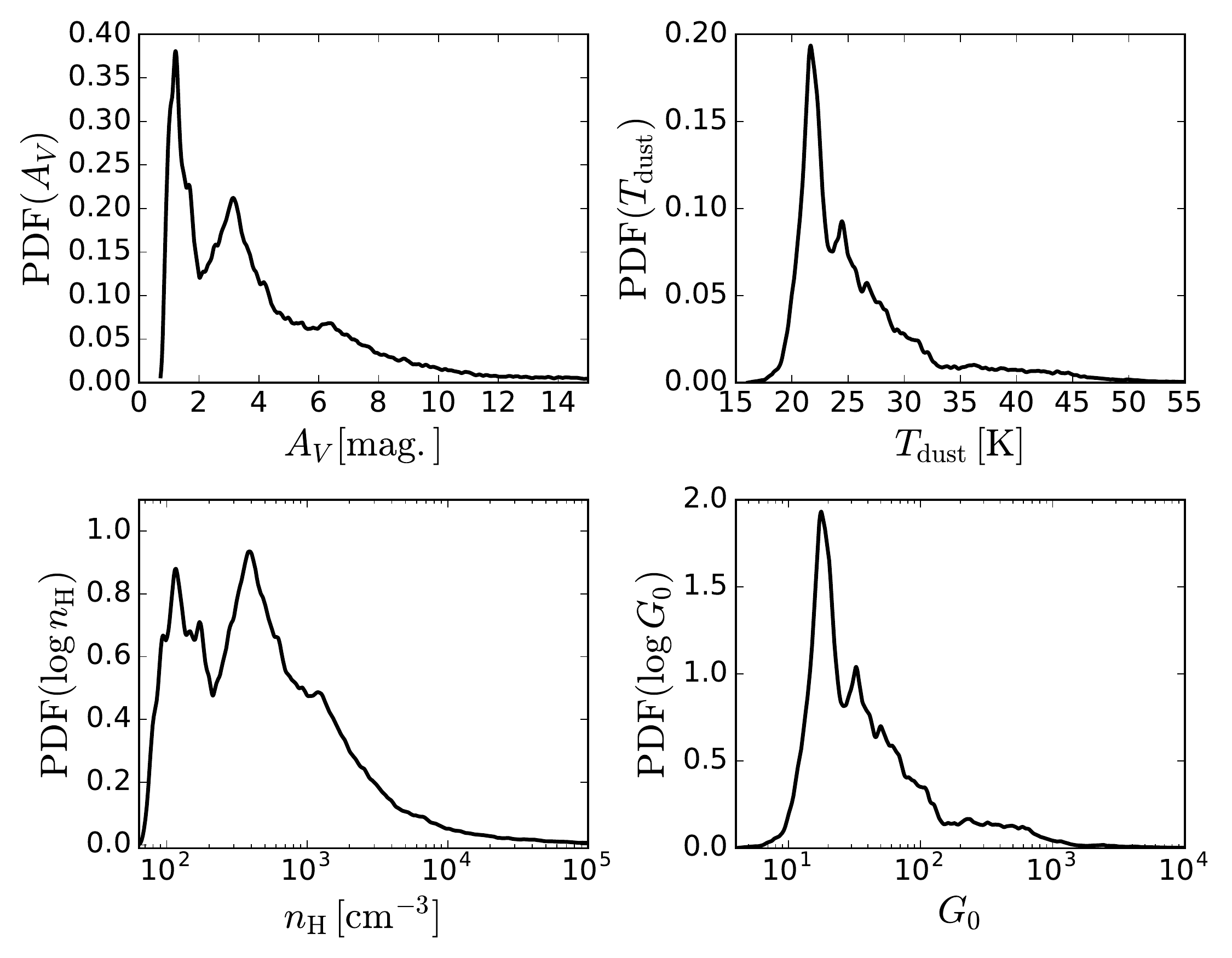}
    \caption{PDFs of the dust visual extinction (top left), the dust
      temperature (top right), the deduced approximate gas volume
        density (bottom left), and the deduced FUV illumination (bottom
      right) in the observed field of view.  }
    \label{fig:DustPDFs}
  \end{figure}
}

\newcommand{\FigClusterCOMap}{%
  \begin{figure*}
    \centering %
    \includegraphics[width=0.49\linewidth]{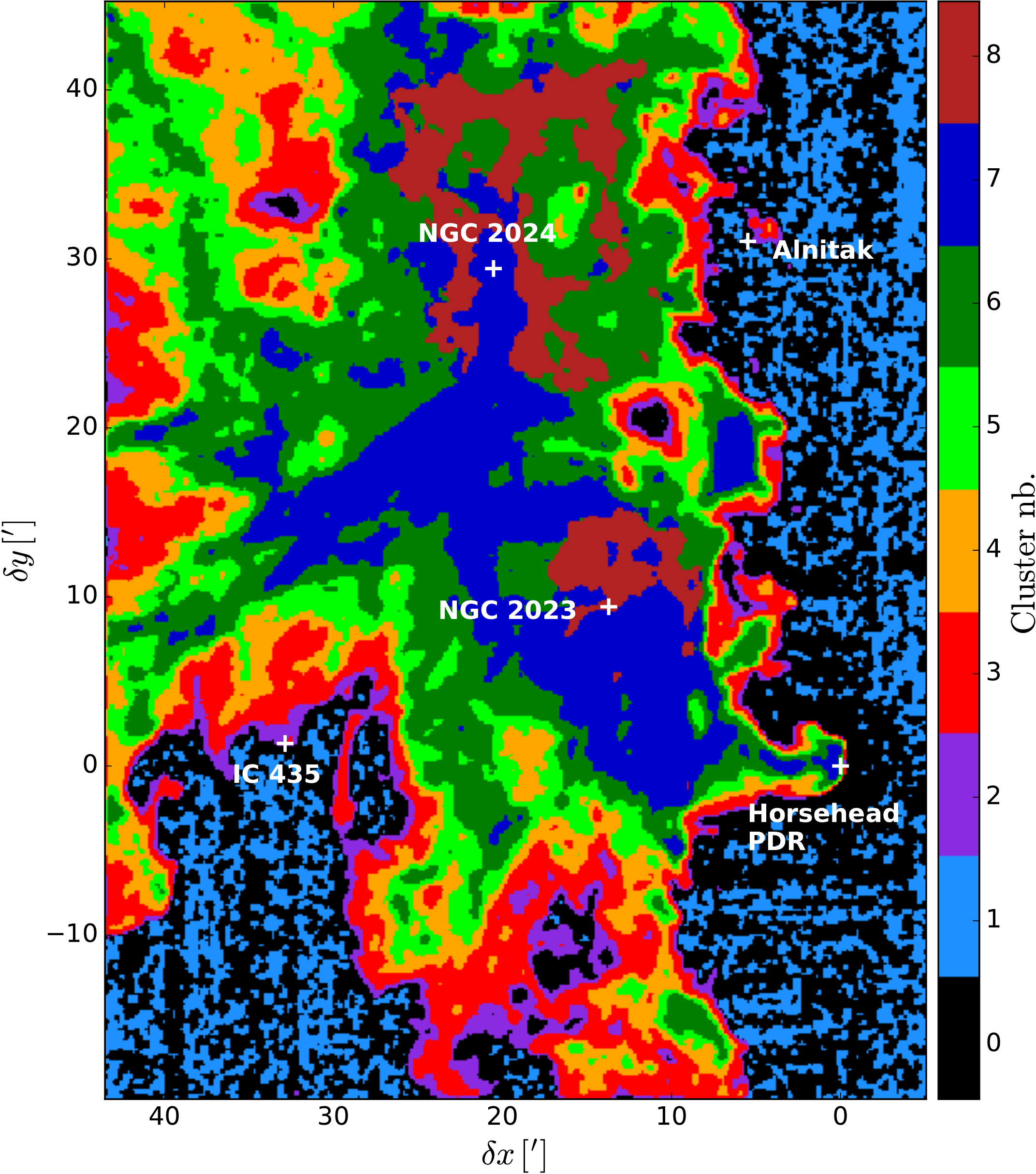}
    \hfill{} %
    \includegraphics[width=0.49\linewidth]{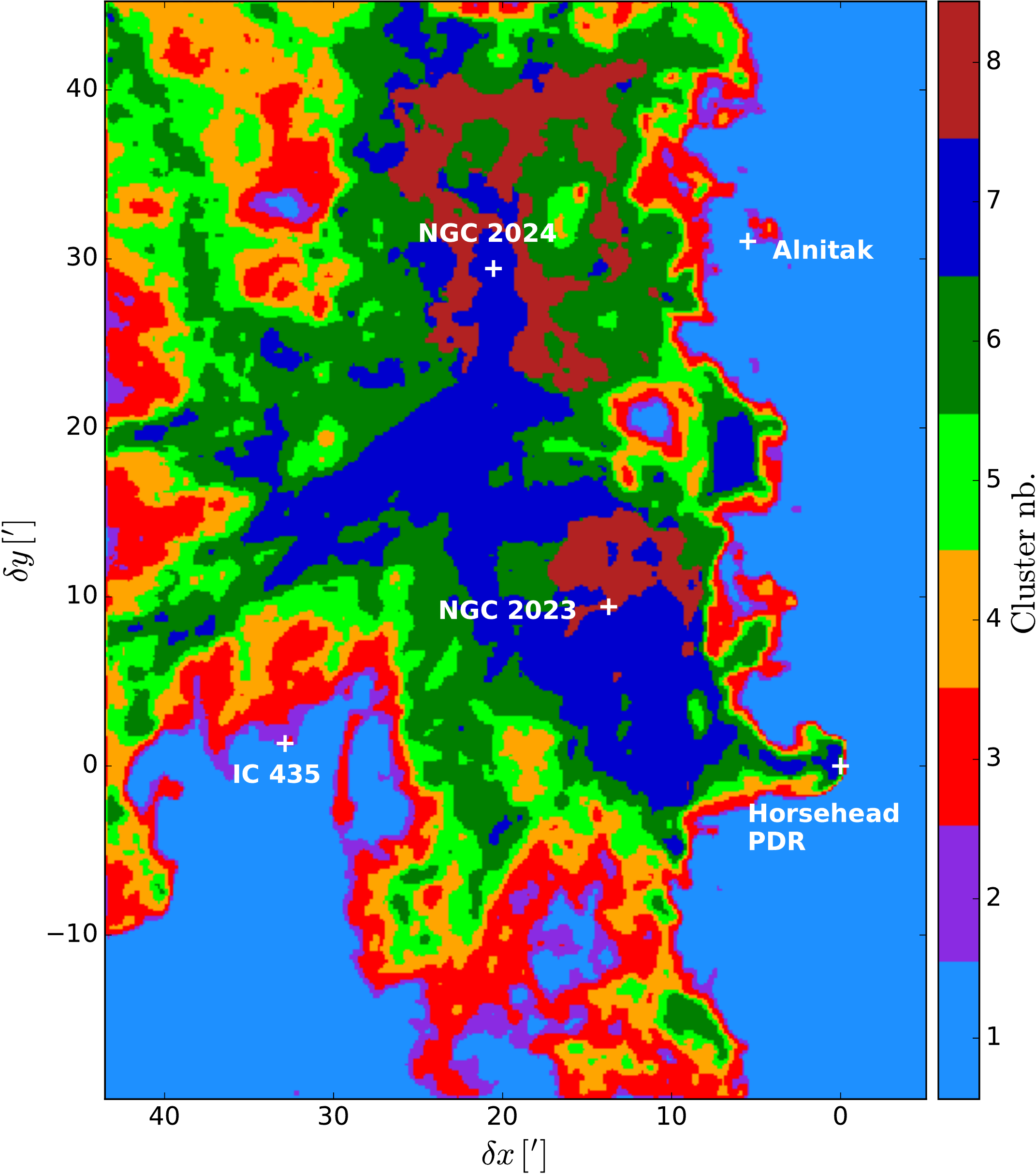}
    \caption{Map of the clusters based on the \twCO, \thCO, and \CeiO{}
      \Jone{} line integrated intensities. White crosses mark the positions
      of remarkable regions. \emph{Left:} Raw results. \emph{Right:}
      Clusters CO-0 and CO-1 are merged into a single cluster named 1.}
    \label{fig:clusterCO_map}
  \end{figure*}
}

\newcommand{\FigClusterCOCompression}{%
  \begin{figure*}
    \centering %
    \includegraphics[width=0.89\linewidth]{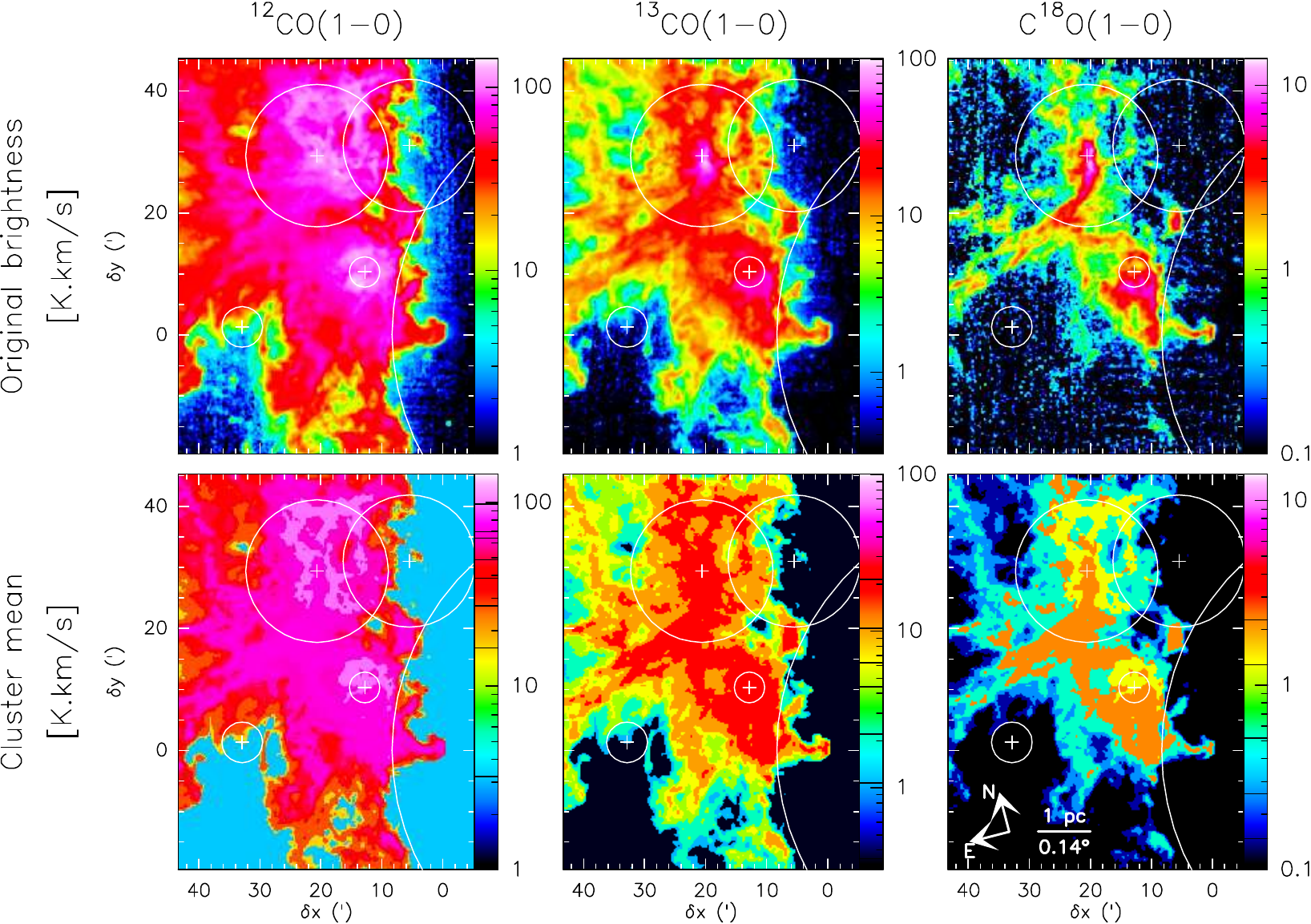}
    \caption{Comparison of the original intensity spatial distribution (top
      row) with the mean intensity computed for each cluster separately
      (bottom row). The colour scales are identical for the top and bottom
      rows, but they differ from one column to another. The levels shown on
      the bottom colour lookup table represents the mean values of the
      cluster intensities. The clusters were defined using the \twCO{},
      \thCO{} and \CeiO{} (1-0) lines. The circles show the typical
      extensions of the H\textsc{ii} regions and the crosses show the
      position of the associated exciting stars~\citep[see][for
      details]{Pety2017}.}
    \label{fig:clusterCO_compression}
  \end{figure*}
}

\newcommand{\FigCOlustersTwoDPDF}{
  \begin{figure*}
    \centering %
    \includegraphics[width=0.49\linewidth]{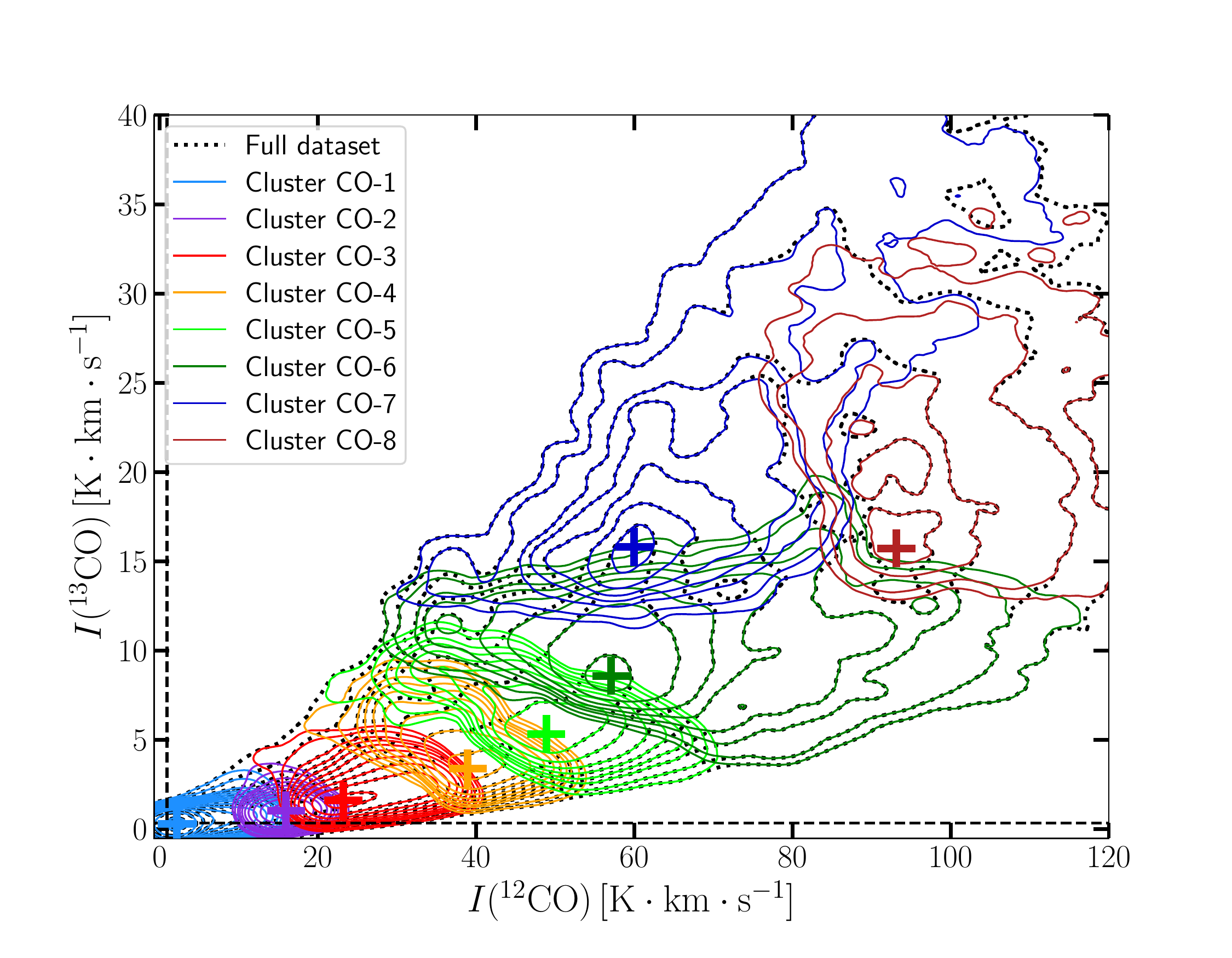}
    \includegraphics[width=0.49\linewidth]{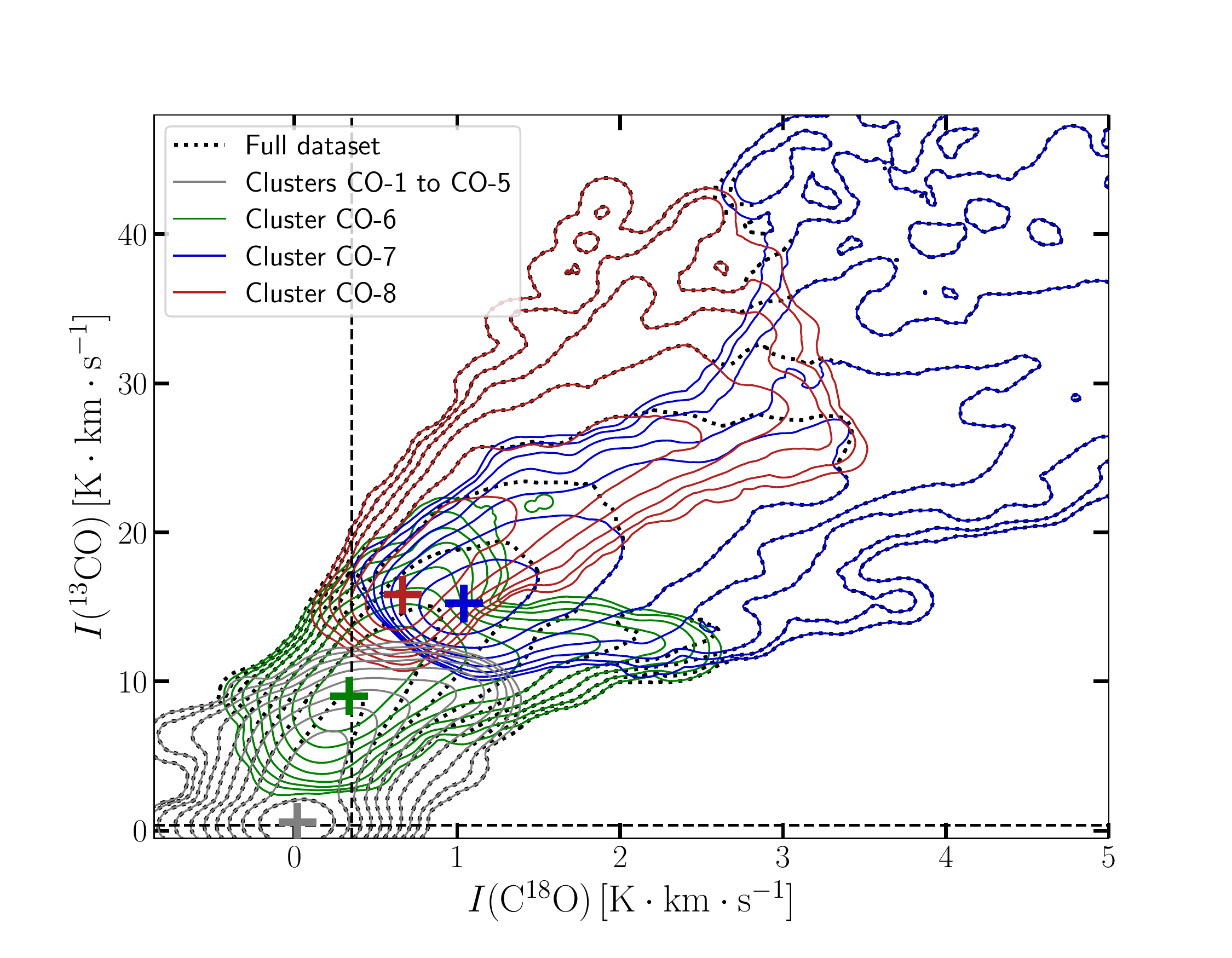}
    \caption{Contour plot of the 2D PDFs of $^{13}$CO vs. $^{12}$CO (left),
      and $^{13}$CO vs. C$^{18}$O (right). The PDFs of the total dataset
      are shown as black dotted contours, while the PDFs of the individual
      clusters are shown as solid contours coloured according to the colour
      coding of clusters in Fig. \ref{fig:clusterCO_map}. On the right
      panel clusters CO-1 to CO-5 have been grouped (grey contours) for
      better readability. The thin vertical and horizontal dashed lines
      show the median 4$\sigma$ noise levels, while the coloured crosses
      show the positions of the PDF maxima for each cluster.}
    \label{fig:COclusters_2D_PDFs}
  \end{figure*}
}

\newcommand{\FigCOClustersCOPDF}{
  \begin{figure}
    \centering %
    \includegraphics[width=1\linewidth]{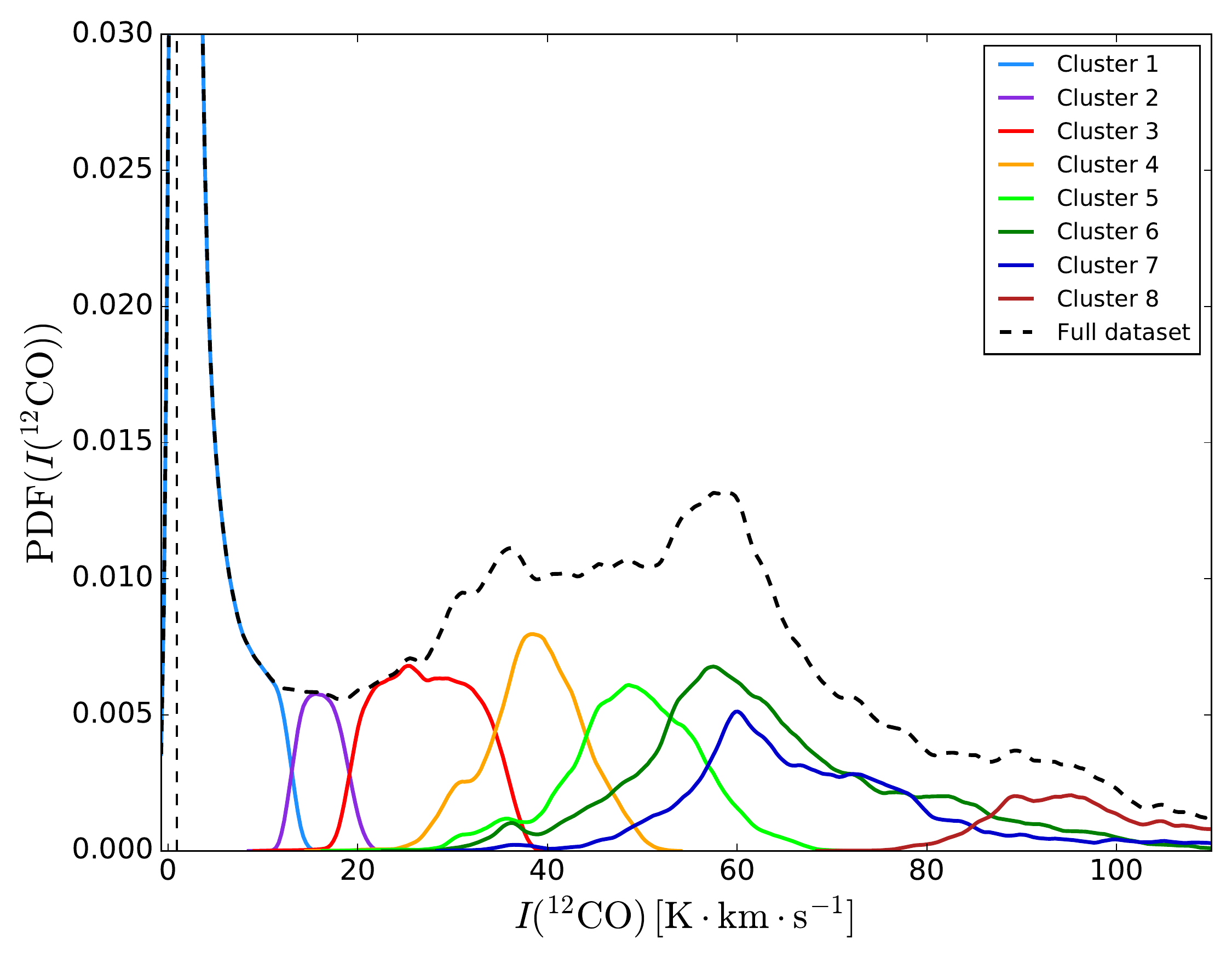}
    \caption{PDF of the $^{12}$CO $J=1-0$ line intensity, comparing the PDF
      of the total dataset (dashed) to the contributions of the different
      clusters (solid contours coloured according to the colour coding of
      clusters in Fig. \ref{fig:clusterCO_map}). The thin vertical dashed
      line shows the median $4\sigma$ noise level.}
    \label{fig:COclusters_12CO_PDF}
  \end{figure}
}

\newcommand{\FigCOClustersDensityPDF}{
  \begin{figure}
    \centering %
    \includegraphics[width=1\linewidth]{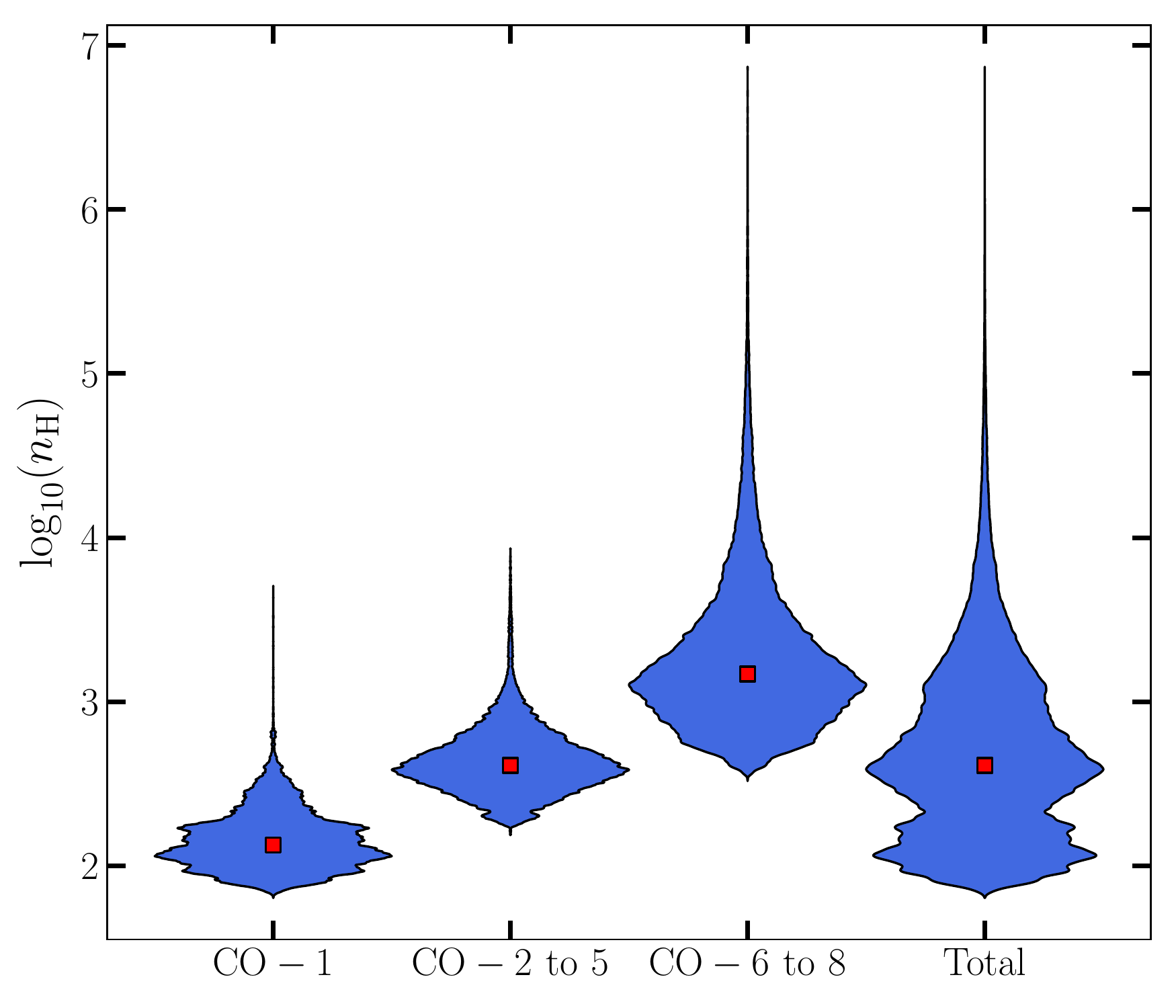}
    \caption{Violin plot showing the PDF of the approximate gas volume
        density $n_H$, comparing the contributions of the three groups of
      CO clusters discussed in the text (CO-1, CO-2 to 5 and CO-6 to 8) to
      the PDF of the total data set.}
    \label{fig:COclusters_nH_PDF}
  \end{figure}
}

\newcommand{\FigCOClustersGProperties}{
  \begin{figure}
    \centering %
    \includegraphics[width=1\linewidth]{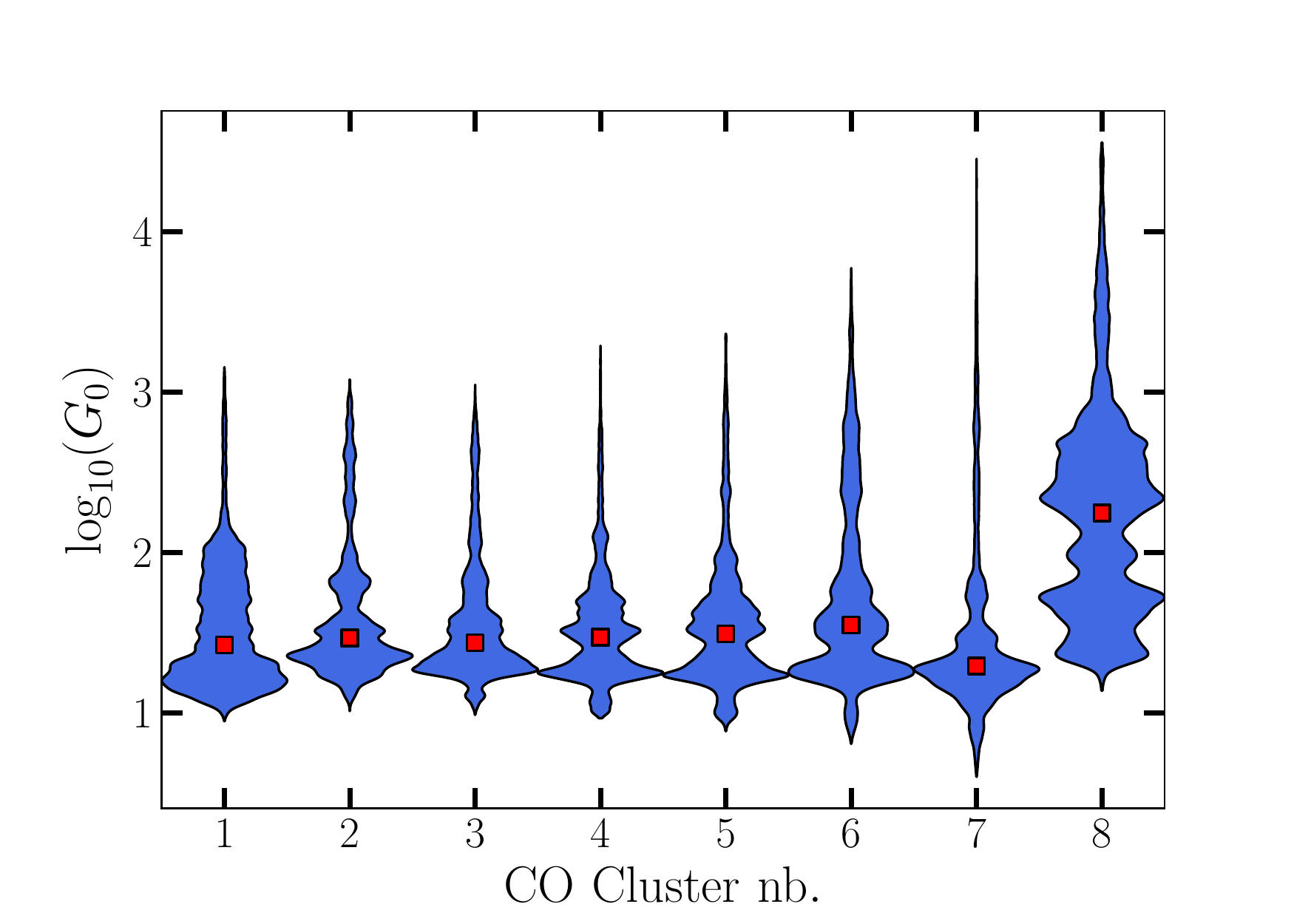}
    \caption{Violin plot showing the PDF of $\log_{10}(G_0)$ for each CO
      cluster (blue profiles) and the median value in each cluster (red
      squares).}
    \label{fig:ClustersCO_G0_properties}
  \end{figure}
}

\newcommand{\FigClusterFUVMapRaw}{
  \begin{figure*}
    \includegraphics[height=10cm]{figures/Cluster_CO_map.pdf}
    \hfill{}
    \includegraphics[height=10cm]{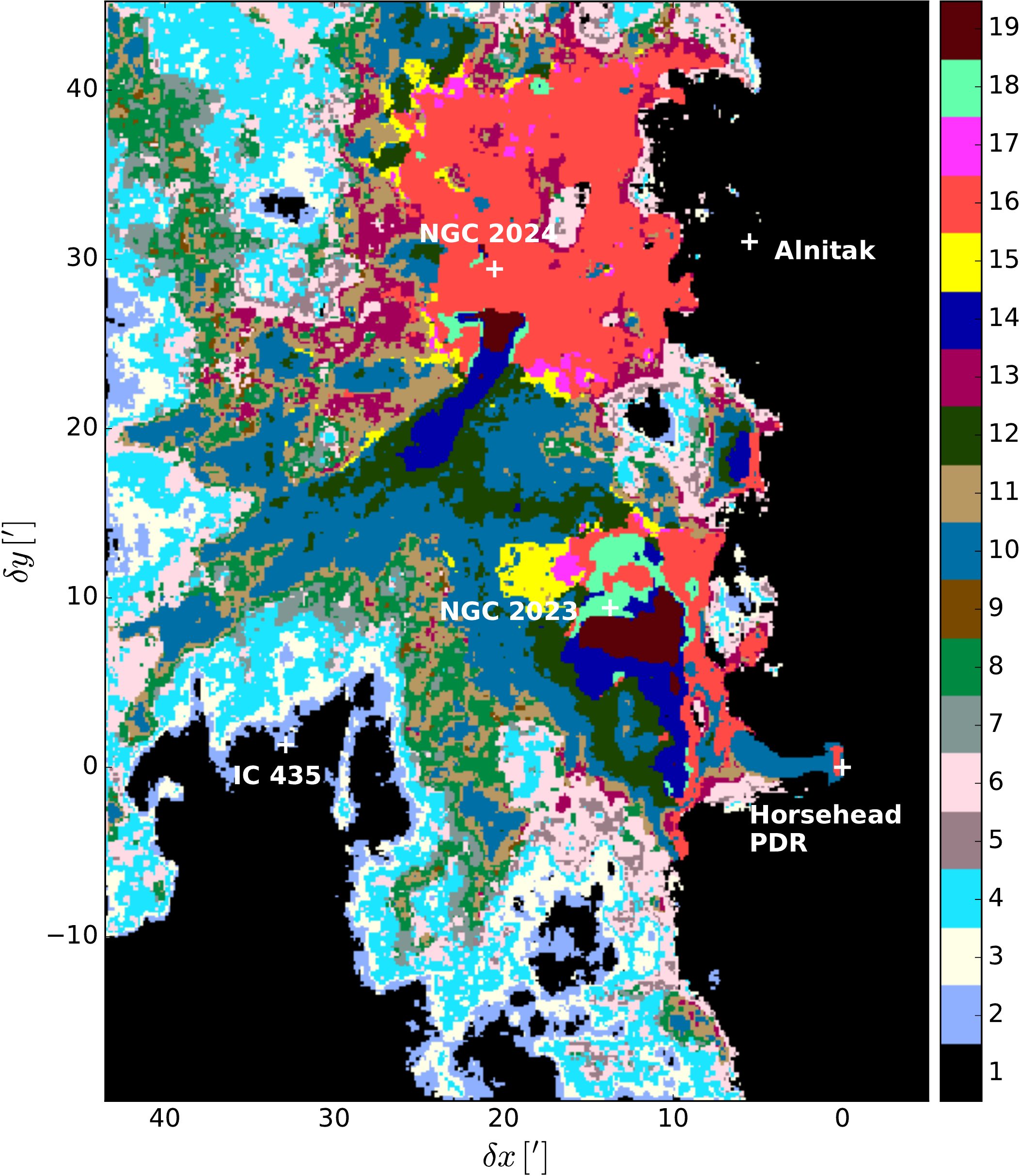}
    \caption{Comparison of the clusters obtained based on the CO
      isotopologues alone (left) and the clusters obtained when associating
      the CO isotopologues to HCO$^+$ and CN (right). In both cases, the
      clusters have been ordered by increasing \twCO{} \Jone{} mean
      intensity. White crosses mark the positions of remarkable regions.}
    \label{fig:cluster_bis_map_raw}
  \end{figure*}
}

\newcommand{\FigClusterFUVCompression}{%
  \begin{figure*}
    \centering %
    \includegraphics[width=\linewidth]{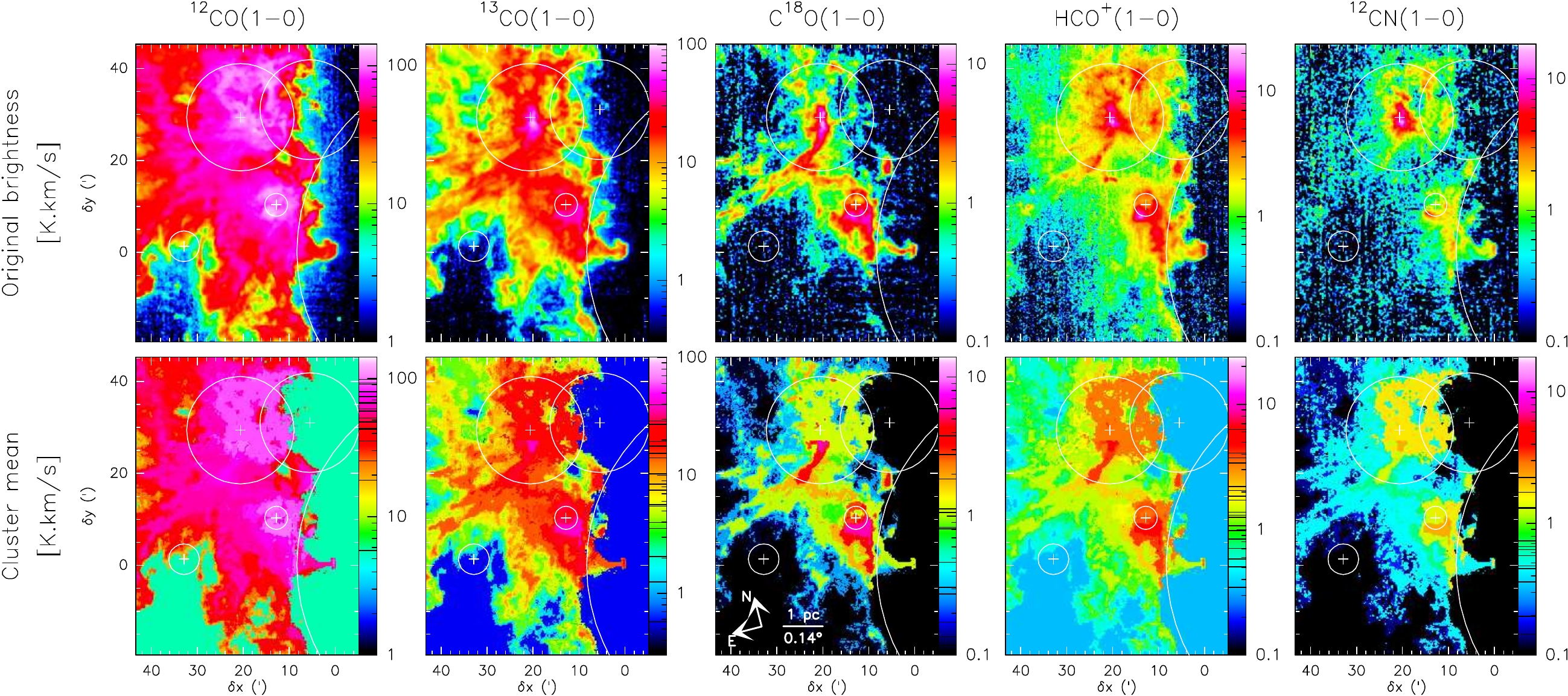}
    \caption{Same as Fig.~\ref{fig:clusterCO_compression}, except that
      clusters were defined on the \twCO, \thCO, \CeiO, CN, and
      \chem{HCO^+} (1-0) lines.}
    \label{fig:clusterFUV_compression}
  \end{figure*}
}

\newcommand{\FigClusterFUVMapGroupedHCOp}{
  \begin{figure*}
    \centering
    \includegraphics[width=0.47\linewidth]{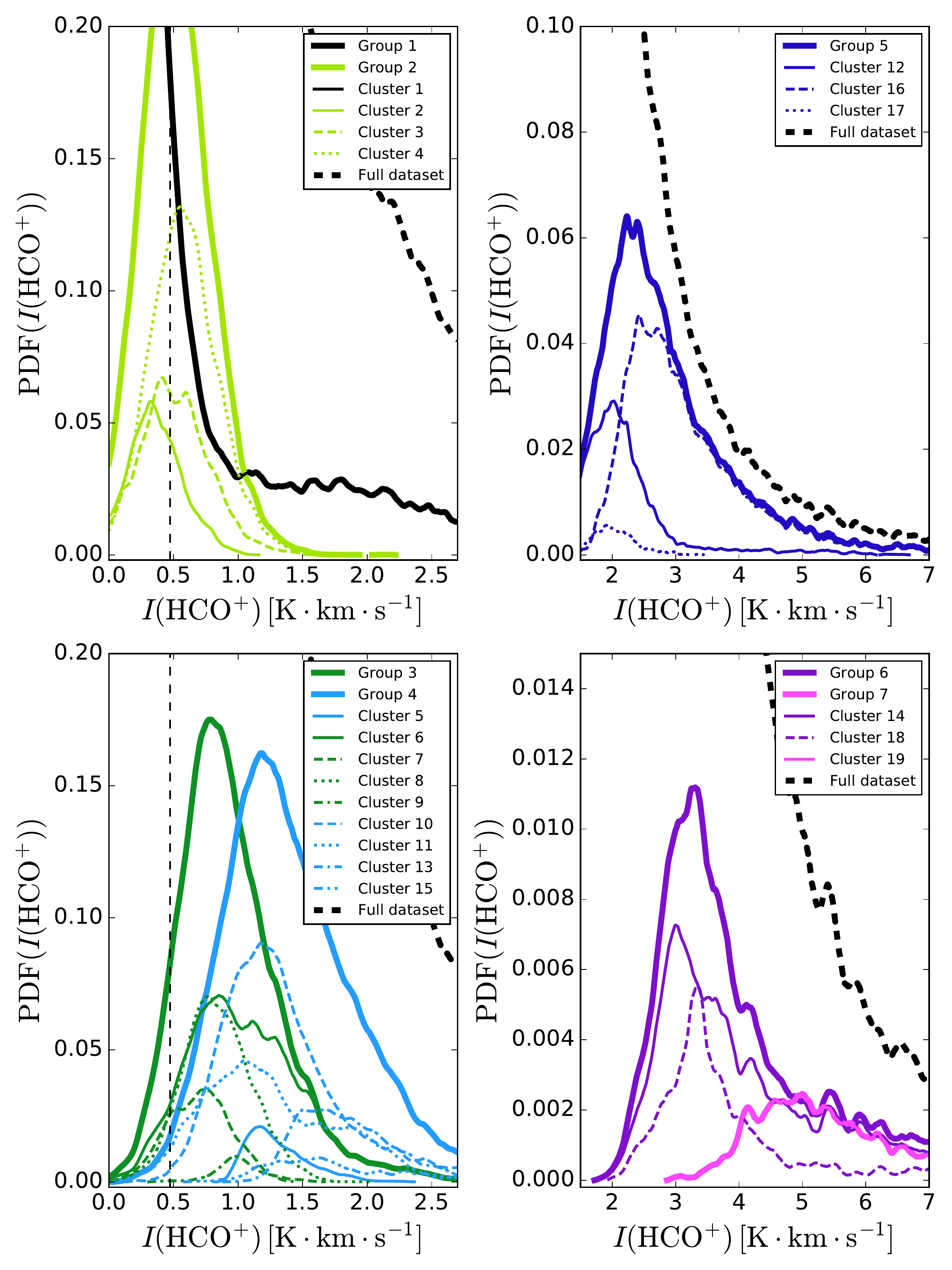}
    \includegraphics[width=0.52\linewidth]{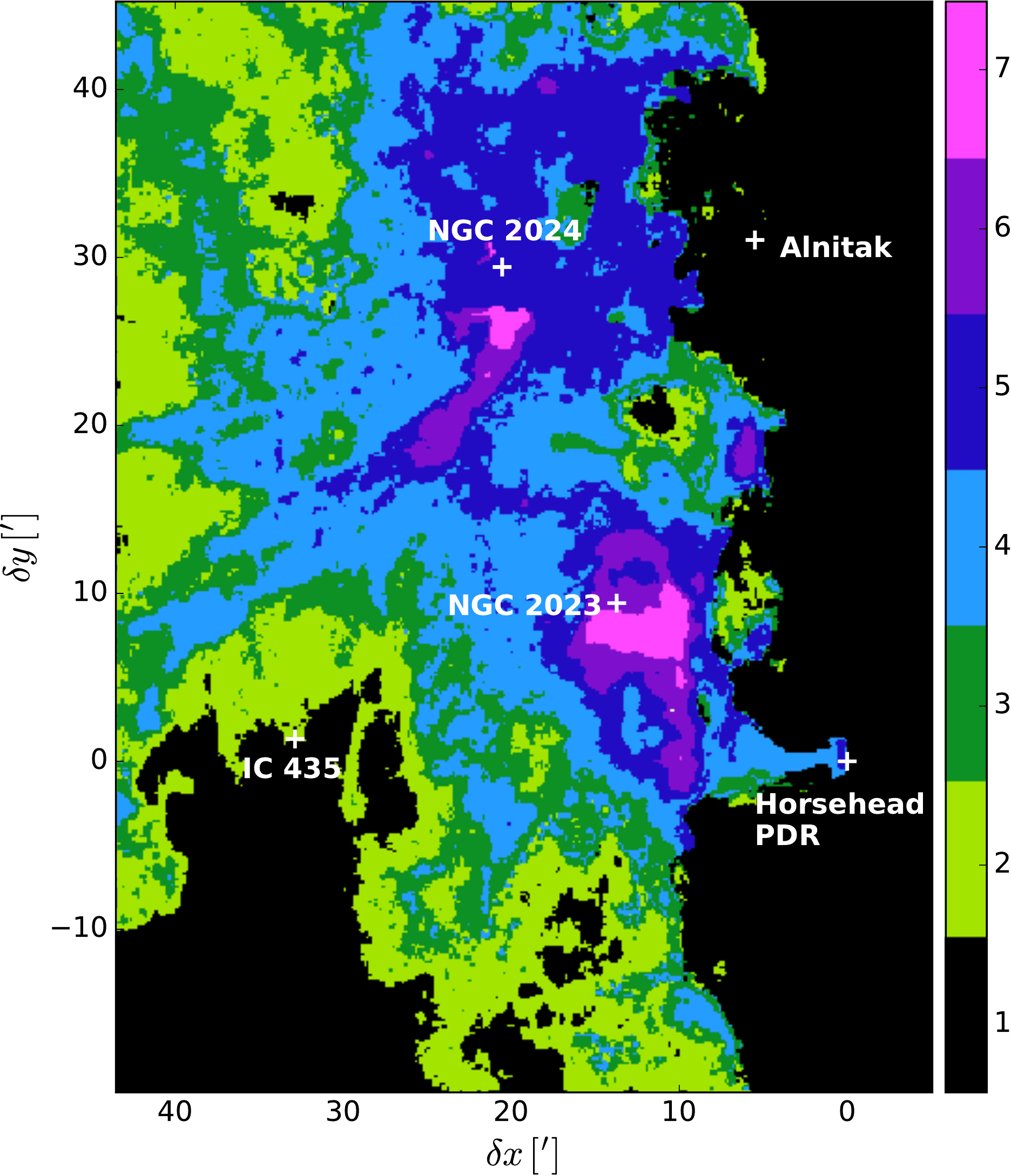}
    \caption{\textbf{Left:} 1D PDF of the HCO+ $(1-0)$ line intensity,
      comparing the full dataset PDF (thick dashed black line), the
      contribution of each of the groups defined in
      Sect. \ref{sec:HCOp-groups} (coloured thick lines), and the
      contribution of each individual cluster (thin coloured lines). The
      cluster contributions are coloured according to the group to which
      they belong. For readability, we have separated groups HCO$^+$-1 to
      HCO$^+$-4 and their constitutive clusters (left panels) and groups
      HCO$^+$-5 to HCO$^+$-7 and their constitutive clusters (right
      panels).  \textbf{Right:} Map of the 7 groups resulting from the
      grouping of consecutive clusters described in the text
      (Sect. \ref{sec:HCOp-groups}).}
    \label{fig:cluster_bis_map_HCOp_grouped}
  \end{figure*}
}

\newcommand{\FigClustersFUVTwoDPDFHCOpGrouping}{
  \begin{figure*}
    \includegraphics[width=0.49\linewidth]{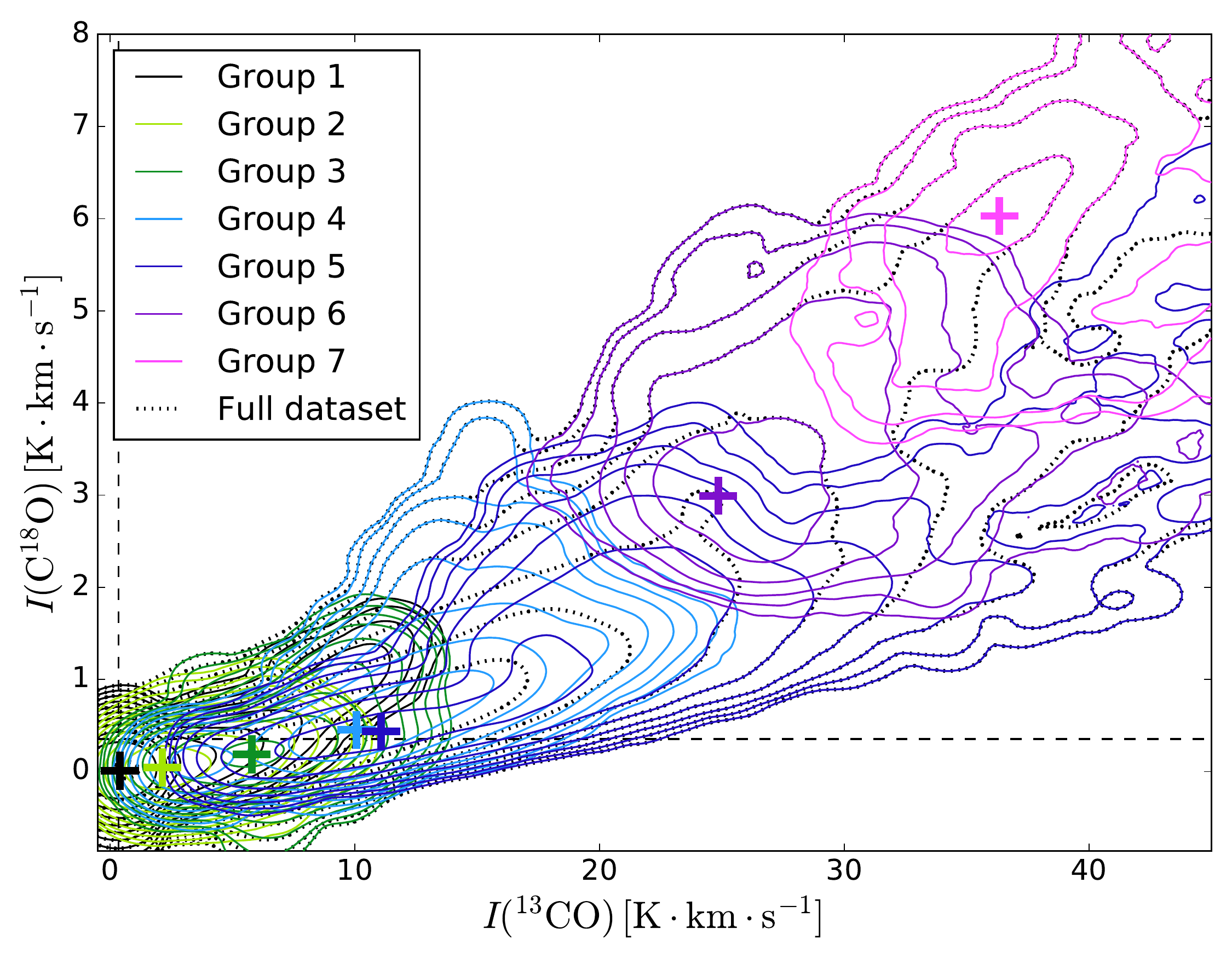}
    \hfill{} %
    \includegraphics[width=0.49\linewidth]{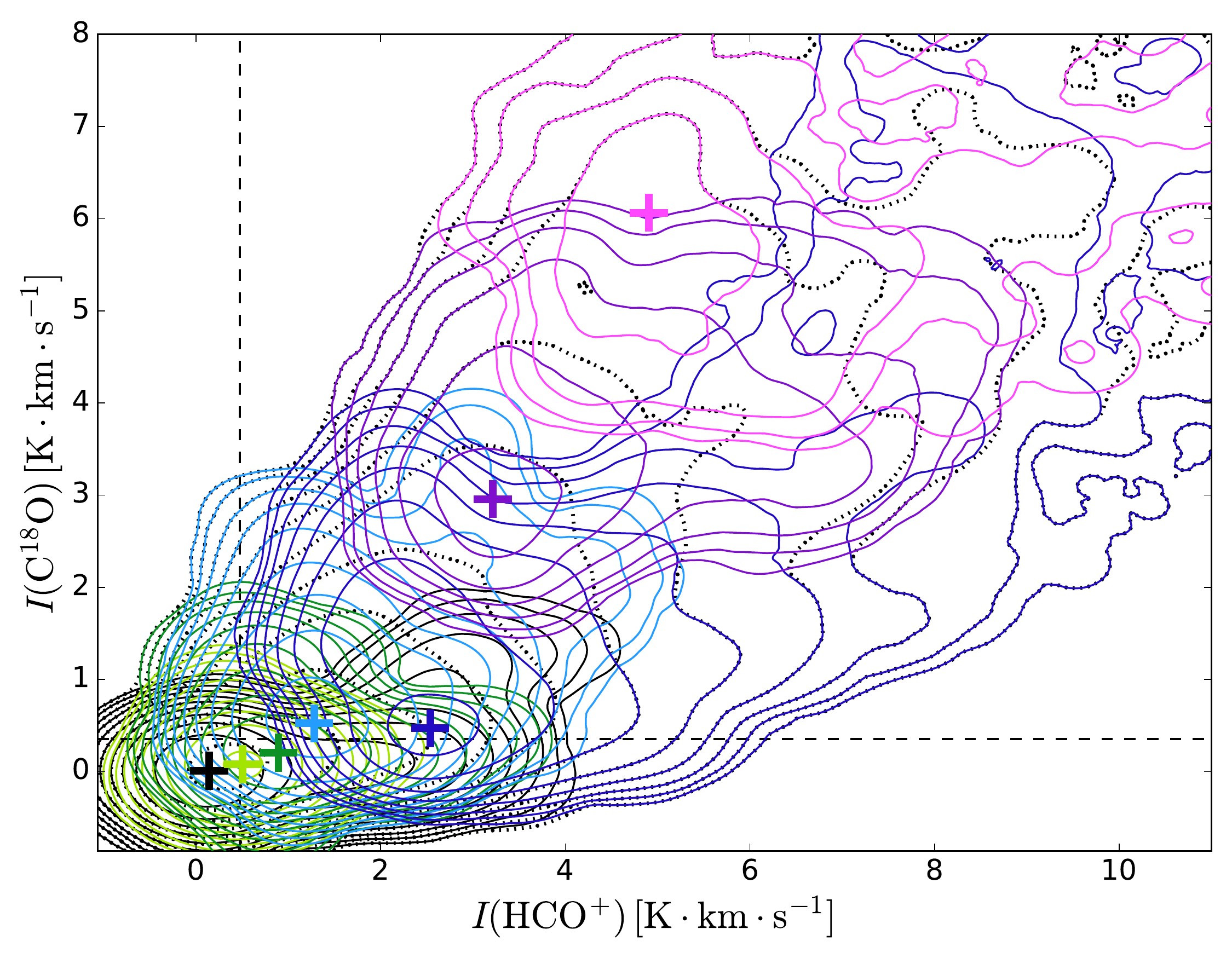}
    \caption{Contour plot of the 2D PDFs of C$^{18}$O vs. $^{13}$CO (left)
      and C$^{18}$O vs. HCO$^+$ (right). The PDFs of the total dataset are
      shown as black dotted contours. The contributions of the 7 groups
      resulting from the grouping discussed in the text are shown in
      contours coloured according to
      Fig. \ref{fig:cluster_bis_map_HCOp_grouped} (right). In addition, the
      PDF maximum of each group is shown as a cross with the same colour as
      the group.}
    \label{fig:clusters_bis_2D_PDFs_HCOp_grouping}
  \end{figure*}
}

\newcommand{\FigClustersFUVdensityPDFHCOpGrouping}{
  \begin{figure}
    \includegraphics[height=7cm]{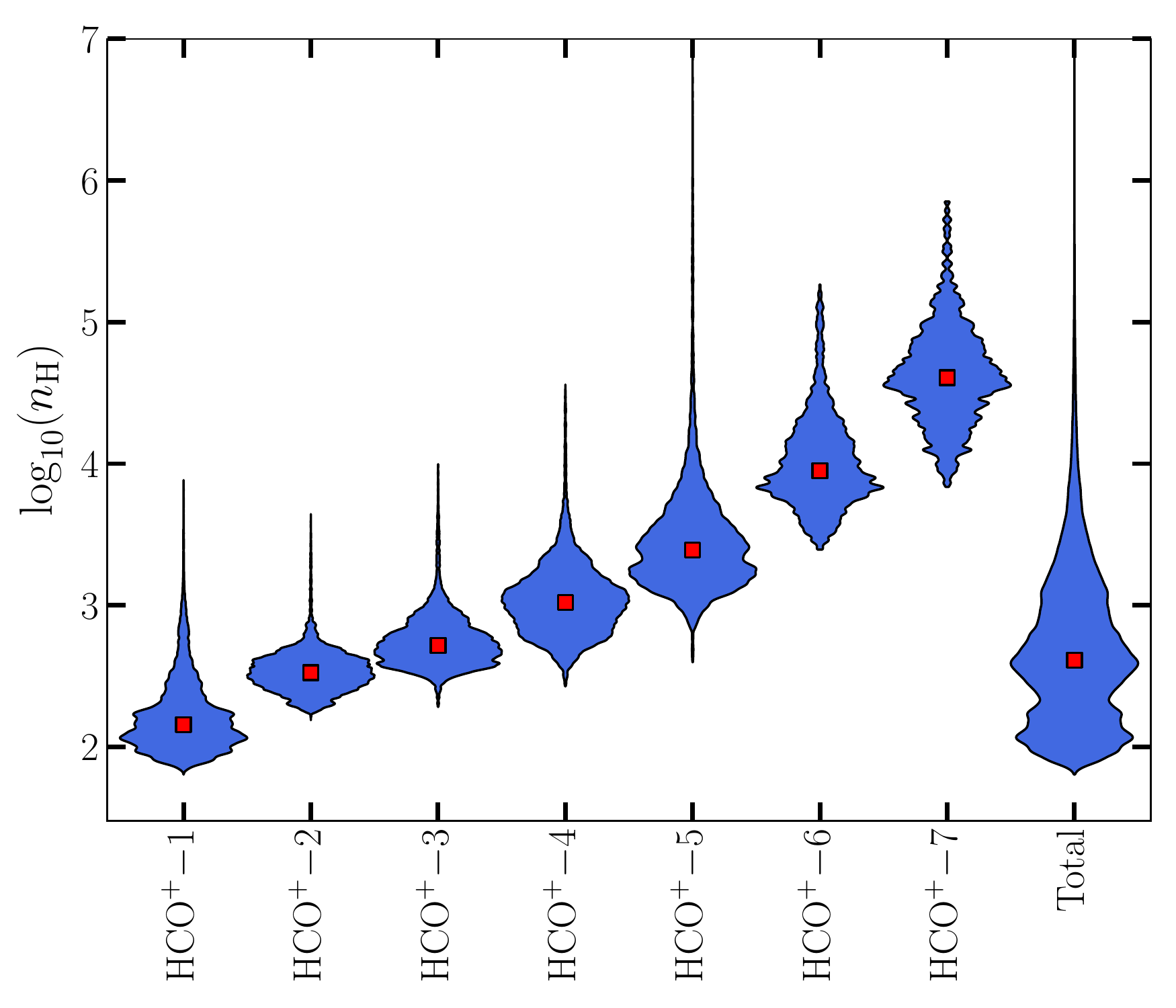}
    \caption{Violin plots showing the PDF (blue profiles) and median values
      (red squares) of the approximate volume density $n_\mathrm{H}$
      for each HCO$^{+}$-group, and for the full map (label "Total").}
    \label{fig:clusters_fuv_density_PDF_HCOp_grouping}
  \end{figure}
}

\newcommand{\FigClustersFUVGzViolinHCOpGrouping}{
  \begin{figure}
    \centering{} %
    \includegraphics[width=\linewidth]{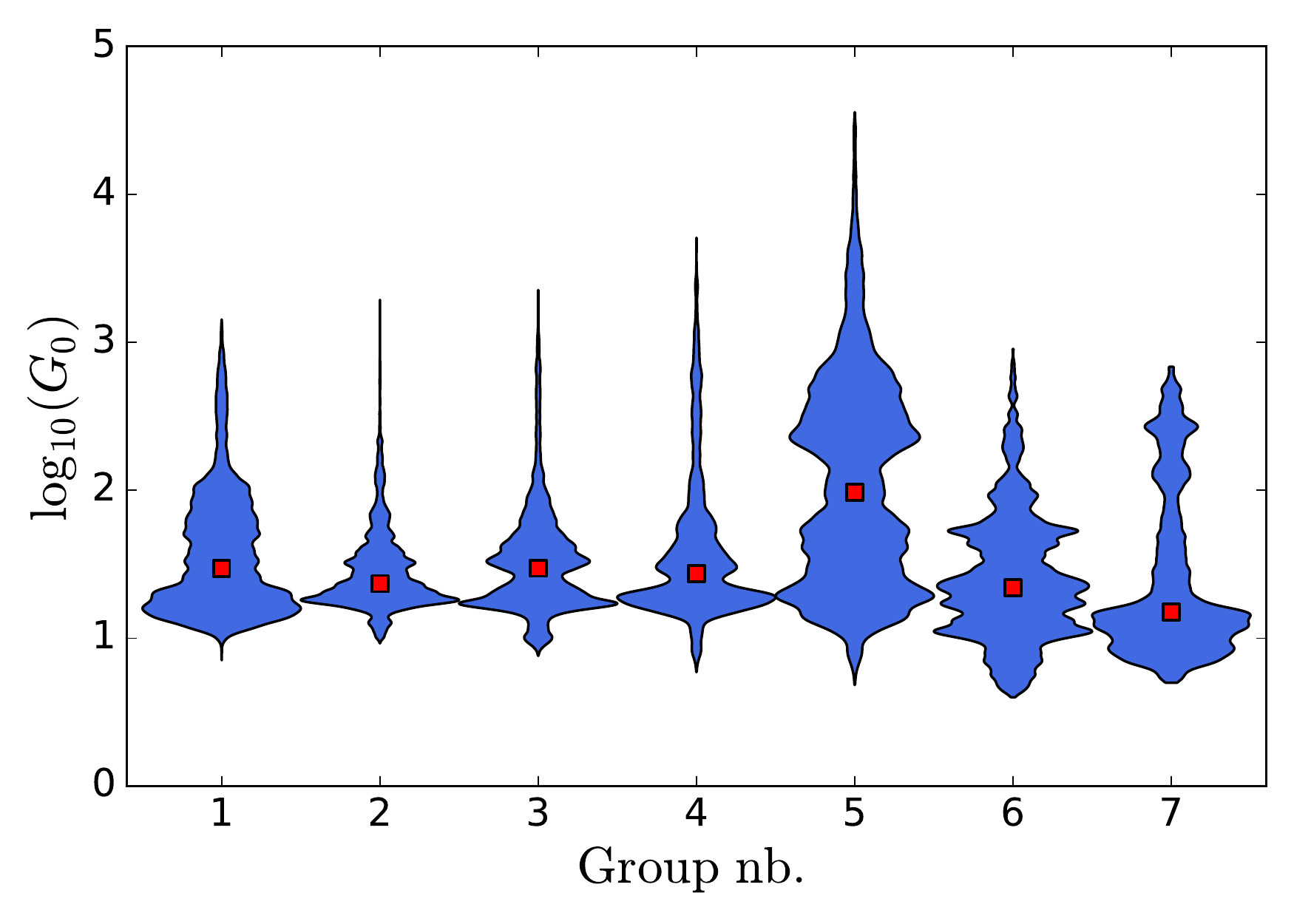}
    \caption{Violin plots showing the PDF (blue profiles) and median values
      (red squares) of the FUV illumination $G_0$ for each
      HCO$^{+}$-group.}
    \label{fig:clusters_fuv_G0_violins_HCOp_grouping}
  \end{figure}
}

\newcommand{\FigClusterFUVMapGroupedCN}{
  \begin{figure*}
    \centering
    \includegraphics[width=0.46\linewidth]{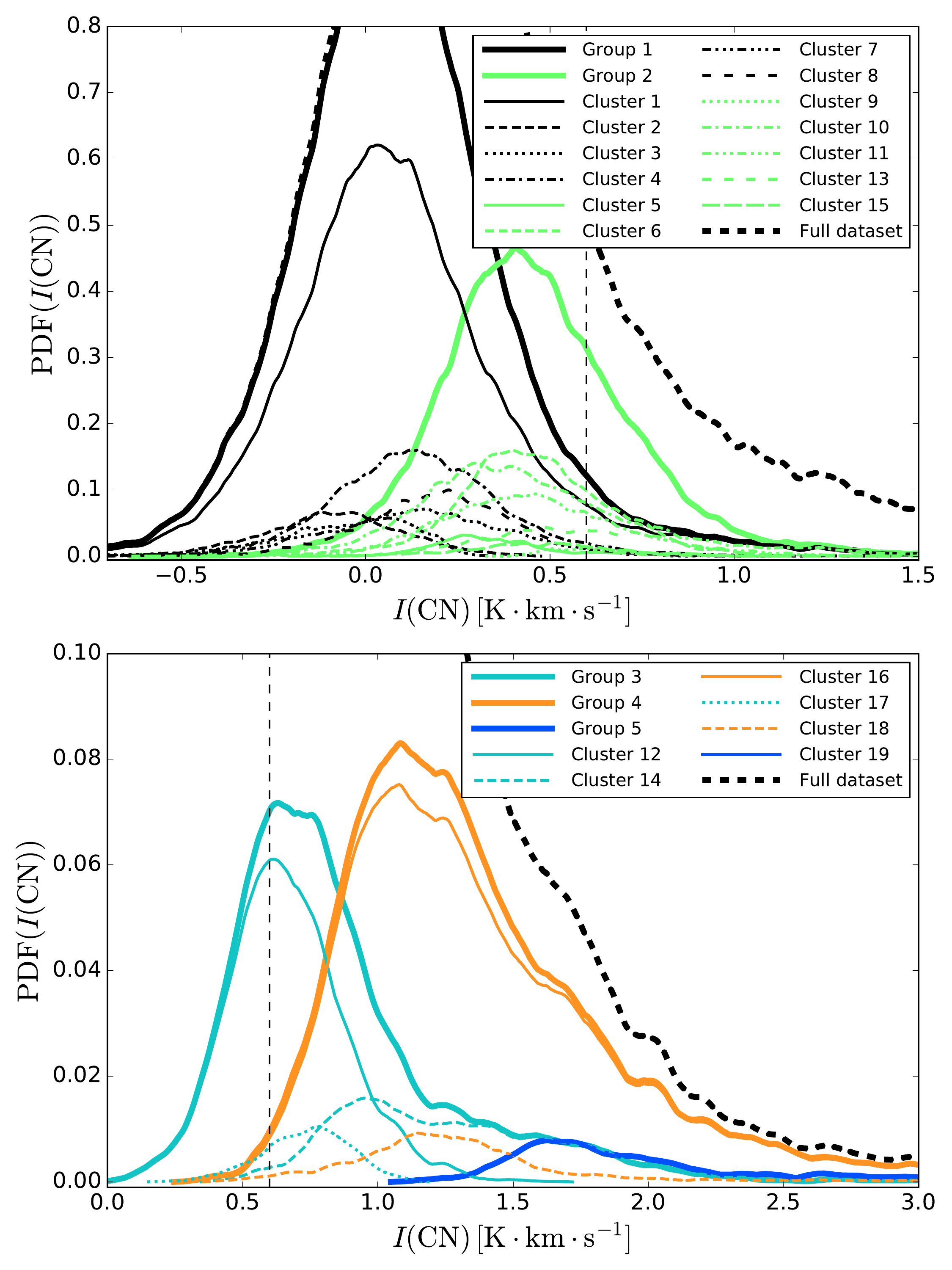}
    \includegraphics[width=0.53\linewidth]{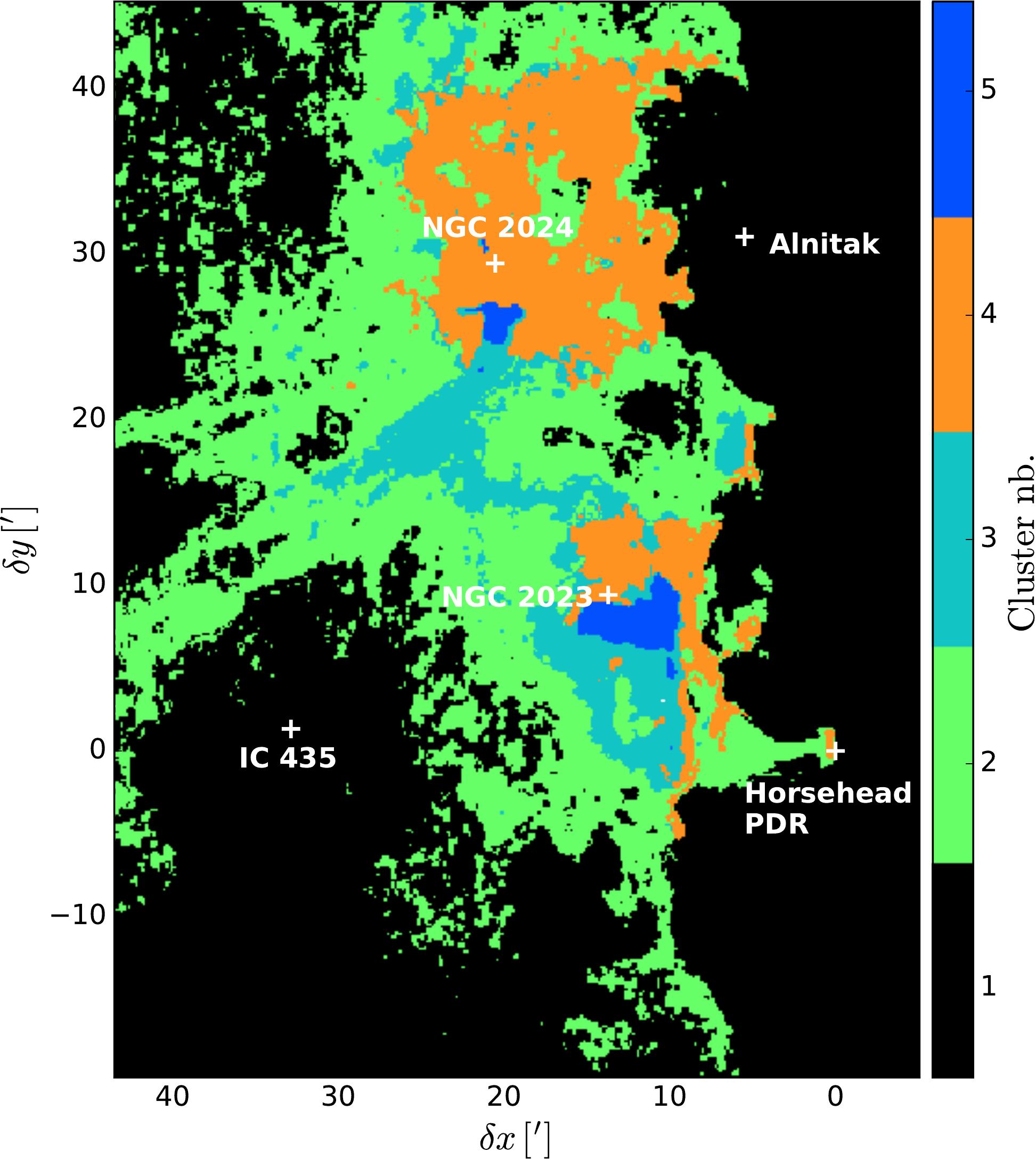}
    \caption{\textbf{Left:} 1D PDF of CN $(1-0)$ line intensity, comparing
      the full dataset PDF (thick dashed black line), the contribution of
      each of the groups defined in Sect. \ref{sec:CN-groups} (coloured
      thick lines), and the contribution of each individual cluster (thin
      coloured lines). The contributions of clusters are coloured according
      to the group to which they belong. For readability, we have separated
      groups CN-1 and CN-2 and their constitutive clusters (top panel) and
      groups CN-3 to CN-5 and their constitutive clusters (bottom panel).
      \textbf{Right:} Map of the 5 groups resulting from the grouping of
      consecutive clusters described in the text
      (Sect. \ref{sec:CN-groups}).}
    \label{fig:cluster_bis_map_CN_grouped}
  \end{figure*}
}

\newcommand{\FigClustersFUVTwoDPDFCNGrouping}{
  \begin{figure*}
    \centering %
    \includegraphics[width=.49\linewidth]{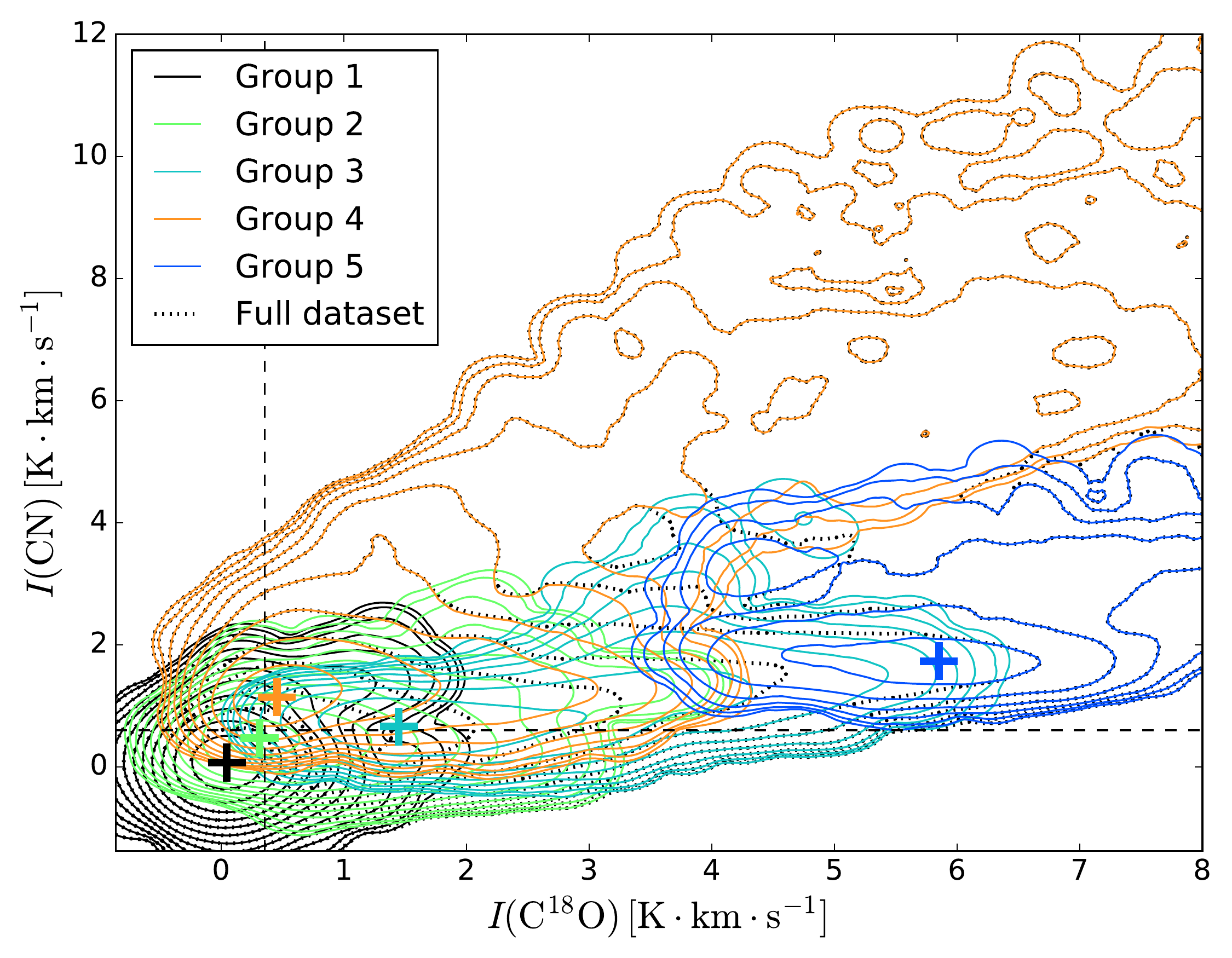}
    \includegraphics[width=.49\linewidth]{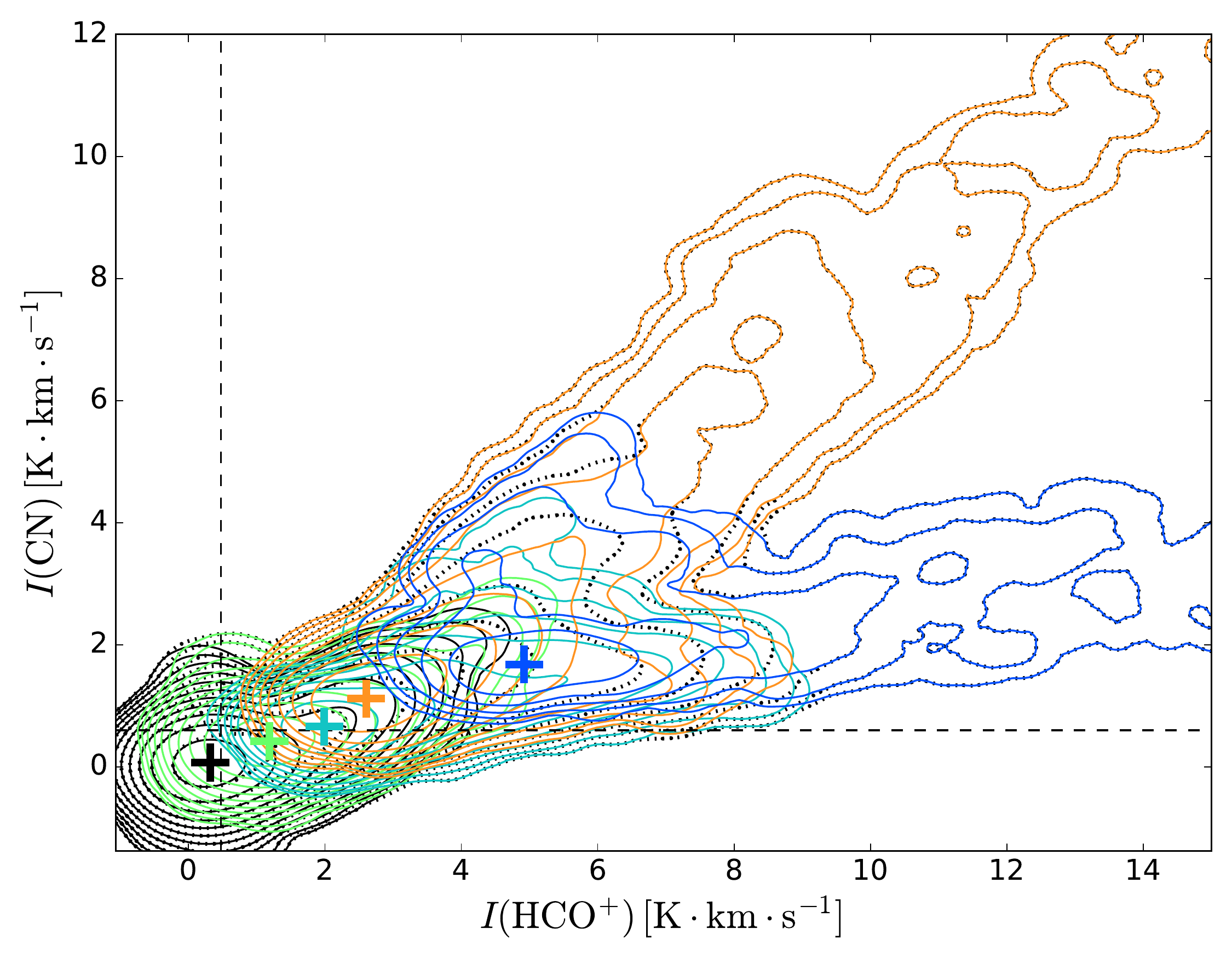}
    \caption{Contour plot of the 2D PDFs of CN vs. C$^{18}$O (left) and CN
      vs. HCO$^+$ (right). The PDFs of the total dataset are shown as black
      dotted contours. The contribution of the 5 groups resulting from the
      grouping discussed in the text are shown coloured according to
      Fig. \ref{fig:cluster_bis_map_CN_grouped} (right). In addition, the
      PDF maximum of each group is shown as a cross with the same colour as
      the group.}
    \label{fig:clusters_bis_2D_PDFs_CN_grouping}
  \end{figure*}
}

\newcommand{\FigClustersFUVDustViolinsCNGrouping}{
  \begin{figure*}
    \centering %
    \includegraphics[width=.49\linewidth]{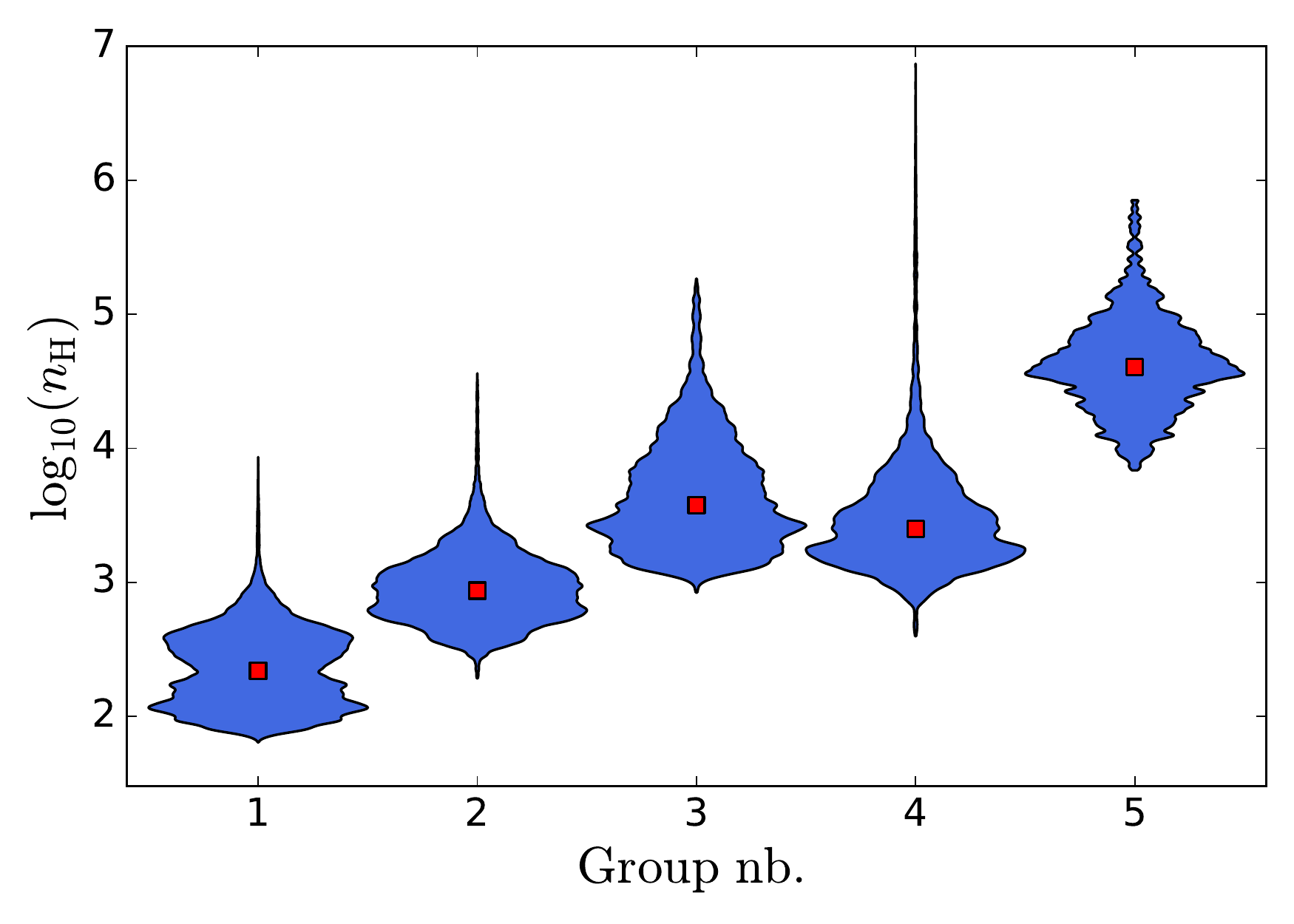}
    \includegraphics[width=.49\linewidth]{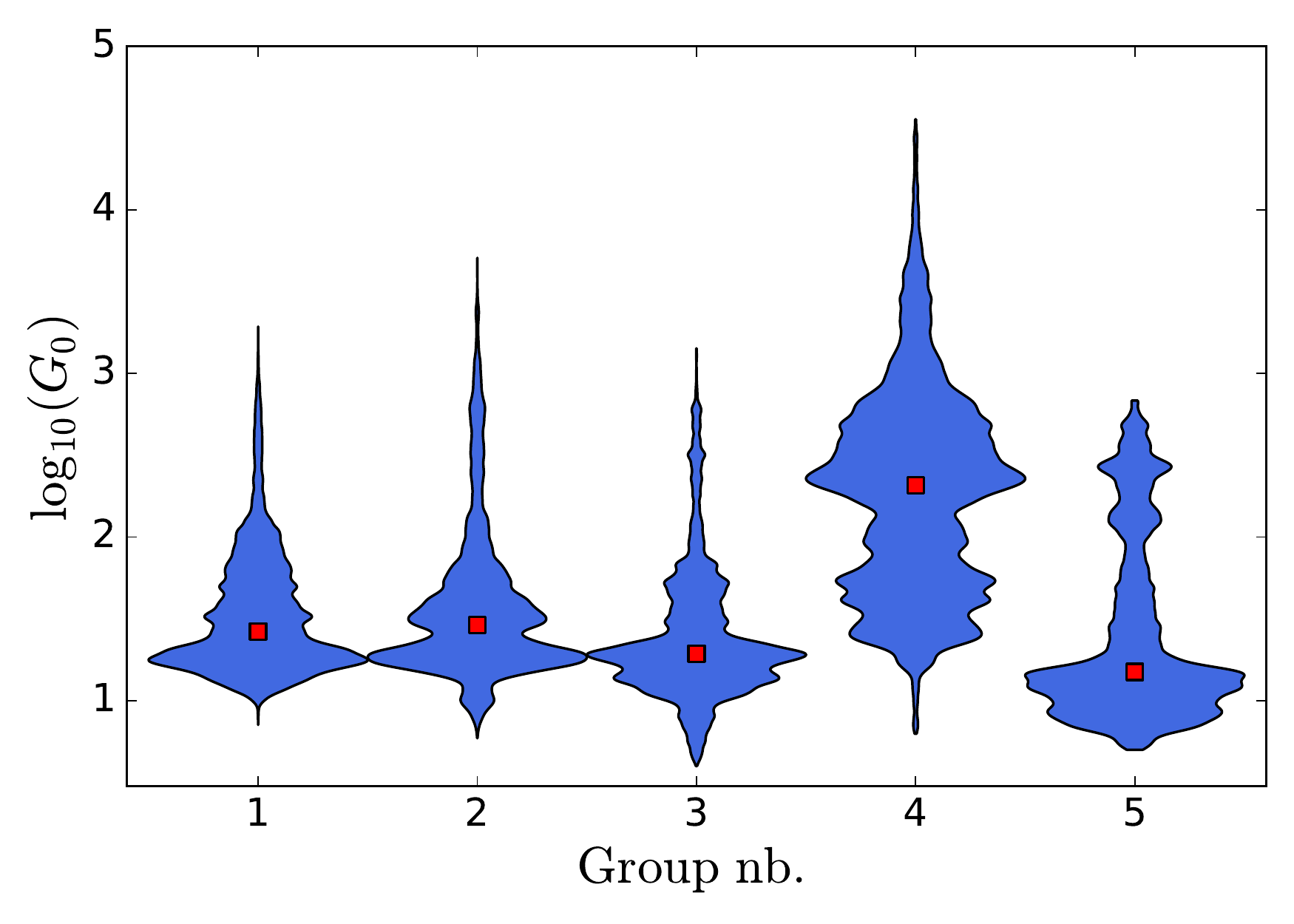}
    \caption{Violin plots showing the PDF (blue profiles) and median values
      (red squares) of the approximate $n_\mathrm{H}$ (left panel) and $G_0$ (right panel)
      in each of the groups CN-1 to 5.}
    \label{fig:clusters_bis_2D_dust_violins_CN_grouping}
  \end{figure*}
}

\newcommand{\FigCOLTE}{
  \begin{figure*}
    \centering %
    \includegraphics[width=1.0\linewidth]{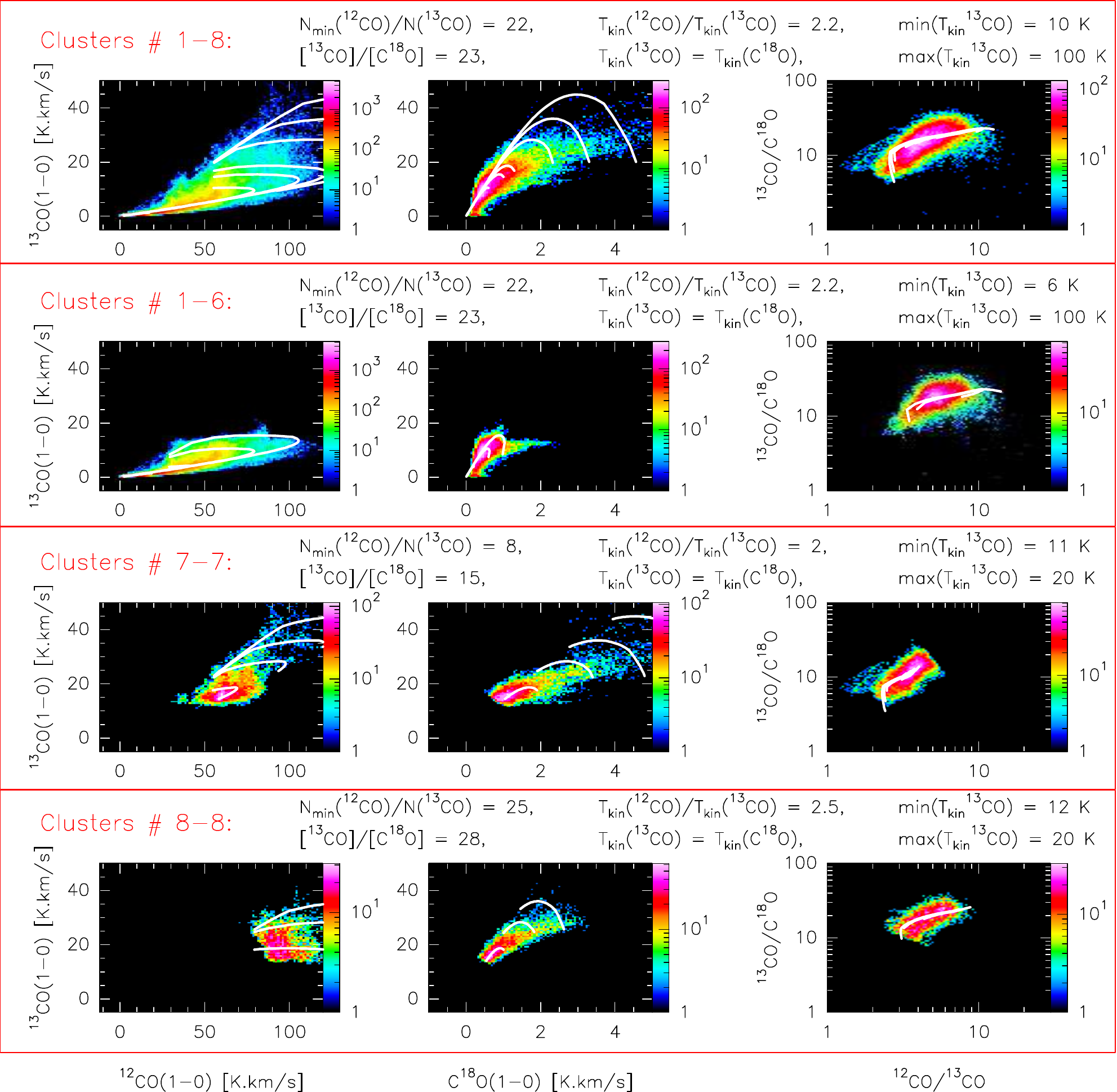}
    \caption{LTE radiative transfer models for the three main CO
      isotopologues. The four rows show the best match between the
      observations and the models for, from top to bottom, the full field
      of view studied here, clusters 1 to 6, and the 7th and 8th
      clusters. The control parameters of the family of models are written
      on top of each associated row. The left and middle columns show the
      joint histogram of the \Jone{} lines of \thCO{} vs. \twCO{}, and of
      \thCO{} vs. \CeiO{}. The right column shows the joint histogram of the
      $\thCO/\CeiO$ vs. $\twCO/\thCO$ intensity ratios. The colour look-up
      tables show the number of lines of sight that fall within a given bin
      of the histogram. The white curves present the LTE intensity
      variations as a function of the \thCO{} kinetic temperature for
      different fixed \thCO{} opacities (0.03, 0.1, 0.3, 0.5, 0.65, 1.2,
      1.75, 2.5).}
    \label{fig:CO:LTE}
  \end{figure*}
}

\newcommand{\FigRadexConstraints}{
  \begin{figure}
    \centering %
    \includegraphics[width=1\linewidth]{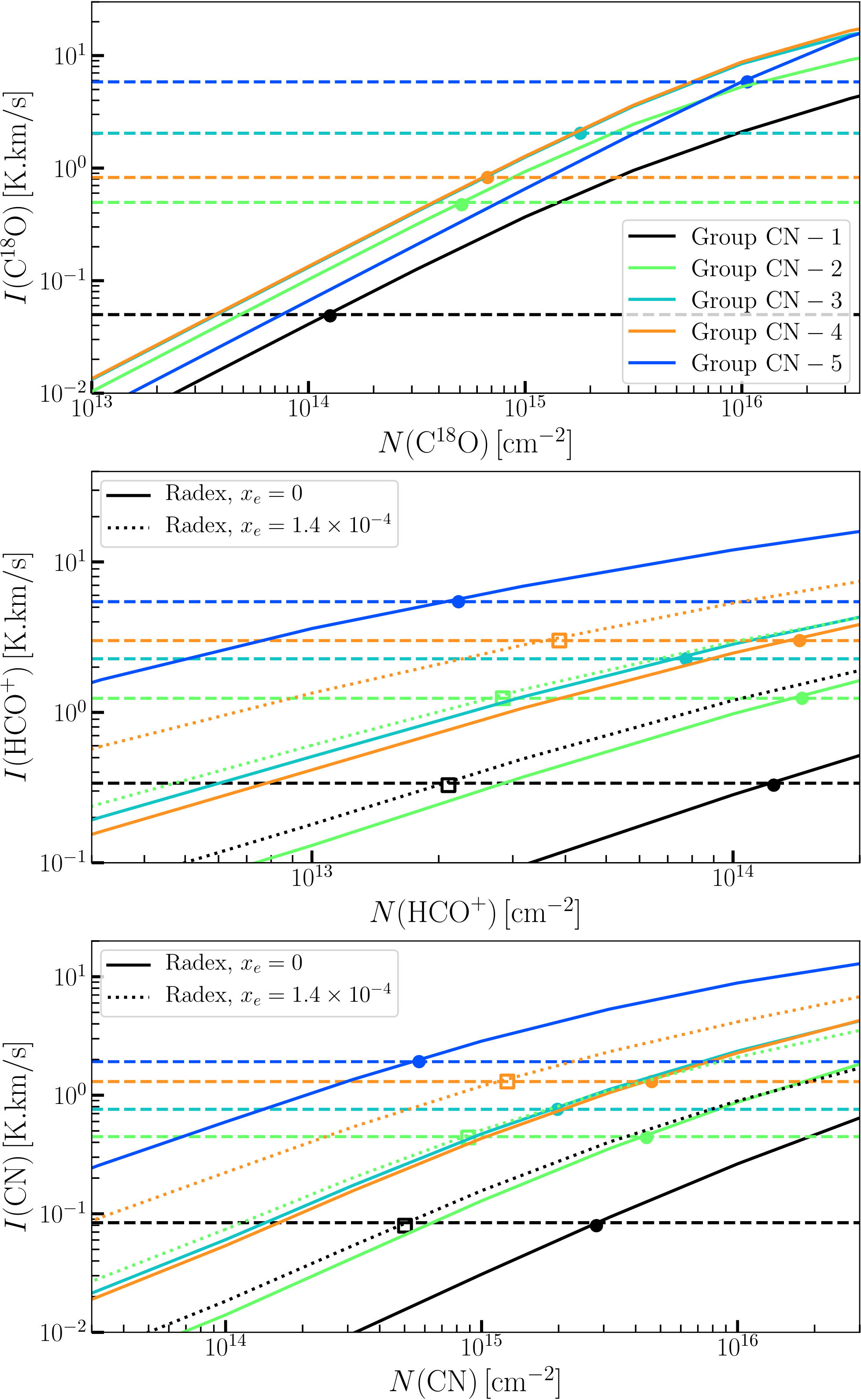}
    \caption{Line intensity as a function of column density for non-LTE
      radiative transfer models at the median gas volume density and
      temperature for each CN group. The full and dotted lines assume a
      ionisation fraction of 0 and $1.4\times10^{-4}$, respectively.  The
      horizontal dashed lines represent the median observed intensity
      computed for each group.  The symbols show the best column density
      for each group according to the RADEX models, with full circles for
      models with $x_e = 0$ and open squares for $x_e = 1.4\times10^{-4}$.}
    \label{fig:RadexConstraints}
  \end{figure}
}

\newcommand{\FigRadexColdenRatios}{
  \begin{figure}
    \centering %
    \includegraphics[width=1\linewidth]{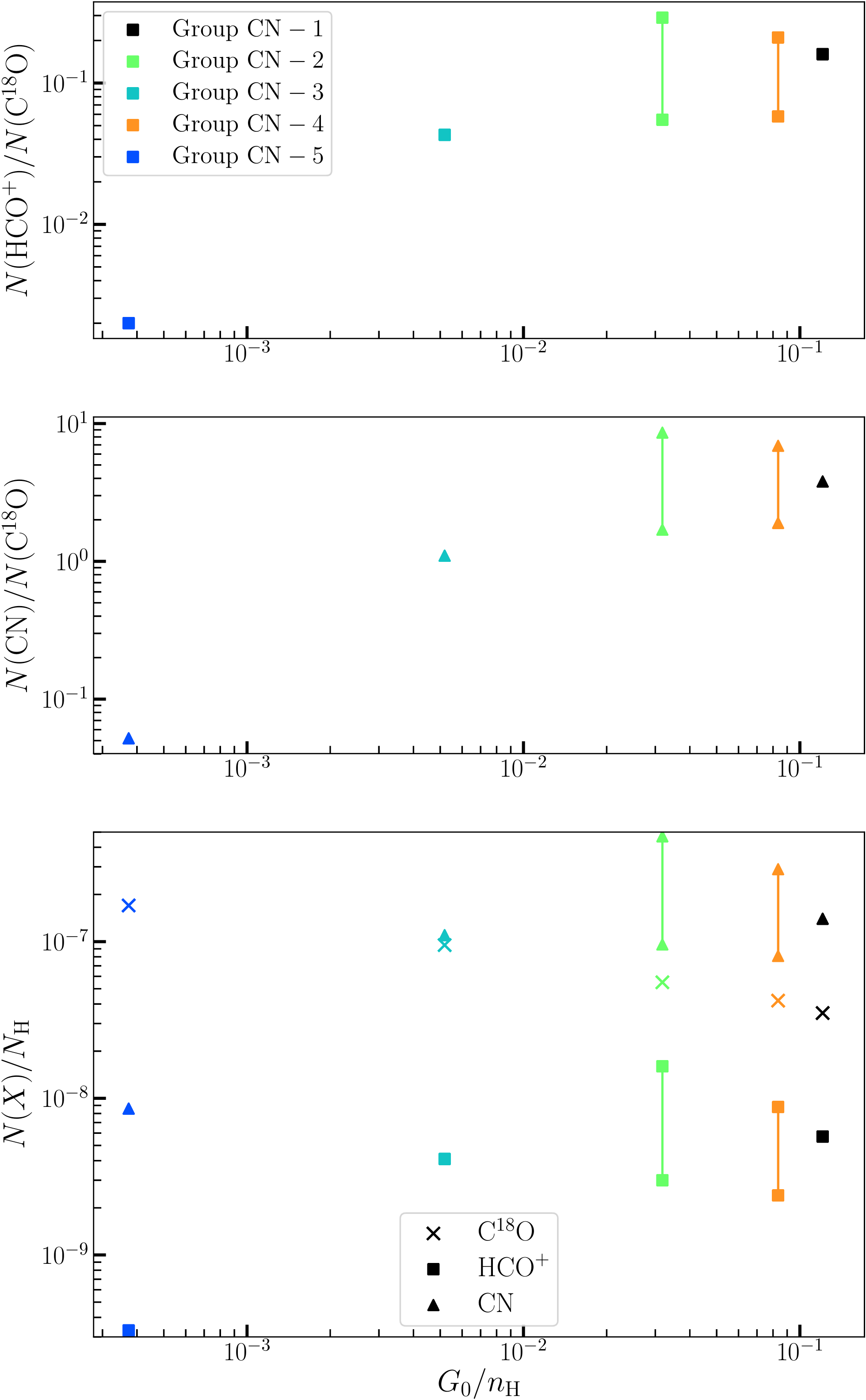}
    \caption{Column density ratios $N($HCO$^+)/N($C$^{18}$O$)$ (top panel)
      and $N($CN$)/N($C$^{18}$O$)$ (middle panel) as a function of the
      $G_0/n_\mathrm{H}$ ratio for the five CN groups. The lower panel
      gives fractional abundances for each of the three species.  A range
      of values is given for CN and HCO$^+$ when the ionisation fraction is
      uncertain (with the limits being $x_e=0$ and $x_e = 1.4\times
      10^{-4}$).}
    \label{fig:RadexColdenRatios}
  \end{figure}
}

\newcommand{\FigNclustersVsKneighborsCO}{
  \begin{figure}
    \centering %
    \includegraphics[width=1\linewidth]{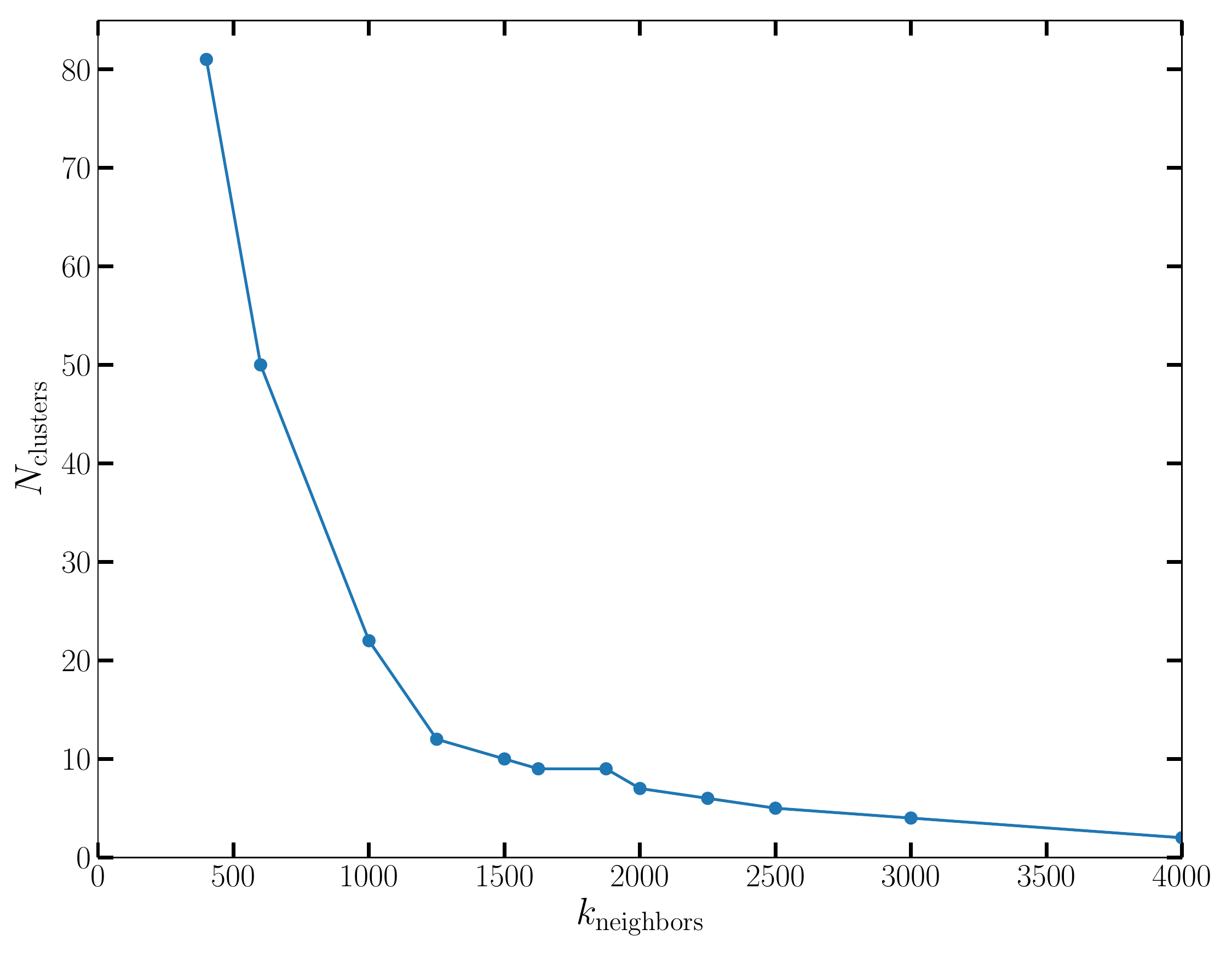}
    \caption{Variation of the number of clusters found
      $N_\mathrm{clusters}$ with the number of neighbours
      $k_\mathrm{neighbors}$ in the adaptive kernel of the Meanshift
      algorithm, when applied to the three CO isotope lines only.}
    \label{fig:Nclusters_vs_Kneighbors_CO}
  \end{figure}
}

\newcommand{\FigDensity}{
  \begin{figure*}
    \centering %
    \includegraphics[width=1\linewidth]{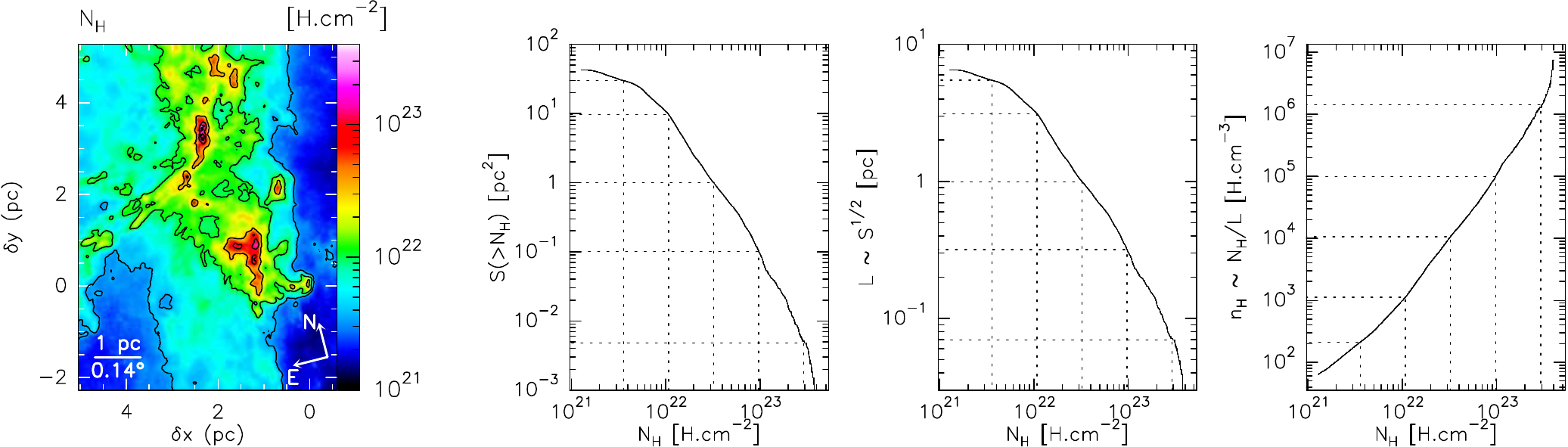}
    \caption{Steps to statistically derive an approximate volume density
      from the column density under the assumptions of isotropy and of
      nested distributions of increasing density.  \textbf{a)} Spatial
      distribution of the column density of gas as deduced from the
      spectral energy distribution of the dust continuum
      emission. \textbf{b)} Surface area of the observed field of view that
      has a column density larger than a lower limit. \textbf{c)} Typical
      depth (square root of the previous surface) associated with a given
      column density. \textbf{d)} Typical volume density associated with a
      given column density. The contour levels in panel a and the dashed
      vertical lines in panels b to d take the column density values
      corresponding to a visual extinction of 2, 6, 18, 54, and 162
      magnitudes.}
    \label{fig:volume-vs-column}
  \end{figure*}
}

\newcommand{\FigDensityComparison}{
  \begin{figure}
    \centering %
    \includegraphics[width=1\linewidth]{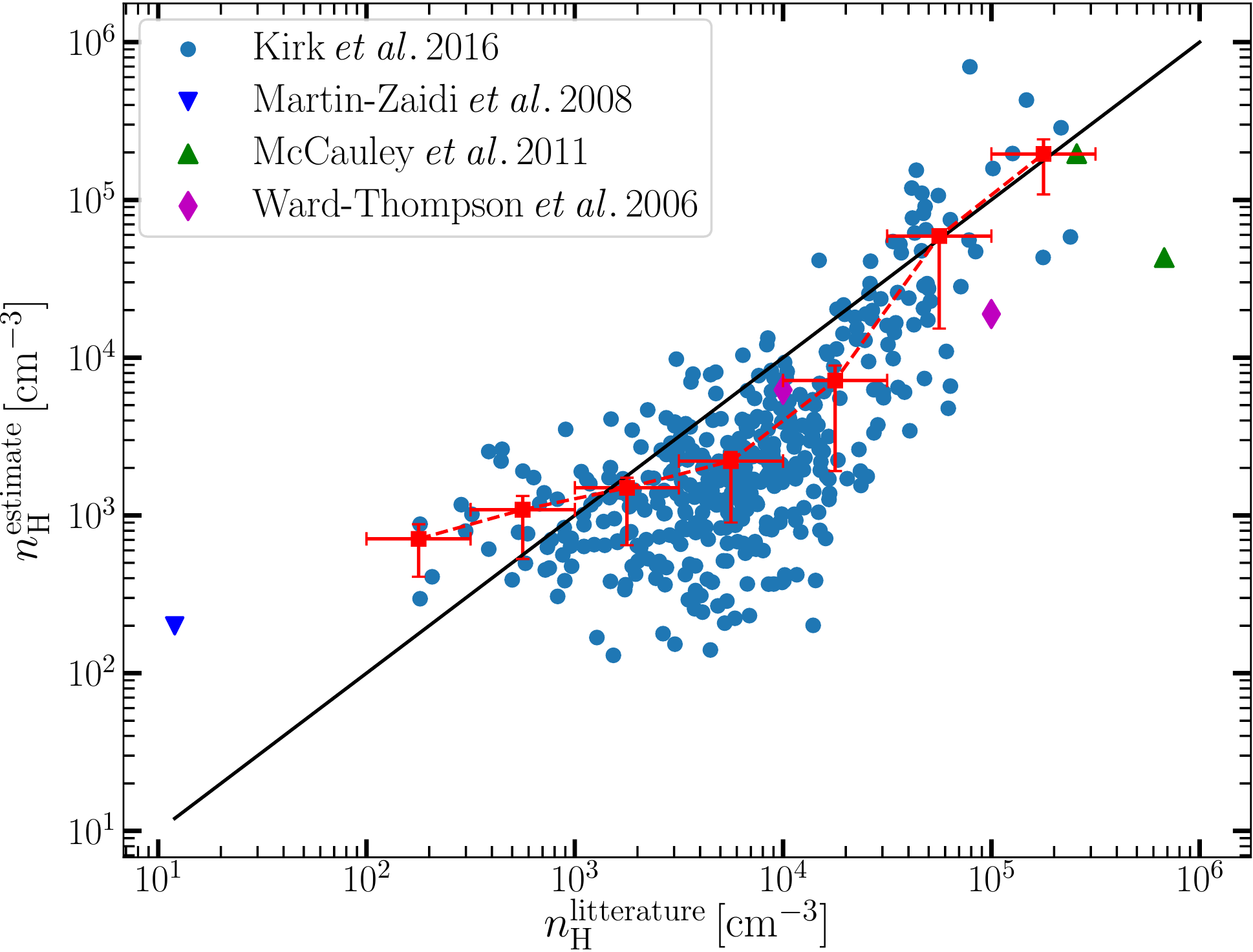}
    \caption{Approximate estimation of the volume density as a function
        of independent estimations from the literature. The solid black line shows a perfect one-to-one
        relationship, while the red points and their error bars show the
        bin averaged relationship. The horizontal and vertical error bars
        indicates the bin size and the interquartile interval,
        respectively.}
    \label{fig:comparaison_densities}
  \end{figure}
}

\newcommand{\FigAnalyticalComparison}{
  \begin{figure}
    \centering %
    \includegraphics[width=1\linewidth]{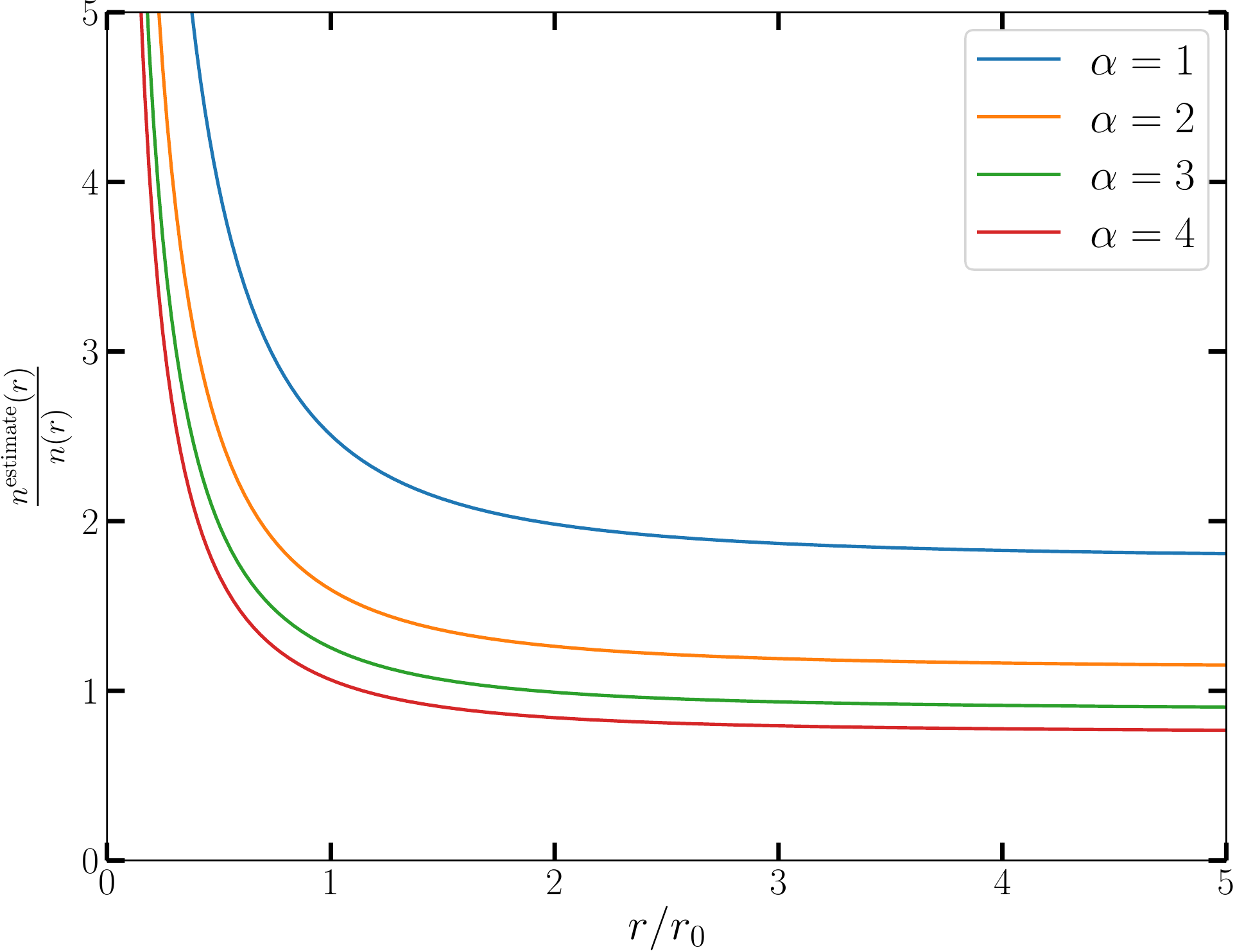}
    \caption{Approximate density estimate divided by the true density
        for different analytical spherical density profiles as a function of
        distance to the centre.}
    \label{fig:comparaison_densities}
  \end{figure}
}

\begin{document}

\title{Clustering the Orion B giant molecular cloud based on its molecular
  emission} %

\author{Emeric Bron\inst{\ref{ICMM},\ref{LERMAMeudon}} %
  \and Chloé Daudon\inst{\ref{LERMA}} %
  \and Jérôme Pety\inst{\ref{IRAM},\ref{LERMA}} %
  \and François Levrier\inst{\ref{LERMA}} %
  \and Maryvonne Gerin\inst{\ref{LERMA}} %
  \and Pierre Gratier\inst{\ref{LAB}} %
  \and Jan H. Orkisz\inst{\ref{UGrenoble},\ref{IRAM},\ref{LERMA}} %
  \and Viviana Guzman\inst{\ref{ALMA}} %
  \and S\'ebastien Bardeau\inst{\ref{IRAM}} %
  \and Javier R. Goicoechea\inst{\ref{ICMM}} %
  \and Harvey Liszt\inst{\ref{NRAO}} %
  \and Karin \"Oberg\inst{\ref{CFA}} %
  \and Nicolas Peretto\inst{\ref{UC}} %
  \and Albrecht Sievers\inst{\ref{IRAM}} %
  \and Pascal Tremblin\inst{\ref{CEA}} }

\institute{%
  ICMM, Consejo Superior de Investigaciones Cientificas
  (CSIC). E-28049. Madrid, Spain. \label{ICMM} %
  \and LERMA, Observatoire de Paris, PSL Research University, CNRS,
  Sorbonne Universit\'es, UPMC Univ. Paris 06, 92190 Meudon,
  France \label{LERMAMeudon} %
  \and LERMA, Observatoire de Paris, PSL Research University, CNRS,
  Sorbonne Universit\'es, UPMC Univ. Paris 06, \'Ecole normale
  sup\'erieure, 75005 Paris, France \label{LERMA} %
  \and IRAM, 300 rue de la Piscine, 38406 Saint Martin d'H\`eres,
  France \label{IRAM} %
  \and Laboratoire d'astrophysique de Bordeaux, Univ. Bordeaux, CNRS, B18N,
  allée Geoffroy Saint-Hilaire, 33615 Pessac, France \label{LAB} %
  \and Univ. Grenoble Alpes, IRAM, 38000 Grenoble,
  France \label{UGrenoble} %
  \and Joint ALMA Observatory (JAO), Alonso de Cordova 3107 Vitacura,
  Santiago de Chile, Chile \label{ALMA}%
  \and National Radio Astronomy Observatory, 520 Edgemont Road,
  Charlottesville, VA, 22903, USA \label{NRAO} %
  \and Harvard-Smithsonian Center for Astrophysics, 60 Garden Street,
  Cambridge, MA, 02138, USA. \label{CFA} %
  \and School of Physics and Astronomy, Cardiff University, Queen's
  buildings, Cardiff CF24 3AA, UK. \label{UC} %
  \and Maison de la Simulation, CEA-CNRS-INRIA-UPS-UVSQ, USR 3441, Centre
  d'étude de Saclay, F-91191 Gif-Sur-Yvette, France. \label{CEA}%
} %

\date{Received 25 August 2017 / Accepted 04 October 2017} %

\abstract%
{Previous attempts at segmenting molecular line maps of molecular clouds
  have focused on using position-position-velocity data cubes of a single
  molecular line to separate the spatial components of the cloud.  In
  contrast, wide field spectral imaging over a large spectral bandwidth in
  the (sub)mm domain now allows one to combine multiple molecular tracers
  to understand the different physical and chemical phases that constitute
  giant molecular clouds (GMCs).} %
{We aim at using multiple tracers (sensitive to different physical
  processes and conditions) to segment a molecular cloud into
  physically/chemically similar regions (rather than spatially connected
  components), thus disentangling the different physical/chemical phases
  present in the cloud.} %
{We use a \emph{machine learning} clustering method, namely the Meanshift
  algorithm, to cluster pixels with similar molecular emission, ignoring
  spatial information. Clusters are defined around each maximum of the
  multidimensional Probability Density Function (PDF) of the line
  integrated intensities. Simple radiative transfer models were used to
  interpret the astrophysical information uncovered by the clustering
  analysis.} %
{A clustering analysis based only on the $J=1-0$ lines of three
  isotopologues of CO proves sufficient to reveal distinct density/column
  density regimes ($n_\emr{H} \sim 100\pccm$, $\sim 500\pccm$, and $>
  1000\pccm$), closely related to the usual definitions of diffuse,
  translucent and high-column-density regions. Adding two UV-sensitive
  tracers, the $J=1-0$ line of HCO$^+$ and the $N=1-0$ line of CN, allows us to
  distinguish two clearly distinct chemical regimes, characteristic of
  UV-illuminated and UV-shielded gas. The UV-illuminated regime shows
  overbright HCO$^+$ and CN emission, which we relate to a photochemical
  enrichment effect. We also find a tail of high CN/HCO$^+$ intensity ratio
  in UV-illuminated regions. Finer distinctions in density classes
  ($n_\emr{H} \sim 7\times10^3\pccm$, $\sim 4\times10^4\pccm$) for the
  densest regions are also identified, likely related to the higher
  critical density of the CN and HCO$^+$ $(1-0)$ lines. These distinctions
  are only possible because the high-density regions are spatially
  resolved.} %
{Molecules are versatile tracers of GMCs because their
  line intensities bear the signature of the physics and chemistry at play
  in the gas. The association of simultaneous multi-line, wide-field
  mapping and powerful machine learning methods such as the Meanshift
  clustering algorithm reveals how to decode the complex information
  available in these molecular tracers.}


\maketitle %

\section{Introduction}

The inter-stellar medium (ISM) is made of several physical/chemical phases:
dense versus diffuse gas, hot versus cold gas, ionised, atomic, or molecular gas,
FUV-illuminated versus FUV-shielded gas, and gravitationally bound versus free-floating
gas. The ISM molecular composition is particularly sensitive to the changes
that affect the gas and dust when they cycle between these different
phases. Wide-field mapping of the line emission of many molecules sensitive
to different physical processes could thus enable one to segment giant
molecular clouds (GMCs) into regions belonging to distinct
physical/chemical phases, so that we can then study the physics and
chemistry of these regions in detail. Moreover, understanding the
prevalence of these different phases inside a given molecular cloud, and
learning how to disentangle their relative contributions to each molecular
tracer is of interest to interpret the spatially unresolved molecular
emission in extragalactic studies.

The advent of wide-band high-spectral-resolution spectrometers associated
to future multi-beam receivers in the \mbox{(sub-)millimetre} domain will
enable radio-astronomers to easily map the emission of many lines from tens
of species over tens of square degrees on the sky.  The pioneer ORION-B
project (Outstanding Radio-Imaging of OrioN-B, PI: J.~Pety and M.~Gerin)
currently uses the IRAM-30m/EMIR spectrometer to image about 4.5 square
degrees of the Southern part of the Orion B molecular cloud at typical
spectral resolutions of $0.6\kms$ and an angular resolution of $26''$ (\ie
$\sim50\unit{mpc}$ or $\sim10^{4}\unit{AU}$ at the distance of Orion B:
400\pc, \citealt{Menten2007,Schlafly2014}) and a typical sensitivity of
$0.1\K$ over almost all of the 3\,mm atmospheric window.

This paper is part of the first series of papers based on the already
acquired dataset that covers about 1 square degree of the Orion B molecular
cloud surrounding the Horsehead nebula, NGC\,2023, and NGC\,2024, in the
$84 - 116\GHz$ frequency range. \citet{Pety2017} introduce the molecular
anatomy of the Orion B GMC, including relationships between line
intensities and gas column density or far-UV radiation field, and
correlations between selected lines and line ratios. They obtain a
dust-traced gas mass that is less than approximately one third of the
CO-traced mass, using the standard $X_\chem{CO}$ conversion factor. The
presence of overluminous CO can be traced back to the dependence of the CO
intensity on the gas kinetic temperature, which in turn is affected by the
FUV illumination (photo-electric heating). While most lines show some
dependence on the UV radiation field, CN and \chem{C_2H} are found to be
the most sensitive. Moreover, dense cloud cores are almost exclusively
traced by \chem{N_2H^+}. Other traditional high-density tracers, such as
HCN(1-0) or HCO$^+$(1-0), are also easily detected in extended translucent
regions at a typical density of about
$500\chem{H}\pccm$. \citet{Gratier2017} propose a first multi-line approach
applying Principal Component Analysis (PCA, \citealt{jolliffe2002}) on 12
of the brightest lines (integrated over a narrow velocity range) to reveal
the pattern of correlations between the different tracers.  This approach
emphasises three clear trends: 1) the line intensities are well correlated
with the column density, that is, the more matter along the line of
sight, the brighter the lines; 2) CCH, CN, HCN are correlated with the FUV
($<13.6$ eV) irradiation (while \chem{N_2H^+} and the CO isotopologues are
anti-correlated); and 3) the PCA method confirms the known anticorrelation
between \chem{N_2H^+} and CO in dense cores.  Finally, the $^{13}$CO
position-position-velocity cube has been used by \cite{Orkisz2017} to show
that solenoidal motions clearly dominate over the observed field of view,
in agreement with the low star formation efficiency measured in
Orion\,B~\citep{Lada1992,Carpenter2000,Megeath2016}.

In this paper, we take a further step to characterise the different ISM
phases from a multi-line wide-field dataset.  The basic idea is similar to
remote sensing in Earth studies \citep[e.g.][]{Inglada2017} which tries to
classify environments (forests, deserts, mountains, oceans, etc.), based on
the dominant colour they emit. In other words, we wish to segment the
dataset into a small (yet unknown) number of classes that have a well
defined physical or chemical meaning, based on their molecular
emission. This goal requires the use of data-mining techniques in order to
go beyond a tracer-by-tracer analysis, and take advantage of the full
information hidden in the joint variations of the different
tracers. Classification techniques are divided into two
categories. Supervised ones use known examples of the desired classes to
learn how to automatically classify new observations.  They thus require
\emph{a priori} independent knowledge of the physical or chemical
properties of the different ISM phases.  These approaches will be explored
in future papers. In contrast, clustering, which is an unsupervised
technique, aims to reveal how the data points naturally group themselves
into distinct clusters of points with similar properties, hinting at the
existence of different physical or chemical regimes. This is the approach
that we adopt in this paper.

Traditional segmentation approaches in GMC studies typically segment the
map of emission of a single tracer into constitutive
clumps~\citep{Stutzki1990,Williams1994,Rosolowsky2006,Colombo2015}.  These
methods use in one way or another the topology (contiguity) of the emission
in the position-position-velocity space, sometimes associated with
additional physical properties such as the virial state. Their goal is thus
to separate the \emph{spatial} components of a GMC.  In contrast, we here
propose to work on multi-dimensional Probability Density Distributions
(PDFs) of the line integrated intensities. The PDF shape can indeed show
distinct components, which can reveal distinct physical/chemical regimes,
and which we want to automatically separate.  For instance, in the
Hertzprung-Russell (HR) diagram, different branches in the colour versus
magnitude plot correspond to different stages of stellar evolution (main
sequence, giant branch, etc.). However, while recognising structure by eye
is possible in two-dimensional (2D) datasets, direct visualisation of the data
becomes difficult in higher dimension.  Simple 2D projections for each pair
of line intensities do not necessary reveal all the existing structure, and
clustering algorithms become necessary.  In the case of the ORION-B
dataset, \citet{Gratier2017} show that understanding the physics and
chemistry underlying the extended molecular line emission requires a
multi-dimensional analysis of the data.  To our knowledge, clustering a GMC
based on its (multi-molecule) molecular emission similarity rather than
spatial (or PPV) contiguity has never been done before.

This paper is organised as follows. In Sect.~\ref{sec:Data}, we present the
data used in our analysis.  We then explain the clustering algorithm that
we chose to use in Sect.~\ref{sec:Method}. This clustering method is first
applied to the most widely observed lines in the millimetre wave domain,
that is, the $J=1\rightarrow0$ lines of three CO isotopologues (\twCO,
\thCO~and \CeiO) in
Sect.~\ref{sec:Results_CO}. Sect.~\ref{sec:Results_HCOp_CN} then discusses
the additional results obtained when adding the ground-state transitions of
HCO$^+$ and CN to the analysis, whose intensities are known to be related
to the FUV illumination. We discuss the benefits and limits of the method
in Sect.~\ref{sec:Discussion_Meanshift}. We present our conclusions in
Sect. \ref{sec:Conclusions}.
  
\section{Data}
\label{sec:Data}

\subsection{IRAM-30m observations}

\citet{Pety2017} present in detail the acquisition and data reduction of
the dataset used in this study. In short, the data were acquired at the
IRAM-30m telescope by the ORION-B project from August 2013 to November
2014. The frequency range from 84 to 116\GHz{} was completely sampled at
200\kHz{} spectral resolution. Such a large bandwidth allowed us to image
over 20 chemical species at a median sensitivity of 0.1\K{} (main-beam
temperature) per channel. As opposed to several small bandwidth mappings,
the spectral lines in this survey are observed in only two tunings covering
16\GHz{} each. They are thus well inter-calibrated, which gives an
unprecedented spectral accuracy for such a large field of view. The
intensity dynamic range reaches $\sim720$.

The field of view presented covers $0.8\degr \times 1.1\degr$
towards the Orion B molecular cloud part that contains the Horsehead
nebula, and the H{\sc ii} regions NGC\,2023, NGC\,2024, IC\,434, and
IC\,435. The angular resolution ranges from 22.5 to $30.5''$.  The
position-position-velocity cubes of each line were smoothed at a common
angular resolution of $31''$ to avoid resolution effects during the
comparison. At a distance of 400\pc{}~\citep{Menten2007}, the sampled
linear scales range from $\sim50\unit{mpc}$ to $\sim8\pc$.

The observations provided a position-position-frequency cube of $315 \times
420 \times 160\,000$ pixels, each pixel covering $9'' \times 9'' \times
0.5\kms$ (Nyquist sampling at the highest frequency, \ie \twCO(1–0) at
115.27\GHz{}).

We here study maps of line integrated intensities. Lines are detected over
quite different velocity ranges. Using a large velocity range would
artificially increase the noise for most of the lines, while adapting the
velocity range to each line could bias the comparisons. We thus focus on a
narrow velocity range where the bulk of the gas emits. While
\citet{Gratier2017} computed for each line the mean of three $0.5\kms$
velocity channels around 10.5\kms, we use here a more common
radioastronomical approach, which is to integrate the line intensity
profile. We here integrate over the $[9,12\kms]$ velocity interval
where the main velocity component of the Orion B cloud sits \citep[see
section 2.5 of][]{Pety2017}. To first order, both sets of maps are
proportional to one another (the respective velocity ranges 
differ slightly)\footnote{The data products associated with this paper are available
  at \url{http://www.iram.fr/~pety/ORION-B}}.

\subsection{Herschel observations}
\label{sec:GouldBeltSurvey_obs}

\FigDustPDFs{} %

In order to get independent constraints on the physical conditions in the
Orion B cloud, we use the dust continuum observations from the
\textit{Herschel} Gould Belt Survey~\citep{Andre2010,Schneider2013} and
from the \textit{Planck} satellite~\citep{Tauber2011}. The fit of the
spectral energy distribution by~\citet{Lombardi2014} gives us access to the
spatial distributions of the dust opacity at $850\mum$ and of the dust
temperature.  As in \cite{Pety2017}, we converted $\tau_{850\mum}$ to
visual extinctions using $A_\mathrm{V} = 2.7\times 10^4 \, \tau_{850} \,
\mathrm{mag}$.  The top panels of Fig.~\ref{fig:DustPDFs} show the PDF of
the dust visual extinction and temperature.

The $A_\mathrm{V}$ PDF shows three distinct peaks, indicating that the field of
view samples three different regimes: $A_\mathrm{V}=1-2$, $2-6$, and $\ge
6$. These regimes are consistent with the usual distinction between
diffuse, translucent, and high-column-density regions~\citep{Snow2006}.  As
in \citet{Pety2017}, we use $N_\mathrm{H}/A_\mathrm{V} =
1.8\times10^{21}\pscm/\mathrm{mag}$ as conversion factor between visual
extinction and hydrogen column density: $N_\mathrm{H} =
N_\mathrm{HI}+2N_\mathrm{H_2}$. In addition, we propose a conversion from
the column density to an approximate volume density
map. The procedure is discussed in detail in
  Appendix~\ref{sec:column-vs-volume-density}. In summary, we assume a
rough isotropy of the cloud (similar dimensions along the line of sight and
in the plane of the sky) to deduce an estimate of the average
hydrogen density along each line of sight as follows. For a given column
density value $x$, we consider the region where $N_\mathrm{H} \geq x$. We
then estimate the line-of-sight depth $l$ of this region from its
plane-of-the-sky surface $S$ as $l\simeq\sqrt{S}$ (using our isotropy
assumption). We finally assign the approximate volume density
$n_\mathrm{H}=x/l$ to the pixels where $N_\mathrm{H} = x$.

The resulting approximate volume density PDF is shown in the
bottom-left panel of Fig.~\ref{fig:DustPDFs}. The three distinct
$A_\mathrm{V}$ regimes correspond to three volume density regimes: one low-density peak close to $10^2\,\mathrm{cm}^{-3}$ corresponds to diffuse gas,
a second peak covering the range $300-800\,\mathrm{cm}^{-3}$ is associated
to the translucent gas, and a third smaller peak slightly above
$10^3\,\mathrm{cm}^{-3}$ with a long tail extending up to a few
$10^6\,\mathrm{cm}^{-3}$ corresponds to denser gas. The values found after
our conversion are consistent with the usual orders of magnitude for
diffuse, translucent, and denser gas. More quantitatively,
  Appendix~\ref{sec:column-vs-volume-density} shows that our estimation of
  the volume density is valid in a statistical way with a bias of a factor of 3 at most 
  and a typical scatter of one order of magnitude, when compared with volume density estimates
  from the literature that make different hypotheses. The deduced values
  of $n_\mathrm{H}$ are rough estimates that should not be trusted beyond
  order-of-magnitude comparisons. However, this
method reproduces the observed range of densities fairly well, indicating
that the shape of the PDF is also approximately correct.

The $T_\mathrm{dust}$ PDF shows a less marked multi-peak structure with a
sharp first peak at $\sim22\K$, a small secondary peak at $\sim25\K$ and a
shallow third peak at $\sim27\K$. A first steep tail extends up to
$\sim33\K$, followed by a second flatter tail (reaching values up to
$100\K$). These two tails are indicative of highly FUV-illuminated regions.
\citet{Pety2017} converted the dust temperature map into an approximate map
of the far ultra-violet radiation field $G_0$ in units of the Habing
Interstellar Standard Radiation Field (ISRF, \citealt{Habing1968}), using
the simple approximation of~\cite{Hollenbach1991}
\begin{equation}
  G_0 = \left ( \frac{T_\mathrm{dust}}{12.2\K} \right )^5.
\end{equation}
\citet{Shimajiri2017} compared this estimation with another estimation
directly using the far infra-red intensities at 70 and $100\mum$. Both
estimates agree within 30\%.  The PDF of $\log(G_0)$ is shown as the bottom
right panel of Fig.~\ref{fig:DustPDFs} and is very similar to the dust
temperature PDF (as the conversion is a simple power law). As mentioned
before, the main peak is close to 20 times the ISRF while the tail extends
up to several thousand times the ISRF. As for our estimate of $n_\mathrm{H}$, the
  deduced values of $G_0$ should only be trusted at order-of-magnitude
  levels.

\section{The Meanshift clustering method}
\label{sec:Method}

From a mathematical viewpoint, the data are a set of points characterised
by their two sky coordinates and the integrated intensities of $D$
molecular lines. The dataset thus lies in a space of $D+2$ dimensions. As
discussed before, we aim to cluster the datapoints based on their molecular
emission only, and not their spatial proximity.  For clarity, we thus
separate the data space into two parts: 1) the usual \emph{position space},
and 2) the \emph{line space} of dimension $D$ spanned by the molecular line
intensities. Clustering will only be done in the line space, ignoring the
location of the pixels in the position space. After a non-exhaustive
discussion of methods that segment the data based on their multi-dimensional
PDF, we describe the Meanshift algorithm and the implementation used in
this paper.

\subsection{The two families of PDF-based clustering methods}

We are interested in clustering methods that aim to separate components in
the (multi-dimensional) PDF of the data. Two families of such methods can
be defined.  The first family assumes that the data PDF can naturally be
decomposed into components of some given functional form, for example, Gaussian
functions, controlled by free parameters. These methods are thus called
parametric approaches.  The complete dataset is assumed to be a mixture of
several components, with the same functional form but different values for
the control parameters. These algorithms are thus usually called mixture
models ~\citep[see][Chap. 9]{Bishop2006}.  This approach has two main
drawbacks. First, the number of free parameters tends to increase quickly
with the dimension of the problem, resulting in a difficult and often
degenerate optimisation. To alleviate this problem, restrictions are
imposed on the free parameters. For instance, assuming Gaussian components
and forcing them to all have the same scalar covariance matrix (thus
forcing spherically symmetric clusters of equal size) yields the K-means
algorithm, one of the most used clustering algorithms. Second, the
assumption of a functional PDF form is a strong \emph{a priori} that can
bias the clustering when this form is inadequate for the studied data. More
flexible functional forms reduce this problem but result in more free
parameters. In this family, a compromise has to be made between the
flexibility of the assumed functional PDF form and the number of free
parameters.

The second family takes a data-driven approach, by defining clusters around
local maxima of the data PDF.  Each cluster is thus a region of high
density in the line space, separated from the other clusters by regions of
lower density. This definition has two advantages: 1) It allows to capture
any shape of the PDF of the clusters (possibly curved and elongated); 2)
The number of clusters is determined automatically from the data. Data
clusters must however create a maximum in the PDF to be detected. A small
group of datapoints blended in the tail of another more common cluster
might thus not be detected.  The two most famous algorithms in this family
take a different approach to finding the high-density regions in the line
space.  First, the DBSCAN algorithm~\citep{Ester96} uses a graph-based
approach to find high-density regions, but it assumes a similar density of
points inside all clusters. Second, the Meanshift
algorithm~\citep{Comaniciu2002} searches for the maxima of the data PDF
using a kernel-based approach.  We choose to use the Meanshift approach, as
it can detect clusters that have both different sizes and different
densities. Moreover, its direct link to the data PDF eases the
interpretation of the clusters. The following Section describes this algorithm in
detail.

\subsection{The Meanshift algorithm}

\subsubsection{General description}

The Meanshift algorithm (see \citealt{Comaniciu2002} for more details)
associates each data point to the closest local maximum of some empirical
estimate of the PDF. The algorithm iteratively climbs up the slope of the
PDF starting from each of the datapoints.  The set of datapoints converging
to the same PDF maximum constitutes a cluster.

The algorithm is based on the same concept as the Kernel Density Estimate
method~\citep{Rosenblatt1956,Parzen1962}, which estimates the PDF of a
random variable (here, the intensities) from one sample realisation. The
kernel density estimator for a given set of $N$ $D$-dimensional data points
$\{\mathbf{x}_i\}_{1\le i \le N}$ is
\begin{equation}
  \label{eq:KDE}
  f_{h,K}(\mathbf{x}) = \frac{C_h}{N}\,\sum_{i=1}^N K\left( \frac{\emr{dist}(\mathbf{x},\mathbf{x}_i)^2}{h^2} \right),
\end{equation}
where $\emr{dist}(\mathbf{x},\mathbf{x}_i)$ is a distance in the line space
between a given a vector $\mathbf{x}$ and the vector $\mathbf{x}_i$
associated to the $i$th datapoint (both are vectors of line intensities),
$K$ is the smoothing kernel (a non-negative decreasing function of
$\mathbb{R}^+$), $h$ is the bandwidth of the smoothing and $C_h$ a
normalisation constant.  The kernel often has a finite support $[0,1]$, so
that the estimation of the PDF at $\mathbf{x}$ is only based on the
datapoints that are closer to $\mathbf{x}$ than the bandwidth $h$.

The Meanshift algorithm avoids the estimation of the PDF itself by directly
estimating the PDF gradient with the same kernel smoothing approach.  By
taking the gradient of Eq.~\ref{eq:KDE} in the case of an Euclidean
distance, and noting $G(x) = - K'(x)$, which is a new kernel, we obtain
\begin{equation}
  \label{eq:KernelGradient}
  \mathbf{\nabla} f_{h,K}(\mathbf{x}) 
  \propto \,f_{h,G}(\mathbf{x}) \, \mathbf{m}_{h,G}(\mathbf{x}),
\end{equation}
\begin{equation}
  \label{eq:MeanshiftVector}
  \mbox{where} \quad
  \mathbf{m}_{h,G}(\mathbf{x}) =
  \frac{\sum_{i=1}^N \mathbf{x}_i\,G\left(\frac{|| \mathbf{x} -
        \mathbf{x_i}||_2^2}{h^2}\right)}{\sum_{i=1}^n G\left(\frac{|| \mathbf{x} -
        \mathbf{x_i}||_2^2}{h^2}\right)} - \mathbf{x}
,\end{equation}
which is called the Mean Shift vector as it gives the shift from the current
position $\mathbf{x}$ to the mean of the datapoints weighted by the kernel
$G$ centered on $\mathbf{x}$. Equation~\ref{eq:KernelGradient} indicates
that this Mean Shift vector gives an estimate of the relative gradient (the
local PDF gradient divided by the local PDF).

The following iterative algorithm is applied, starting from each of the
datapoints:
\begin{itemize}
\item Compute the Mean Shift vector $\mathbf{m}_{h,G}(\mathbf{x})$ at the
  current estimate $\mathbf{x}$ of the searched local maxima.
\item Modify the current estimate by shifting it by the Mean Shift vector.
\end{itemize}
This algorithm converges to points where the PDF gradient estimate is zero,
and that usually are local maxima due to its hill climbing
nature. Convergence points that are closer to each other than the bandwidth
$h$ are then merged, and clusters are defined as the sets of datapoints
that have converged to the same extremum.  Data points lying close to local
minima can sometimes stay stuck due to the associated null gradient, but
the resulting unwanted clusters can easily be recognised by the very small
number of datapoints they contain, and removed by assigning their
datapoints to the closest clusters.

\subsubsection{The FAMS implementation}

In this article, we used the Fast Adaptive Mean Shift (FAMS) code described
in \citet{Georgescu2003}, and provided by the authors on their
webpage\footnote{
  \url{http://coewww.rutgers.edu/riul/research/code/AMS/index.html}}.  This
code implements several modifications to the Meanshift principle in order
to improve the quality of the gradient estimate in low-density regions.

The quality of the PDF gradient estimate depends on the number of
datapoints present inside the smoothing kernel. Consequently, using a fixed
bandwidth results in insufficiently sampled gradient estimates in the tails
of the data PDF. This leads to many artificial maxima being found in these
tails, unless the bandwidth is increased. But the central parts of the PDF
then risk being overly smoothed and significant maxima can be lost. To
alleviate this problem, \citet{Comaniciu2001} proposed an
adaptive-bandwidth Meanshift algorithm, in which a different bandwidth
$h_i$ is assigned to each data point $\mathbf{x}_i$ ($h$ is replaced by
$h_i$ in Eq. \ref{eq:MeanshiftVector}). The Adaptive Meanshift Algorithm
uses the simplest method to choose the bandwidth for each datapoint: taking
the distance to the $k^{\mathrm{th}}$-nearest neighbour to the datapoint as
the bandwidth, so that there is always roughly the same number of
datapoints inside the kernel. This number of neighbours $k$ becomes a
parameter of the method to be adjusted by the user (instead of the
bandwidth value in the classical Meanshift algorithm). In other words,
there is a trade-off between the sampling variance and the smoothing factor
to estimate the PDF: an adaptive smoothing is applied so that low-density
regions are smoothed on a larger scale than high-density regions.  All
estimates of the PDF gradient are then equally well sampled.

Second, the FAMS implementation uses a $L_1$ distance
($\emr{dist}(\mathbf{x},\mathbf{y}) = \sum_{i=1}^{D} | x_i - y_i |$)
instead of the usual Euclidean distance ($\emr{dist}(\mathbf{x},\mathbf{y})
=\sqrt{ \sum_{i=1}^{D} (x_i- y_i)^2 }$).  This allows for an additional
optimisation in high dimension~\citep{Georgescu2003}; but it is only an
approximation of the Meanshift algorithm, as Eq.
\ref{eq:MeanshiftVector} assumes a Euclidean distance. Finally, the kernel
used in this implementation is $G(x) = (1 - x)^2$.

\subsubsection{In practice}

The main control parameter of the Meanshift algorithm is either the
bandwidth value $h$ when using the fixed bandwidth implementation or the
number of nearest neighbours $k$ when using the adaptative bandwidth
version. This parameter controls the scale on which the PDF is smoothed in
both cases. We use here the adaptive bandwidth version. In each studied
case, $k$ was varied from a few hundred to a few thousand, and adjusted to
increase or decrease the number of clusters found. Any cluster that
contains a number of pixels that is very small compared to $k$ is merged with the
closest cluster. This can happen when the Meanshift algorithm is stuck into
a local minimum of the PDF.

The smoothing part uses a spherical kernel, so that the applied smoothing
is isotropic, that is, it has the same absolute bandwidth in all
directions. It is thus necessary to ensure that the variability of the
dataset along the different dimensions (that is the intensity dynamic of
each line) is comparable. Without a linear rescaling, either the variations of the faint
lines would be smoothed out by a bandwidth adapted to the bright
lines or the bright lines would drive the segmentation into many small
clusters when using a bandwidth adapted to the faint lines. We thus
standardised the dataset (we ensured that the intensity PDF of each line
has a unit standard deviation) before applying the Meanshift
algorithm. Additional non-linear transformations applied to the
dataset before clustering it with Meanshift would in general modify the
number and positions of PDF maxima and thus affect the results. As we wish
to check the amount of physical/chemical information encoded in the line
intensities, we chose to only linearly standardise the data.

The clustering of our $\sim 10^5$ data points with $D=5$, and $k=500-2000$,
typically takes 10 to 20 hours of computation on a single standard CPU in
2017. Taking into account that our data has some redundancy (Nyquist
sampling), we tested the method on a decimated dataset before getting the
final results on the full dataset.

\citet{Comaniciu2001} showed that adding spatial coordinates in addition to
intensities when clustering images may smooth the resulting clusters. We
did not use the spatial information present in our dataset, as our focus is
on grouping pixels where the intensities are similar rather that pixels
belonging to the same spatial structure. We will however use the spatial
coherence of the clusters found as a consistency check of the results
because some amount of physical/chemical similarity is expected between
neighbouring pixels.

When discussing the results, we visualise the 2D PDFs of pairs of
lines (comparing the contributions of the obtained clusters) using kernel
density estimation \citep{Rosenblatt1956,Parzen1962} from the
\texttt{scikit-learn} Python package~\citep{Pedregosa2011}, with an
Epanechnikov kernel ($\propto 1 - x^2$, optimal in terms of mean squared
error, \citealt{Epanechnikov1969}).  This implementation uses a fixed
bandwidth.  This is however only used as a visualisation tool, and is
independent of our Meanshift clustering analysis.

\section{The CO isotopologue emission enables us to separate the diffuse,
  translucent, and denser gas regimes}\label{sec:Results_CO}

\FigClusterCOMap{} %
\FigClusterCOCompression{} %

The \twCO, \thCO, and \CeiO{} $J=1-0$ lines are amongst the most observed
radio lines in molecular clouds. At constant elemental ratios of the carbon
isotopes, the naive chemical interpretation suggests that the relative
abundances of these three species should be identical in all lines of sight
of a GMC. Moreover, the critical densities of these three
lines for collisional excitation with H$_2$ are similar
($\sim2\times10^3\,\mathrm{cm}^{-3}$), implying similar excitation
conditions. Differences in optical depths should therefore be the main
factor governing the intensity ratios. In this Section, we ask whether
or not clustering the intensities of these lines can distinguish physical regimes
known to happen in the studied field of view. We thus first apply the
Meanshift algorithm to a dataset consisting of the maps of the $J=1-0$
lines of $^{12}$CO, $^{13}$CO and C$^{18}$O only ($N=141050$, $D=3$).

After several trials for the number of neighbours in the adaptive kernel
width, we chose a compromise between avoiding picking up sampling
fluctuations in the PDF as artificial maxima and smoothing out physical
maxima of interest (see Appendix \ref{sec:Kneighbors_choice} for a
discussion of this choice).  We settled on $k=1900$ neighbours. In order to
understand and interpret the clustering, we present in the following the
spatial distribution of the clusters as well as PDFs of 1D or 2D
projections of the data (that is, PDFs of single lines or pairs of lines).

\subsection{Spatial distribution}

We find nine clusters, whose spatial distribution is shown in
Fig.~\ref{fig:clusterCO_map}.  For the following discussion, we name these
clusters CO-0 to CO-8.  While our use of the Meanshift clustering does not
take into account the spatial contiguity, they show very consistent spatial
distributions, except for clusters CO-0 and CO-1 which share the outer
region of the cloud and display a noise-like pattern. A visual inspection
of the CO isotopologue PDFs shows that this distinction most likely comes
from noise properties. We thus merge cluster CO-0 with cluster CO-1. We
call the new cluster CO-1 and represent it in light blue. The resulting
cluster map is shown on the right panel of Fig.~\ref{fig:clusterCO_map}.

Clusters CO-1 to CO-7 show a nested pattern highlighting successive layers
from the surface to the inner parts of the cloud.  These clusters are
present both on the eastern and western sides that surround the inner
region of the imaged field of view. However, an asymmetry between the two
sides is visible in the much smaller thickness of the surface layers
highlighted by the transition from clusters CO-2 to CO-5 on the western FUV-illuminated side. This is a consequence of the much steeper intensity
gradients for all three CO lines on this side of the cloud. In other words,
high FUV illumination has a much stronger impact on the CO intensity
gradients than on the intensities themselves. Cluster CO-8 differs as it
only appears towards the two H\textsc{ii} regions, NGC\,2023 and NGC\,2024,
which are embedded in the south-western part of Orion B.

One way to check the quality of the clustering is to compare the spatial
distributions of the original line intensities with the line intensities
averaged in each cluster. This is somehow a test of the ability of the
clustering method to compress the information contained in the line
intensity maps while retaining the most important aspects.  Taking the mean
of the line intensities for each cluster conserves the total flux in the
output images.  Figure~\ref{fig:clusterCO_compression} presents such a
comparison.  The mean intensities (and other characteristic intensity
values) of each line in each cluster are listed in
Table~\ref{tab:clustersCO_intensities}.  Most of the \twCO{} and \thCO{}
spatial features are preserved in the clustered images, while the
representation of the \CeiO{} image is not as good: some spatial features
appear and others disappear.  This comparison also highlights that cluster
CO-8 is characterised by an increased \twCO{} intensity compared to its
surrounding. We can note that the Horsehead pillar and the other dense
clumps that emerge from the IC\,434 H\textsc{ii} region belong to cluster
CO-7 which is associated with relatively dense gas (as discussed in the
following subsections).

\subsection{Projected PDFs}

\FigCOClustersCOPDF{} %
\FigCOlustersTwoDPDF{} %

Figure~\ref{fig:COclusters_12CO_PDF} shows the \twCO{} 1D PDF computed for
the full dataset and for each individual cluster. The full dataset PDF is
clearly multi-peaked with a main peak at low intensities ($\sim 2\,\Kkms$),
two other peaks around 37 and 60\Kkms{} separated by a plateau, and several
minor peaks, one of them corresponding to very bright \twCO{} around
90\Kkms. In contrast, the 1D PDF of \thCO{} and \CeiO{} (not shown here)
are mostly mono-modal (only one clear peak).  A good correspondence between
the maxima of the \twCO{} PDF and most of the clusters can be seen. This
indicates that \twCO{} plays a major role in the definition of the
clusters.  Cluster CO-1 corresponds to the highest and narrowest,
low-intensity peak, clusters CO-3 and CO-4 constitute the 37\Kkms{} peak,
cluster CO-5 contributes to the plateau between the 37 and the 60\Kkms{}
peaks, cluster CO-6 and CO-7 corresponds to the 60\Kkms{} peak, and cluster
CO-8 corresponds to the clear bump in the high-intensity tail of the PDF
(around 90\Kkms{}).

However, significant overlap between the clusters can be seen, and two
clusters share the 60 \Kkms{} peak. Both facts highlight the influence of
the other two isotopologues on the clustering. To understand the role of
\thCO{} and \CeiO{} in the clustering, Fig.~\ref{fig:COclusters_2D_PDFs}
shows the 2D PDFs of \thCO{} versus \twCO{}, and \thCO{} versus \CeiO{} as
contour plots. In each case, the PDF of the full dataset is shown as black
dotted contours, while the PDFs of the different clusters are coloured
according to the cluster colours presented in
Fig.~\ref{fig:clusterCO_map}. All clusters have clear separations in the
\thCO{} versus \twCO{} plane, so that the overlap seen in
Fig.~\ref{fig:COclusters_12CO_PDF} is only a projection effect. Clusters
CO-1 to CO-5 clearly follow the ridgeline (the line
connecting the cluster maxima, similar to the ridgeline of a mountain ridge
connecting the summits) of the 2D PDF .  The maxima to which they are associated are small
bumps along this ridge. The separations of the basins of attraction of each
of these maxima thus lie roughly orthogonally to the direction of this
ridge.  This is probably why the shape of CO-5 cluster looks like an
anti-correlation. Clusters CO-2 to CO-5 are associated to relatively small
fluctuations of the PDF along the ridge line.  While statistically
significant, these fluctuations might be too weak to be each attributed a
physical meaning: these four clusters might thus represent a single
physical category. In the \thCO{} versus \CeiO{} space, the PDFs of clusters
CO-1 to CO-5 are nearly indistinguishable because they lie below the
\CeiO{} detection limit. We thus grouped them into a single PDF (grey
contours) for better readability.

While clusters CO-6 and CO-7 are undistinguishable on the \twCO{} 1D PDF,
they are clearly separated by their $^{13}$CO intensities (cluster CO-7
having $I_{^{13}\mathrm{CO}} \gtrapprox 13\,\Kkms$). Cluster CO-8 is
distinguished both by its high \twCO{} intensity and by higher
\thCO/\CeiO{} ratios at similar \CeiO{} intensity than lines of sight
belonging to cluster CO-7.  Clusters CO-7 and CO-8 thus correspond to a
separation of the \thCO{} versus \twCO{} ridge line into two distinct ones at
high intensities. Cluster CO-6 is an intermediate cluster that probably
lies around the intersection of these two ridge lines. This explains why it
is so extended along the \twCO{} axis and narrow along the \thCO{} axis.
We discuss these cluster shapes in relation to LTE calculations in
Sect.~\ref{sec:CO-LTE}.

\subsection{Relation with volume density and FUV illumination}

\FigCOClustersDensityPDF{} %
\FigCOClustersGProperties{} %

Figure~\ref{fig:COclusters_nH_PDF} shows how the clusters contribute to the
PDF of the approximate volume density (see
Section~\ref{sec:GouldBeltSurvey_obs}) in the form of a violin plot: for
each group of clusters, the blue profiles show the volume density PDF
(normalised to an identical width) and the median values are shown as red
squares.  We find a close correspondance between our clusters and the three
peaks of the density PDF discussed in Sect.~\ref{sec:GouldBeltSurvey_obs}:
Cluster CO-1 corresponds to diffuse gas ($n_\mathrm{H} \sim 100\pccm$), the
group of clusters CO-2 to CO-5 are associated to translucent gas
($n_\mathrm{H} \sim 500\pccm$), while clusters CO-6, 7, and 8 correspond to
denser gas ($n_\mathrm{H} \ga 1000\pccm$). The CO clusters thus reveal
underlying density regimes.

Figure~\ref{fig:ClustersCO_G0_properties} compares the distributions of the
FUV illumination for the different clusters. Clusters CO-1 to CO-6 have
similar median values of $G_0 \sim 30 - 35$. Moreover, a higher $G_0$ wing
is present for all these clusters. In contrast, cluster CO-7 has a
significantly lower value of $G_0 \sim 20$. This cluster thus tracks gas
relatively shielded from the FUV illumination.  In addition, the high-$G_0$
wing is negligible for this cluster. Finally, cluster CO-8 has a much
larger median value of $G_0 \sim 180$, and its PDF has two broad components
at typical values of $G_0 \sim 50$ and $300$, consistent with the presence
of the NGC\,2023 and NGC\,2024 \textsc{Hii} regions.

In summary, clustering of the CO isotopologues allows us: 1) to
distinguish three different regimes of column/volume density (diffuse,
translucent, and higher density), and 2) to start distinguishing
FUV-illuminated from FUV-shielded gas; but only for relatively dense gas.

\subsection{Interpretation: nested CO isotopologues and higher
  [\thCO]/[\CeiO] abundance ratios in FUV-illuminated dense
  gas} \label{sec:CO-LTE}

\FigCOLTE{} %

We wish to understand the physical and chemical processes that determine
the variations in line intensities detected by the clustering method. We
thus make the simplest possible radiative transfer model that will allow us
to match the observed line intensities and ratios of the CO isotopologue
lines.We then interpret the astrophysical information uncovered
by the clustering of the CO isotopologues.

\subsubsection{Modeling principles}

The typical density over the studied field of view is $\sim
300\,\mathrm{H}_2\pccm$, and the density of at least 25\% of the
field~\citep{Pety2017} is larger than $2\times10^3\,\mathrm{H}_2\pccm$ (the
typical critical density of CO $J=1-0$). We thus chose to use LTE models,
as we are mainly interested by the denser parts of the field of view.  The
clustering analysis taught us that the CO isotopologue intensities trace
different ranges of hydrogen column density, and consequently of
approximate volume density (See
Sect.~\ref{sec:GouldBeltSurvey_obs}).  We have thus chosen to model
intensity curves for several fixed values of the CO isotopologue column
densities, varying the kinetic temperature.

Figure~\ref{fig:clusterCO_compression} indicates that the structure
of the dense inner parts of the cloud is well delineated
by the \CeiO{} \Jone{} emission, still visible in the
\thCO{} \Jone{} emission, and mostly hidden in the \twCO{} \Jone{}
emission. The usual interpretation is that the \twCO{} line is so optically
thick that the outer, less dense layers along the line of sight can already
produce a saturated \twCO{} emission. This means that the \thCO{} and
\CeiO{} \Jone{} lines would be sensitive to denser, cooler gas
more deeply embedded along the line of sight while the \twCO{} \Jone{} line
would to first order be mostly sensitive to the foreground, more
FUV-illuminated and thus warmer gas. We thus propose to use two different
values for the excitation temperature: a high value for \twCO{} and a lower
value for \thCO{} and \CeiO{}. This in turn implies that the model \twCO{}
column density will be a lower limit to the total CO column density as it
only represents the warm gas.

The detailed parametrisation of our modelling is described in
Appendix~\ref{sec:CO-LTE:details}.

\subsubsection{Observations and modeled curves}

Figure~\ref{fig:CO:LTE} shows the modeled curves over the joint histograms
of the \thCO{} versus\ \twCO{} emission (left column), of the \thCO{} versus\
\CeiO{} emission (middle column), and of the \thCO/\CeiO{}
versus \twCO/\thCO{} line ratios (right column). The ratio versus\ ratio
histograms allow us to check how the models take care of the co-variations
of the three studied CO lines.

The first row presents the observations for the full field of view, while
the next three rows present the observations for different sets of CO
clusters (from 1 to 6, the 7th one, and the 8th one, respectively).  The
sets of input parameters described above each row were chosen to deliver
the best visual match between the modelled curves and the three associated
histograms.

In all cases, only the lines of sight where the isotopologue lines
  considered have intensities above $4\sigma$ are used to compute the
histogram.  For each \thCO{} opacity (i.e. along each white curve), the
\thCO{} and \CeiO{} excitation temperature increases clockwise and
counter-clockwise for the \thCO{} versus\ \twCO{}, and \thCO{} versus\ \CeiO{}
histograms, respectively.  On the ratio versus\ ratio histograms (right
column), the \thCO{} and \CeiO{} excitation temperature increases from left
to right.

On the line versus line histograms (left and middle columns), the higher the
\thCO{} opacity, the more opened the corresponding model curve.  In
contrast, the dependency on the column density is reduced in the ratio versus
ratio histograms, as indicated by the fact that all curves for different
\thCO{} opacities almost overlap. This is linked to the fact the line
intensity is proportional to the column density to lowest order.  And
therefore, line ratios remove this trend.

\subsubsection{Global results}

The first row of Fig.~\ref{fig:CO:LTE} shows the best visual match between
observations and models for the full field of view. The FWHM of the lines
(2\kms) is the median value measured over the field of view on the
10.5\kms{} main component of the \thCO{} and \CeiO{} \Jone{} lines. The
line emission of the modelled curves is integrated over 3\kms{} as in the
observations.

The range of \thCO{} \Jone{} opacities runs from optically thin lines
(minimum: 0.03) to moderately saturated lines (maximum: 2.5). The \CeiO{}
\Jone{} line is always optically thin as expected from the fact that
\CeiO{} \Jone{} shows an excellent correlation with the visual
extinction~\citep{Pety2017}. The \twCO{} \Jone{} line is almost always
optically thick.

Kinetic temperatures of up to $\sim100\K$ are required to explain the low
intensity part of the \thCO{} versus \twCO{} histogram. The $[\thCO]/[\CeiO]$
abundance ratio is larger than the expected elemental ratio value of $\sim
8$~\citep{Wilson1994}. Both inferences can be explained by the significant
FUV illumination in the observed field of view, with contributions from
external and embedded H\textsc{ii} regions. The joint histogram of \thCO{}
versus \twCO{} intensities, especially the range of \thCO{} intensities at a
given \twCO{} intensity, can only be explained if \twCO{} and \thCO{} have
different kinetic temperatures. In contrast, the observation space
can be understood with similar kinetic temperatures of the gas that emits
the \thCO{} and \CeiO{} \Jone{} lines.

\subsubsection{Results per CO clusters}

In order to better understand how the CO clustering can distinguish
different regimes of density and FUV illuminations, we now discuss the
input parameters that deliver the best visual match between the modelled
curves and the three histograms for three different subsets of the CO
clusters.

Starting with the histograms computed for clusters CO-1 to CO-6, shown on
the second rows, we obtain the same set of parameters as for the entire
field of view (first row).  There are only two exceptions. First, we need
relatively low \thCO{} opacities (from 0.03 to 0.5), confirming that we
deal with low-opacity lines of sight.  The horizontal edge between clusters
CO-6 and CO-7/8 in the \thCO{} versus \twCO{} histogram closely follows a
constant column density curve, confirming that it separates two column
density regimes (the cut corresponds to a \thCO{} opacity of
$\sim0.5$). Second, the minimum kinetic temperature is slightly lower than
for the global fit as we are less constrained by the lower edge of the
\thCO{} versus \CeiO{} histogram. It is unclear whether this fact is
significant.

The third and fourth rows show our best matches for clusters CO-7 and CO-8,
respectively. In both cases, only the high \thCO{} opacity (0.65 to 2.5)
curves are displayed. This confirms that we are in the high-column-density
regime. In both cases, we need to restrict the \thCO{} kinetic temperature
range from 11-12 to 20\K. The other parameters differentiate the two
clusters. First, the \twCO{}/\thCO{} kinetic temperature ratio is higher in
cluster CO-8 than in cluster CO-7. This confirms the idea that the outer
layers of the CO-8 cluster are more exposed to the FUV illumination than
those of the CO-7 cluster. Finally, the $[\thCO]/[\CeiO]$ abundance ratio
is much closer to the expected elemental abundance ratios for cluster CO-7
than for cluster CO-8. This is consistent with the idea that most of the
gas in cluster CO-7 is well shielded, in complete contrast with the gas in
cluster CO-8, as discussed in the following section.

\subsubsection{Discussion}

In all our models, we need to distinguish the kinetic temperature of the
gas that emits the \twCO{} \Jone{} line on the one hand, and the \thCO{}
and \CeiO{} \Jone{} lines on the other. The effect is the most
pronounced in cluster CO-8 that is highly FUV illuminated and then in
clusters CO-1 to CO-6 that contain diffuse and translucent gas. This implies
that the \twCO{} and \thCO{}/\CeiO{} emissions have different spatial
  extents along the line of sight (as \twCO{} emission quickly saturates 
and thus only traces a limited surface layer). It is thus impossible to use the flux
ratio of these lines to try to infer the [\twCO]/[\thCO] and
[\twCO]/[\CeiO] abundance ratios.

In the high column/volume density clusters, $^{12}$CO is very optically
thick~\citep[the saturation of the $^{12}$CO versus\ $A_\mathrm{V}$ relation
is noticeable starting from $A_\mathrm{V} \sim 5$ as shown by][]{Pety2017},
and the variations of integrated intensities are largely caused by
variations in the CO excitation temperature, which is close to the gas
kinetic temperature. Cluster CO-8 shows significantly higher $^{12}$CO
intensities ($\sim 90\,\Kkms$) than clusters CO-6 and CO-7 ($\sim
60\,\Kkms$) and thus traces a warmer dense gas regime. This is consistent
with its location around the NGC~2024 and NGC~2023 star forming regions.

Cluster CO-8 is also distinguished from cluster CO-7 by higher
$^{13}$CO/C$^{18}$O intensity ratios: the median ratio is $\sim 18$ in
cluster CO-8, while it is $\sim 11$ in cluster CO-7. This difference is a
sign of FUV illumination as the $^{13}$CO abundance in PDRs is tightly
coupled to $^{12}$CO by the isotopic fractionation reaction
\citep{Langer1984}
\begin{equation}
  ^{12}\mathrm{CO} + ^{13}\mathrm{C}^+ \rightarrow ^{13}\mathrm{CO} + ^{12}\mathrm{C}^+ + 35\,\mathrm{K},
\end{equation}
which, at the gas temperatures of PDRs, does not favour any enrichment, but
ensures a strong coupling between the abundances of \twCO{} and \thCO{}.
In contrast, C$^{18}$O in FUV-illuminated regions is formed separately
from pure carbon chemistry followed by reactions of small hydrocarbons such
as CH, CH$_2$ or C$_2$H with $^{18}$O. As a result, $^{13}$CO indirectly
benefits from $^{12}$CO self-shielding while \CeiO{} is easily
dissociated. This might explain the large $[\thCO]/[\CeiO]$ abundance ratio
compared to the value expected from elemental abundances.

Clusters CO-7 and CO-8 thus highlight a separation of the global intensity
PDF into two different tails at high column density that correspond to warm
illuminated dense regions around massive star forming regions (cluster
CO-8) and shielded dense gas (cluster CO-7).

\FigClusterFUVMapRaw{} %

\section{Adding HCO$^+$ and CN to get a better clustering of high-density
  and high-FUV-illumination regimes}\label{sec:Results_HCOp_CN}

Clustering ability is limited by the information contained in the tracers
input to the algorithm.  Our first application of the Meanshift clustering
algorithm to the CO isotopologues proved its ability to reveal several
distinct density regimes. It also hinted at a first distinction of FUV
illumination regimes. However, using only  the three CO isotopologues is
insufficient to clearly distinguish FUV illumination effects.  We thus now
include in the clustering analysis the HCO$^+$ and CN ($1-0$) maps together
with the three CO isotopologues maps. Indeed, \cite{Pety2017} and
\cite{Gratier2017} have shown that HCO$^+$ and CN were sensitive to FUV
illumination.  We chose CN rather than small hydrocarbons (C$_2$H or
c-C$_3$H$_2$) as the latter are detected at a lower signal-to-noise ratio,
making the clustering noisier.  For simplicity, we only used the brightest
hyperfine component of the CN \Jone{} transition. In addition, these two
lines have significantly higher critical densities ($\sim 2 \times
10^5\,\mathrm{cm}^{-3}$ for HCO$^+$ and $\sim 2 \times
10^6\,\mathrm{cm}^{-3}$ for CN).

We used in this case the adaptive bandwidth method with 425 neighbours.
This number is a compromise between eliminating artificial clusters coming
from sampling fluctuations of the PDF and retaining sufficiently fine
cluster subdivisions to find the interesting physical distinctions.

\subsection{Resulting spatial distribution}

\FigClusterFUVCompression{} %

Figure~\ref{fig:cluster_bis_map_raw} compares the clusters obtained based
on the CO isotopologues alone with the clusters obtained by adding CN and
HCO$^+$.  Several striking facts are visible. First, the CO clustering
brought 8 clusters while the new clustering identifies 19 clusters.  The
increase in number of clusters is related to the finer physics we wish to
reveal through the increase of the dimension of the intensity space from 3
to 5 lines, at a constant number of pixels.  We name the new clusters FUV-1 to FUV-19. 
For convenience they have been numbered by order of
increasing mean $^{12}$CO intensity.

Second, the spatial edges of the FUV clusters that appear from the
north-eastern to the south-western corners appear noisier. Indeed, mostly
translucent gas is present in these regions, implying that CN and HCO$^+$
are barely detected there. Third, a clear East-West asymmetry is now seen
in the distributions of the clusters. Some clusters, such as FUV-2, 3, 4,
7, 8 or 11, appear mostly on regions less exposed to FUV illumination,
while cluster FUV-16 is clearly associated with the NGC\,2024, NGC\,2023,
and IC\,434 \textsc{Hii} regions. In the previous clustering, cluster CO-8
does not tag the PDRs associated with IC\,434 (the Horsehead PDR, for
instance). The western edges of the CO-3 to CO-5 clusters are mostly merged
now in the cluster FUV-1 that mainly contains diffuse gas. For instance,
the envelope of the Horsehead nebula has been merged into FUV-1, only
leaving the less familiar silhouette of the denser parts of the Horsehead
visible.

Figure~\ref{fig:clusterFUV_compression} compares the spatial distributions
of the original line intensities with those of the line intensities
averaged per cluster. Comparing with Fig.~\ref{fig:clusterCO_compression},
we see that the FUV clustering reproduces the \twCO{} \Jone{} faint
intensity regimes $(\le 5\Kkms)$ less well, but it much better samples the
\twCO{} \Jone{} high intensity regime $(\ge 50\Kkms)$ and the \thCO{} and
\CeiO{} median intensity regimes (between 10 and $50\Kkms$, and between 2
and $6\Kkms$, respectively). The better sampling of the median- to high-intensity regimes of the CO isotopologue \Jone{} lines is linked to the
detection of several clusters at relatively high HCO$^+$ integrated
intensity $(\ge 2\Kkms)$. This is particularly clear on the \CeiO{} and
HCO$^+$ compressed maps that emphasise dense regions extending south of
  NGC\,2024 and surrounding NGC\,2023. In contrast, the high-CN-integrated-intensity end $(\ge 2\Kkms)$ is not well sampled by the new clustering.

In summary, this clustering seems to provide most of the FUV illumination
contrast between East and West. On the one hand, the FUV clustering thus
provides a better data compression in the inner dense parts and in
the FUV-illuminated regions; on the other, the shapes of the Horsehead
and of the western illuminated edge are less well reproduced.

\FigClusterFUVMapGroupedHCOp{} %

While the 19 clusters are statistically significant, interpreting all of
them is difficult. Indeed, clusters with extreme behaviour have relatively
clear physico-chemical interpretations but they are separated in the line
space by clusters with intermediate properties that reflect subtler,
second-order distinctions.  The major physical distinctions brought forward
by this clustering are thus best discussed in terms of groups of clusters.
Moreover, we wish to understand the first-order roles of the HCO$^{+}$ and
CN \Jone{} lines in the classification. We thus group the clusters in two
ways. We group together clusters with similar most probable intensities of
HCO$^{+}$ or CN respectively, ordered by increasing values.  Our goal is to
keep the minimum number of groups needed to visualise the physico-chemical
regimes first brought forward by each line.

\subsection{HCO$^+$-based grouping and high-density
  regimes}\label{sec:HCOp-groups}

We first constitute groups based on the HCO$^+$ intensities in each cluster
following the numerical recipe explained in the previous section.

\subsubsection{One-dimensional PDFs and spatial distribution}

The left panels of Fig.~\ref{fig:cluster_bis_map_HCOp_grouped} show the
HCO$^+$ PDF of each individual cluster, and the way we grouped them: the
clusters whose most probable values (1D-PDF peaks) gather at similar
integrated intensities are grouped. We end up with seven groups, named
HCO$^+$-1 to HCO$^+$-7, whose mean integrated intensities regularly
increase from 0.3 to 6.3\Kkms{}.  Cluster FUV-1 alone constitutes group
HCO$^+$-1 as it mostly traces diffuse gas surrounding the molecular
cloud. We nevertheless note that it has a wing between 1 and 3\Kkms{} that
corresponds, for instance, to the Horsehead envelope.  Group HCO$^+$-2
contains clusters FUV-2, 3 and 4, Group HCO$^+$-3 clusters FUV-6, 7, 8 and
9, group HCO$^+$-4 clusters FUV-5, 10, 11, 13 and 15, group HCO$^+$-5
clusters FUV-12, 16 and 17, group HCO$^+$-6 clusters FUV-14 and 18, and
finally, group HCO$^+$-7 contains only cluster FUV-19. The characteristic
intensity values (median, mean, standard deviation) of the lines in each
group are listed in Table~\ref{tab:clustersHCOp_intensities}.

The spatial distribution of these groups is displayed as the right panel of
Fig.~\ref{fig:cluster_bis_map_HCOp_grouped}. We see that the resulting
groups have a faint mean HCO$^+$ intensity in the outer part of the cloud
and that this intensity increases towards the densest parts.  In comparison
to the CO clustering, the CO-6 to CO-8 bright clusters are now distributed
over groups HCO$^+$-3 to HCO$^+$-7, and clusters CO-1 to CO-5 (faint CO
intensity) are distributed over the groups HCO$^+$-1 and HCO$^+$-2.  This
suggests that the HCO$^+$ \Jone{} line is better at discriminating
higher-density regimes than the CO \Jone{} lines, even though about half of
the HCO$^+$ flux over the observed field of view is coming from diffuse and
translucent regions~\citep{Pety2017}. Finally, the boundaries of some
groups are close to the CO clustering results, indicating that CO
isotopologues still play an important role in defining some of the groups.
For instance, the boundary between the HCO$^+$-4 and HCO$^+$-5 groups is
similar to the boundary between clusters CO-7/8 and CO-6.

\subsubsection{Two-dimensional PDFs}

\FigClustersFUVTwoDPDFHCOpGrouping{} %

To further understand the relative roles of the CO isotopologues and
HCO$^+$, we show in Fig.~\ref{fig:clusters_bis_2D_PDFs_HCOp_grouping} the
2D-PDFs of $^{13}$CO versus C$^{18}$O (left) and HCO$^+$ versus C$^{18}$O
(right), which we found to be the most informative among the possible pairs
of lines. The first striking impression is that the groups overlap
considerably in both 2D PDFs. However, some groups clearly separate in one
of the 2D PDFs but not in the other. For instance, while groups HCO$^+$-4
and 6 strongly overlap in the (HCO$^+$ vs.\ C$^{18}$O) PDF, they are
cleanly separated in the ($^{13}$CO vs.\ C$^{18}$O) PDF. Finding the right
2D projection to reveal cluster separations quickly becomes
impossible. Moreover, this 2D projection might not even exist when the
clusters are not linearly separable; for example, when one cluster is completely
surrounded by another one. We have to rely on the Meanshift algorithm to
reveal information about the morphology of the complete PDFs (maxima and
their associated basins of attraction) that we cannot otherwise directly
visualise.

The groups form a sequence that mostly follows a single trend with
increasing line intensities. This is better visualised when trying to
connect the crosses that represent the group mean intensities. Group
HCO$^+$-5 only partially follows this trend: a part of it (actually mostly
cluster FUV-16) is overluminous in HCO$^+ $ at constant \CeiO{}
intensity. This is linked to the sensitivity of the HCO$^+$ \Jone{} line to
the FUV illumination (cf.\ the following section).

\subsubsection{Link with volume density and FUV illumination}

\FigClustersFUVdensityPDFHCOpGrouping{} %
\FigClustersFUVGzViolinHCOpGrouping{} %

As shown in \cite{Gratier2017}, the main underlying parameter contributing
to intensity variations across our maps is the gas column density. The
single trend highlighted here could thus to first order be associated with
column density variations and thus approximate volume density
variations as discussed in Section~\ref{sec:GouldBeltSurvey_obs}.
Figure~\ref{fig:clusters_fuv_density_PDF_HCOp_grouping} shows the PDF and
median value of the volume density in each group (violin plots). We indeed
see that the HCO$^+$-groups correspond to increasing ranges of volume
densities. On the one hand, the first five HCO$^+$-groups contribute to the
three main peaks of the approximate volume density PDF: the
HCO$^+$-1 group corresponds to the diffuse gas peak, the HCO$^+$-2 and 3
groups dominate the translucent gas peak, and the HCO$^+$-4 and 5 groups
contribute to the denser gas peak. On the other hand, groups HCO$^+$-6 and
7 are located in the high-density tails.

While part of the distinction is linked to the underlying existence of the
three density regimes, the distinction between groups HCO$^+$-6 and 7 hints
at the existence of higher-density regimes. Their rarity makes them only
barely noticeable as bumps in the density PDF. This distinction is probably
the result of an excitation effect. Indeed, their typical volume densities
(probably underestimated as they are averaged along the line of sight) are
$7\times10^3$, and $4\times10^4\pccm$, respectively.  These values approach
the critical density of HCO$^+$ for collisional excitation with H$_2$
($\sim10^5\pccm$). We thus probably experience a transition from a weak
excitation regime \citep{Liszt2016} towards a regime closer to the usual
thermalised excitation.

Figure~\ref{fig:clusters_fuv_G0_violins_HCOp_grouping} shows the PDF and
median value of the FUV illumination for each group. In contrast to the
density, no clear separation of the groups in terms of FUV illumination is
visible, except for the fact that HCO$^+$-5 has a much broader $G_0$
distribution than the other groups. The HCO$^{+}$ grouping thus does not
cleanly capture distinctions only related to FUV illumination, even though
the complex behaviour of HCO$^+$-5 in the 2D PDFs of the line intensities is
likely related to the presence of a mixture of FUV illumination (varying by
more than one order of magnitude) in this group. This is not a property of
the initial 19 clusters but of the grouping, as is shown in the following Section.

\subsection{CN-based grouping and FUV illumination
  regimes}\label{sec:CN-groups}

We now present the second grouping of the clusters, based on their CN
intensities.

\subsubsection{One-dimensional PDFs and spatial distribution}

\FigClusterFUVMapGroupedCN{} %

The left panels of Fig.~\ref{fig:cluster_bis_map_CN_grouped} show how the
CN PDF of the individual clusters contributes to the PDF of their
group. After sorting them by increasing CN mean intensity, we merged the
first six clusters into group CN-1 (clusters FUV-1, 2, 3, 4, 7, 8) because
they correspond to regions where CN is not detected. We then merged the
next seven clusters into group CN-2 (clusters FUV-5, 6, 9, 10, 11, 13,
15). The CN line is barely detected in these clusters. Groups CN-3 and CN-4
gather the next three (FUV-12, 14 and 17) and two clusters (FUV-16 and 18),
respectively. The last cluster (FUV-19) is significantly brighter in CN. It
thus has its own group CN-5. The characteristic intensity values (median,
mean, standard deviation) of the lines in each group are listed in
Table~\ref{tab:clustersCN_intensities}.

The most striking distinction revealed by the resulting spatial
distribution (see the right panel of
Fig.~\ref{fig:cluster_bis_map_CN_grouped}) is the separation of the central
regions of the cloud between groups CN-3 and CN-4; contrary to the previous
cases, this separation does not show a nested pattern.  The CN-3 group
covers dense regions in the inner parts of the cloud. The CN-4 group
appears towards the interfaces between the molecular cloud and the
NGC\,2024, NGC\,2023, and IC\,434 \textsc{Hii} regions.  This distinction
thus seems to separate FUV-shielded dense gas (group CN-3) and
FUV-illuminated dense gas (group CN-4).  Group CN-5 highlights smaller
regions at the interface between groups CN-3 and CN-4. Groups CN-1 and 2
represent outer regions of the cloud.

\subsubsection{Two-dimensional PDFs}

\FigClustersFUVTwoDPDFCNGrouping{} %

Figure~\ref{fig:clusters_bis_2D_PDFs_CN_grouping} shows the 2D PDFs of CN
versus\ C$^{18}$O (left panel), and CN versus\ HCO$^+$ (right panel); it compares
the full dataset PDF (dashed line) with the contributions of our five
groups (solid lines). Two distinct trends in the tail of the PDF are
obvious on the 2D PDF of CN versus\ HCO$^+$: a low CN/HCO$^+$ ratio
corresponding to group CN-5, and a high CN/HCO$^+$ ratio corresponding to
group CN-4. The other groups lie in a low-intensity region where the two
trends are blended.

This dual trend is already seen on the (CN vs.\ C$^{18}$O) 2D-PDF: for
groups CN-2, CN-3, and CN-5, CN increases very slowly with C$^{18}$O, in a
mostly linear way. In contrast, group CN-4 has higher CN intensities than
the other groups, with the CN intensity increasing much faster with the
C$^{18}$O one; although this trend has a larger scatter.

Finally, while there is a large overlap region between groups CN-4 and CN-5
in the PDF of CN versus\ HCO$^+$, these two groups are clearly separated in
the PDF of CN versus\ C$^{18}$O. Group CN-4 is thus observationally
distinguished by an overly bright CN emission.

\subsubsection{Link with FUV illumination (and volume density)}

\FigClustersFUVDustViolinsCNGrouping{} %

As a species easily detected in diffuse clouds, CN is a good tracer of FUV-illuminated gas~\citep{Snow2006}.  We thus interpret the dual trend as a
separation between regions where the photo-chemistry is active and regions
of dense FUV-shielded molecular gas. The spatial consistency of groups CN-3
with regions of dense FUV-shielded gas and CN-4 with FUV-illuminated gas
strengthens this interpretation, all the more so that spatial information
is not used in the clustering analysis.

More quantitatively,
Fig.~\ref{fig:clusters_bis_2D_dust_violins_CN_grouping} shows the
approximate volume densities (left panel) and FUV illuminations
(right panel) found for the different groups. Groups CN-3 and CN-4 have
very similar volume densities (median densities of $\sim 3\times 10^3$ and
$\sim 4\times 10^3 \pccm$, respectively), while the CN-1, CN-2, and CN-5
correspond to distinct ranges of volume densities (respective median
densities of $\sim 2\times 10^2$, $\sim 8\times 10^2$ and $\sim 4\times
10^4\pccm$). In contrast, the $G_{0}$ distributions show that group CN-4
clearly has higher FUV illumination (median $G_0 \sim 210$) than all other
groups (median $G_0$ between 15 and 30), in particular groups CN-3 and CN-5
(median $G_0 \sim 20$ and $G_0 \sim 15$, respectively). This confirms our
interpretation that group CN-4 corresponds to FUV-illuminated (relatively)
dense gas.

Groups CN-1 and CN-2 have lower approximate volume densities than
groups CN-3 and CN-4.  However the distribution of $T_\mathrm{dust}$ and
therefore $G_{0}$ in groups CN-1 and CN-2 overlaps with that of both groups
CN-3 and CN-4. This shows that it is more difficult to separate the
influence of the radiation field for low-density regions.  This is related
to the fact that the envelope of the Horsehead nebula has been merged into
group CN-1, leaving only the less familiar silhouette of the denser parts
of the Horsehead visible.

The highest-density group (CN-5) is found only in the immediate vicinity of
the two star-forming regions NGC~2023 and NGC~2024 and their H\textsc{ii}
regions.  This might be a signature of compression of the molecular gas by
the expansion of the H\textsc{ii} regions: \cite{Tremblin2014} have indeed
found this process to cause bimodality in the column density PDF (on
spatial scales of a few pc). The gas kinematics in these two regions was in
addition found to be dominated by compressive (rather than solenoidal)
motions by \cite{Orkisz2017}, in contrast to the rest of the field of view.

\subsubsection{Interpretation: Enrichment of HCO$^+$ and CN with respect to
  \CeiO{} in FUV-illuminated gas}\label{CN-group-interpretation-Radex}

The PDFs of HCO$^+$ versus\ C$^{18}$O and CN versus\ C$^{18}$O show a dual regime
at high C$^{18}$O \Jone{} intensity (cf.\
Fig.~\ref{fig:clusters_bis_2D_PDFs_HCOp_grouping}, right panel, and
Fig.~\ref{fig:clusters_bis_2D_PDFs_CN_grouping}, left panel), with one
regime where HCO$^+$ and CN are overluminous relative to \CeiO{}. This
latter regime is clearly associated with high FUV illumination at high
volume density. We here check whether this is the sign of a chemical
enrichment of HCO$^+$ and CN in FUV-illuminated regions, using a non-LTE
code (RADEX, \citealt{VanDerTak2007}) to estimate the column densities of
these species.

The modeling details and the derived column densities are described in
Appendix~\ref{Appendix-radex-models}. For simplicity, we only model the
radiative transfer for typical conditions in each group (median gas volume
density, median kinetic temperature, and median line integrated
intensities). As an estimate of the kinetic temperature, we take a
combination of the dust temperature and the $^{12}$CO excitation
temperature as in the Appendix A of \cite{Orkisz2017}: we take the
\twCO~excitation temperature when it is above 60\K, and the maximum of the
dust temperature and \twCO~excitation temperature otherwise.

\FigRadexColdenRatios{} %

Figure~\ref{fig:RadexColdenRatios} shows the behaviour of the derived
abundance ratios [HCO$^+$]/[C$^{18}$O], and [CN]/[C$^{18}$O] as a function
of the $G_0/n_\mathrm{H}$ parameter, which is expected to be the dominant
controlling parameter of the physics and chemistry in a
PDR~\citep{Hollenbach97}.  An increase by about two orders of magnitude of
the HCO$^+$ and CN abundances relative to C$^{18}$O can be seen when
increasing $G_0/n_\mathrm{H}$.  The abundances relative to \CeiO{} seem to
reach a constant value at high $G_0/n_\mathrm{H}$ values in both cases.

Fractional abundances for each of the three species are also computed
relative to the total (dust-derived) column density $N_H$. These abundances
are shown in the lower panel of Fig.~\ref{fig:RadexColdenRatios} as a
function of $G_0/n_\mathrm{H}$.  Both the HCO$^+$ and CN abundances steeply
increase with $G_0/n_\mathrm{H}$ with $G_0/n_\mathrm{H}$ at first (up to
$G_0/n_\mathrm{H} \sim 10^{-2}$), before saturating at a nearly constant
value up to $G_0/n_\mathrm{H} \sim 10^{-1}$. The \CeiO{} abundance smoothly
decreases with increasing $G_0/n_\mathrm{H}$ over the full range of values.
This marked difference of chemical behaviour therefore explains the
difference in line intensities identified by the MeanShift algorithm.  The
abundances of CN and HCO$^{+}$ remain at a high but nearly constant value
in strongly or mildly FUV-illuminated regions, while the \CeiO{} abundance
decreases with increasing FUV illumination.  The overbright HCO$^+$ and CN
regime found in FUV-illuminated regions is thus caused by a combination of
photochemical enrichment in CN and HCO$^+$ and photodissociation of
\CeiO{}.

\section{The Meanshift algorithm, an interesting clustering method: biases
  and data requirements}
\label{sec:Discussion_Meanshift}

We first present why and how clustering and principal component analysis
are complementary. We then discuss the effects (noise, sampling,
dimensionality) that can alter our results.

\subsection{On the complementarity of clustering and principal component
  analyses}

We chose here to use a clustering approach in order to analyse the
structure of the multi-dimensional PDF of several line intensities, based
on the idea that this structure can reveal interesting insights into the
physics and chemistry at play.  As soon as we use a dataset with more than
two dimensions, visualising the structure of its PDF becomes difficult, and
specific methods must be used.  Each of these methods is usually focused on
highlighting a particular kind of structure, and applying different methods
to the same dataset thus provides complementary results.

For instance, clustering provides a complementary approach to Principal
Component Analysis (PCA). On the one hand, PCA highlights the
non-sphericity of the data by revealing the axes of strong covariance or
correlation. However, PCA cannot capture non-linear patterns of
co-variations between the intensities. Moreover, PCA highlights variations
around a centre of the dataset (usually the mean), which might not be
relevant if the data points are gathered in several natural clusters, with
different centres (that is, if the PDF is multimodal).  On the other hand,
clustering algorithms aim at revealing any grouping of the data points in
different regions of intensity space, that is, to reveal multimodality in
the (multi-dimensional) PDF of the data.

By applying PCA to this dataset, \cite{Gratier2017} showed that column
density, volume density, and FUV illumination are some of the underlying
parameters controlling the intensity variations. And they listed the
tracers that are the most affected by each of these parameters. The
clustering analysis we have performed in the present study reveals, in
addition, a multimodality of the line intensity PDF with modes related to
the density and the FUV illumination. In the case of the CO clustering, the
modes of the CO isotopologue PDF are directly related to modes of the
column/volume density PDF, thus revealing the existence of
distinct density regimes in the Orion B cloud. In our clustering analysis
including HCO$^+$ and CN, the transition to a photon-dominated chemistry
leads to a separate mode corresponding to dense PDRs. In this case,
clustering reveals a transition between two different physical/chemical
regimes.

Moreover, both approaches can be used as data compression methods, in order
to reduce the volume of data before applying some other very time-consuming
data analysis. PCA compresses the dataset by reducing the number of
variables characterising each data point, while clustering can be used as a
segmentation method, discretising the possible values of each
variable. Finally, these two approaches could be combined in future
work. One possibility would be to decompose a multi-line PDF into separate
components with simpler structure before applying PCA to each of these
components.  Alternatively, a PCA analysis could be performed first to
eliminate irrelevant components of the data (e.g. noise), followed by a
clustering analysis restricted to the relevant features deduced by the PCA.

\subsection{What is the impact of noise?}\label{sec:noise_effects}

The effect of the measurement noise present in our line-intensity maps on
the results of the Meanshift algorithm can be understood in two ways: 1)
Its effect on the data PDF, used by the Meanshift algorithm to define
clusters, and 2) its effect on the attribution of a given pixel to one of
the clusters.

Assuming an identical noise rms, $\sigma$, on all datapoints~\citep[a good
approximation for this dataset, see][]{Pety2017}, the addition of Gaussian
noise to the true variables replaces the underlying PDF by an observed PDF
that is the convolution of the underlying PDF with a Gaussian of standard
deviation $\sigma$. This is equivalent to a Gaussian smoothing of the
PDF. This effect will mostly tend to merge some maxima if their separation
is too small compared to the smoothing scale (that is, the noise level),
rather than creating artificial maxima.  The smoothing effect can also
slightly shift the position of the extrema. But the existence of the
clusters will be unaffected as long as their PDF maxima are well separated
compared to the smoothing scale.

Noise also alters the boundaries between the clusters.  If the true
intensity values of a given pixel place it close enough to a boundary
between clusters in the line space (typically closer than the noise level),
adding noise can move this pixel across the border, and thus change the
cluster to which it belongs. As a result, noise on the line intensities
tends to make the spatial boundaries between clusters appear noisy (not
forming a regular curve on the map). This effect is more pronounced in
regions of the map where intensity gradients are small. In this case,
pixels relatively far away from the cluster spatial boundary can still be
close to the cluster boundary in the line space and thus be transferred to
another cluster. On the contrary, if intensity gradients are steep at the
spatial boundary, even pixels located just one or two pixels away from the
spatial boundary can be far enough from the cluster boundary in the line
space so that noise is unlikely to transfer them to another cluster. This
effect is at play when we include CN and HCO$^+$ which have low
signal-to-noise ratios on significant fractions of the map. Relatively
noisy boundaries can be seen on the eastern side of the cloud while the
boundaries on the western edge remain sharp (cf.\
Fig.~\ref{fig:cluster_bis_map_raw}). The intensity gradients are indeed
much steeper on the western side of the cloud than on the eastern side.

\subsection{What is the impact of limited sampling (field of view)?}
\label{sec:sampling_noise}

The Meanshift algorithm estimates the PDF gradient in the line space in
order to find the PDF maxima, and it needs to estimate this gradient from a
finite sample (the observed dataset). There are two different aspects
here. First, the observed field of view may be biased towards some values
of the parameters that control the physics or the
chemistry. \citet{Pety2017} showed that this is the case here, as the
studied field of view has a large FUV illumination compared to the
Interstellar Standard Radiation Field because it includes several
H\textsc{ii} regions.  In the ORION-B project, we will increase the
observed field of view towards regions of lower FUV illumination to
circumvent this limitation. In the meantime, our clustering analysis must
be interpreted with this limitation in mind.

Second, the gradient estimate can be affected by sampling noise: a
different dataset drawn from the same underlying PDF (corresponding to this
specific field of view) would yield slightly different gradient estimates,
and thus converge towards slightly different maxima.  We chose to use an
adaptive bandwidth (so that the kernel always includes the same number of
datapoints), rather than a fixed bandwidth (which would give better sampled
gradient estimates close to the major PDF peaks than in the tails) to
reduce the impact of this effect. This ensures that the sampling noise is
similar for all gradient estimations, and it avoids finding
sampling-noise-induced artificial maxima in the tail of the PDF. Having at
least a few hundred datapoints in the kernel generally ensures that the
sampling noise has negligible effect.

However, implicitly increasing the kernel bandwidth in low-density regions
of the PDF means a decreased capacity to resolve small-scale features in
the PDF. As a result, PDF maxima corresponding to a small number of pixels
(compared to the kernel size) might be smoothed out unless these pixels
have intensities widely different from all other pixels. This means that
our analysis is likely to miss specific physical or chemical regimes if
they occur on too small a region of the map. This was the case in our tests
where we found that dense cores are difficult to capture as a cluster with
the Meanshift algorithm even when including specific dense core chemical
tracers such as N$_2$H$^+$.

\subsection{Choice of the number of molecular lines included in the
  analysis}

In contrast to our PCA study~\citep{Gratier2017}, we limited the clustering
analysis to a moderate number of lines (5 at most). This choice was driven
by several considerations.

The first reason is practical. As discussed in
Sect.~\ref{sec:noise_effects}, while including low signal-to-noise-ratio
(SNR) line maps may change the total number of clusters, the presence of
such low-SNR data always degrades the quality of the cluster boundaries.
Well defined clusters with clear interfaces are only obtained in the
regions where all lines have high SNR. A consequence of using several lower-brightness lines at relatively constant noise level (our observing case) is
therefore a reduction of the usable pixels to the regions where all the
lines have high SNR, that is the regions of highest column densities. This
requirement of significant SNR for all line maps limits the usefulness of
the Meanshift algorithm when applying it to a large set of lines of varying
SNR. We thus restricted our study to lines showing extended emission with
high SNR on a large fraction of the map.

The second reason is more fundamental.  In this paper, we try to understand
the physical and chemical processes that regulate the intensities of the
lines used in the clustering algorithm.  The clustering of the CO
isotopologue ground-state line maps, complemented by the HCO$^+$ and CN
lines, showed two trends. First, the number of significant clusters
increases with the number of lines because we add lines that exhibit
different sensitivities to the physical or chemical processes at
work. Second, the interpretation of a large number of clusters is difficult
because the associated clusters have less data points implying a lower
statistical significance of the trends. Moreover, it is difficult to get a
good appreciation of the full distribution of the data over which
the Meanshift algorithm operates in dimensions larger than two through
standard 2D PDFs.

Our experience is thus that it is better to start clustering in low
dimension to understand the sensitivity of the different lines to the many
underlying physical and chemical processes at play in the ISM.  It will
then be possible to cluster higher-dimension data to get a finer
segmentation that will depend on the underlying properties that need to be
emphasised.

\section{Conclusions}
\label{sec:Conclusions}

In this paper, we present a segmentation of the Orion B molecular cloud
into regions of similar molecular emission, in order to reveal the
different physical and chemical phases constitutive of molecular clouds.
We have applied the Meanshift algorithm, a PDF-based (unsupervised) clustering
algorithm defining clusters around the maxima of the PDF, to the
(high-dimensional) multi-line PDF of our dataset.  This is the first
application of a clustering analysis based on molecular emission properties
only (and not spatial proximity between pixels) to ISM data.

We first applied the clustering analysis to the maps of the three main CO
isotopologue lines only.  While the clustering did not take the spatial
distribution of the CO emission into account, it highlighted a nested
pattern from the outer edges to the innermost parts of the Orion B
cloud. Comparison with an approximate volume density map showed
that the clusters have increasing typical volume densities with significant grouping at densities of 100, 500, $>1000\pccm$. The CO
isotopologue maps alone were thus found to be sufficient to reveal the
existence of the diffuse, translucent and high-column-density regimes.
Simple LTE radiative transfer modelling implies that the gas
  emitting the \twCO{} \Jone{} line is more extended than the gas emitting
  the \thCO/\CeiO{} \Jone{} lines. It is thus impossible to use the flux
ratio of these lines to try to infer the [\twCO]/[\thCO] and
[\twCO]/[\CeiO] abundance ratios.

In the densest regime, an additional separation of the PDF in two distinct
tails was found, which we could associate to FUV-illumination
effects. Comparison with LTE radiative transfer models shows that this
distinction is related to the presence of a warmer \twCO-traced surface
layer and higher than usual [\thCO]/[\CeiO{}] ratios, which can both be
explained by the increased FUV illumination caused by the nearby H\textsc{ii}
regions.  It however proved insufficient to get a satisfactory separation
of the FUV illuminated regions.

We thus added two FUV-sensitive tracers (the $(1-0)$ lines of HCO$^+$ and
CN) to the CO isotopologues, and performed a second clustering
analysis. This analysis revealed a similar separation into increasing
density regimes, but captured finer distinctions at higher density
($n_\mathrm{H}\sim 10^4$ and $5\times10^4\pccm$) due to the high critical
density of the added tracers. \cite{Pety2017} have however shown
  that about half of the HCO$^+$ flux over the observed field of view is
coming from diffuse and translucent regions, implying that the use of the
HCO$^+$ line intensity as a tracer of high density gas $(\ga 10^4\pccm)$ in
unresolved GMC observations is questionable.

Moreover, the clustering also revealed the existence of another clear
separation of the data at high column density. On the one hand, part of the
data presents a CN and HCO$^+$ \Jone{} emission that is overly bright with
respect to the \CeiO{} \Jone{}. This data also shows a high CN/HCO$^+$
intensity ratio. The associated lines of sight form the dense PDR regions
around the star forming regions NGC\,2023 and NGC\,2024, and on the Orion B
western edge illuminated by $\sigma$ Ori\ (including the Horsehead
PDR). On the other hand, other high-column-density lines of sight have low
CN/\CeiO, HCO$^+$/\CeiO, and CN/HCO$^+$ intensity ratios. These lines of
sight correspond to the FUV-shielded, dense regions in the inner
  parts of Orion B. Non-LTE models show that this distinction is related
to a clear increase of the [HCO$^+$]/[\CeiO{}] and [CN]/[\CeiO{}] abundance
ratios with $G_0/n_\mathrm{H}$.

Our clustering analysis based on the \Jone{} lines of the CO isotopologues,
HCO$^+$ and CN, thus managed to both capture finer density categories in the
densest regions, and to reveal the existence of two distinct chemical
phases (characterised by different abundance ratios) corresponding to
FUV-induced photochemistry and shielded-gas chemistry.  This exposes the
wealth of physical and chemical information that can be inferred from
molecular tracers when powerful statistical methods (as the Meanshift
algorithm) are applied to large amounts of data. One of the next steps in
the ORION-B project is to stack the spectra inside each of the clusters
found here, to better characterise the molecular content of each regime
using the whole information available in the 3 mm band.

\begin{acknowledgements}
  The authors thank the referee for his constructive comments, useful
  suggestions, and his very fast refereeing.
  EB post-doctoral position during this work was funded by the ERC grant
  ERC-2013-Syg-610256-NANOCOSMOS.  This work was in part supported by the
  Programme National “Physique et Chimie du Milieu Interstellaire” (PCMI)
  of CNRS/INSU with INC/INP, and co-funded by CEA and CNES. We thank the
  CIAS for its hospitality during the three workshops devoted to this
  project. JRG thanks the Spanish MINECO for funding support through grant
  AYA2012-32032. NRAO is operated by Associated Universities Inc. under
  contract with the National Science Foundation. This paper is based on observations carried out at the IRAM-30 m
single-dish telescope. IRAM is supported by INSU/CNRS (France), MPG
(Germany) and IGN (Spain). This research also used data from the Herschel
Gould Belt survey (HGBS) project (http://gouldbelt-herschel.cea.fr). The
HGBS is a Herschel Key Programme jointly carried out by SPIRE Specialist
Astronomy Group 3 (SAG 3), scientists of several institutes in the PACS
Consortium (CEA Saclay, INAF-IFSI Rome and INAF-Arcetri, KU Leuven, MPIA
Heidelberg), and scientists of the Herschel Science Center (HSC).
\end{acknowledgements}

\bibliographystyle{aa} %
\bibliography{Clustering} %

\appendix{}

\section{Column density and approximate volume density}
\label{sec:column-vs-volume-density}

\FigDensity{} %

In this Section, we explore whether the assumptions of isotropy and of
a nested distribution of volume density (density smoothly increasing from the outer regions to the
inner parts of the cloud) can be used to derive an approximate
volume density from the spatial distribution of the column density. After a
discussion of the spatial and statistical distributions of the column
density, we use the above assumptions to derive an approximate volume
density. We then compare the derived values to previously published
measures of the density in this field of view to understand the limits of
the method and to estimate its accuracy.

Figure~\ref{fig:volume-vs-column} shows the spatial (panel a) and
statistical (panel b) distributions of the gas column density deduced from
the dust continuum emission (see Sect.~\ref{sec:GouldBeltSurvey_obs}). The
targeted region exhibits a large range of column densities. There is
slightly more than a factor of 100 between the minimum and maximum
values. A hypothesis of nearly constant gas volume density would imply that
the high-column-density regions should be about one hundred times deeper
than the low-column-density ones, which would require an unrealistic cloud geometry.  
Moreover, the spatial distribution of the column density
shows a nested pattern. Indeed the highest-column-density contours are
surrounded by the smaller-column-density ones. This is also easily seen on
the column density PDF (Fig.~\ref{fig:DustPDFs}), which shows a faint tail
at high column density.  Higher-column-density regions must thus be less extended
along the line of sight (according to our isotropy hypothesis), and as a consequence be associated with higher volume densities, at least in the statistical
sense.

Assuming the simplest possible hypotheses about the spatial distribution of
volume density, that is the hypothesis of no privileged direction and
nested increasing volume density, we estimate the typical lengthscale
  $l$ of regions with a given column density value $x$ as the square root
  of the projected surface area where $N_\mathrm{H} \ge x$.  Pixels with
$N_\mathrm{H}$ are then given the volume density $n_\mathrm{H} =
N_\mathrm{H}/l$.  This procedure is illustrated in the last three panels of
Fig.~\ref{fig:volume-vs-column}.  This reasoning assumes a one to one
correspondence between a column density and the approximate volume
density. We only expect this relationship to hold in a statistical sense:
it may not be valid pixel by pixel but we expect it to correctly represent
the range of volume densities at a given column density.

\FigDensityComparison{} %
\FigAnalyticalComparison{} %

In order to estimate the accuracy of this estimate, we gathered volume
density estimations from the literature at positions within our field of
view, and derived from a variety of different methods. From the catalogue of
cores of \cite{Kirk2016} (SCUBA dust emission observations), we derived
core masses from their Eq. 3 (using the background-substracted fluxes),
and we complemented the mass value with their measured size to yield
  volume densities. We completed this sample with a few density
  estimates at particular positions. The volume density of the diffuse
foreground gas of HD~38087 was derived by \cite{Martin-Zaidi2008} from
H$_2$ UV absorption lines through the use of PDR models.  The volume
  densities for two positions in NGC2024 was inferred from H$_2$CO
  emission lines~\citep{McCauley2011}. And the volume densities in the
  B33-SMM1 and B33-SMM2 cores in the Horsehead were derived from dust
  emission observations~\citep{Ward-Thompson2006}.
  Figure~\ref{fig:comparaison_densities} shows our volume density estimates
  compared to the ones from the literature. In addition, the red points and
  their error bars show the average of our density estimate in log-spaced
  bins of the literature density values.

On average, our density estimate is within a factor of 3 of the densities
from the literature, with a typical scatter of one order of magnitude.  For
the diffuse medium data point of \cite{Martin-Zaidi2008}, we find a density
$\sim10$ times larger. This is consistent with the typical scatter, but could
also come from a bias of our estimate at low densities, as our lengthscale
estimate is limited to the size of our field of view while diffuse medium
might span larger scales. We note that our estimate is however
qualitatively correct by predicting a diffuse-medium-like density at this
position.

As an additional check, we consider a simple analytical example: a
spherically symmetrical cloud whose volume density profile is
\begin{equation}
  n(r) = n_0 \left(
    1 + \frac{r^2}{r_0^2} \right)^{-\frac{1+\alpha}{2}}
    \quad \mathrm{with} \quad
  \alpha \ge 1  
,\end{equation}
(used for instance in \citealt{Krco2016}).
This corresponds to an observed column density profile
\begin{equation}
  N(r) =  \frac{n_0 r_0\Gamma(\frac{\alpha}{2})\sqrt{\pi}}{\Gamma(\frac{\alpha+1}{2})} \left( 1 + \frac{r^2}{r_0^2} \right)^{-\frac{\alpha}{2}}.
\end{equation}
Our simple estimate of the volume density would yield at a distance $r$
  from the cloud centre
\begin{equation}
  n^{\mathrm{estimate}}(r) = \frac{N(r)}{\sqrt{\pi r^2}}.  
\end{equation}
The ratio of the two values,
\begin{equation}
  \frac{n^{\mathrm{estimate}}(r)}{n(r)} 
  = \frac{r_0}{r}\sqrt{ 1 + \frac{r^2}{r_0^2}}\frac{\Gamma(\frac{\alpha}{2})}{\Gamma(\frac{\alpha+1}{2})}, 
\end{equation}
is shown on Fig.~\ref{fig:comparaison_densities} for different values
  of $\alpha$. Our estimation ranges from 0.8 to 2.5 times the true density
  value for all estimation scales larger than the typical scale $r_0$ and
  it diverges when the estimation scale is much smaller than $r_0$. 
  This comes from the fact that the surface estimate describing the scale of
the inner region goes to zero close to the centre. However, the spatial
resolution of our data (60\,mpc) avoids the regime $r \ll r_0$. Indeed, our
comparison to literature values does not show any dramatic overestimation
for the high-column-density pixels.

In conclusion, the method proposed here provides a reasonable estimate
in a statistical sense with a bias of a factor 3 at most and a typical scatter
of one order of magnitude.  It can
thus be trusted for order-of-magnitude comparisons. Moreover, this
estimation of the volume density map is completely independent from the
clustering analysis presented in this paper, and is only used to help in
the interpretation of the clustering analysis.

\section{Impact of the number of neighbours on the number of
  clusters}\label{sec:Kneighbors_choice}

\FigNclustersVsKneighborsCO{} %

In the Adaptive Meanshift algorithm, $k_\mathrm{neighbors}$ controls the
size of the adaptive bandwidth of the smoothing kernel: for each datapoint
the bandwidth is automatically adjusted to include its
$k_\mathrm{neighbors}$ nearest neighbours.  As a result, this parameter
controls the degree of smoothing applied to the PDF gradient estimate,
while ensuring equal sampling in the peaks and tails of the PDF.

Most of the time, increasing $k_\mathrm{neighbors}$ will result in a
decrease in the number of clusters found by the algorithm as local maxima
are merged by the increased smoothing.  As discussed in
Sect. \ref{sec:sampling_noise}, too low a value of $k_\mathrm{neighbors}$
results in finding artificial maxima caused by the sampling noise. A very
large value will smooth the data PDF to a single peak with the shape of the
kernel. When decreasing $k_\mathrm{neighbors}$ from this large value, the
most well separated maxima of the PDF will be distinguished first, then
maxima that are weaker or close to the highest maxima.

The first maxima that appear are thus likely to correspond to the strongest
physical distinctions, and the following ones to subtler and subtler
distinctions (until artificial maxima caused by sampling noise start to
appear).  As a result, the choice of $k_\mathrm{neighbors}$ is mostly a
choice of the level of detail we want in our analysis, as long as it is
large enough to get rid of sampling noise effects.

Figure \ref{fig:Nclusters_vs_Kneighbors_CO} shows the variation of the
number of clusters found as a function of $k_\mathrm{neighbors}$, for the
clustering analysis of the three CO isotopologue lines
(cf. Sect. \ref{sec:Results_CO}).  We see a flat plateau close to 10
clusters starting from $k_\mathrm{neighbors}>1200$, indicating that these
clusters are more strongly separated than the ones appearing at lower
values of $k_\mathrm{neighbors}$ (they do not disappear with increased
smoothing until much larger smoothing). We thus chose a value corresponding
to this plateau.

\section{Characteristic intensities in the
  clusters}\label{appendix:charac_intensities}

  \begin{table*}
    \caption{Characteristic line intensities in each CO cluster: 
      PDF maximum of the cluster, median value, mean value and standard 
      deviation. All values are in \Kkms{}.}
    \label{tab:clustersCO_intensities}
    \begin{center}
      \begin{tabular} {cc|cccccccc|c}
        \hline 
        & cluster & CO-1 & CO-2 & CO-3 & CO-4 & CO-5 & CO-6 & CO-7 & CO-8 & global\\
\hline
 Nb. of pixels &  & 47440 & 4943 & 14352 & 13336 & 14109 & 24769 & 15625 & 6476 & 141050 \\
\hline
 $^{12}$CO & center & 1.3 & 13.1 & 20.8 & 37.7 & 45.4 & 56.7 & 59.6 & 93.5 &  \\
 & median & 2.1 & 16.1 & 27.3 & 38.8 & 49.2 & 61.5 & 66.0 & 96.1 & 35.5 \\
 & mean & 3.2 & 16.1 & 27.4 & 38.6 & 48.9 & 64.1 & 69.3 & 97.9 & 36.4 \\
 & std & 2.9 & 1.9 & 4.8 & 5.0 & 7.2 & 14.6 & 15.6 & 10.5 & 30.3 \\
\hline
 $^{13}$CO & center & 0.2 & 0.8 & 1.3 & 3.1 & 5.0 & 7.8 & 13.8 & 15.2 &  \\
 & median & 0.3 & 1.0 & 2.1 & 4.2 & 6.1 & 10.4 & 18.1 & 20.8 & 3.6 \\
 & mean & 0.4 & 1.1 & 2.2 & 4.4 & 6.3 & 10.4 & 21.5 & 21.5 & 6.6 \\
 & std & 0.3 & 0.4 & 0.8 & 1.4 & 1.6 & 2.3 & 9.7 & 5.2 & 8.2 \\
\hline
 C$^{18}$O & center & 0.0 & 0.0 & 0.0 & 0.1 & 0.1 & 0.3 & 0.8 & 0.6 &  \\
 & median & 0.0 & 0.0 & 0.1 & 0.1 & 0.2 & 0.5 & 1.6 & 1.1 & 0.2 \\
 & mean & 0.0 & 0.0 & 0.1 & 0.1 & 0.3 & 0.5 & 2.3 & 1.3 & 0.4 \\
 & std & 0.1 & 0.1 & 0.1 & 0.2 & 0.2 & 0.3 & 1.7 & 0.6 & 0.9 \\
        \hline 
      \end{tabular}
    \end{center}
  \end{table*}
{} %

Table~\ref{tab:clustersCO_intensities} lists characteristic intensity
values of the three CO isotopologues for the clusters derived from CO
isotopologues only. The values given correspond to the mean, median,
standard deviation inside each cluster, as well as the most represented
intensity of the cluster (\ie, the PDF local maxima). The cluster's most
represented intensity is generally fainter than the cluster median and
mean. This is a natural consequence of the asymmetry of the PDFs that have
extended high intensity tails.

Similarly, Table~\ref{tab:clustersHCOp_intensities} and
\ref{tab:clustersCN_intensities} give the characteristic intensity values
(median, mean and standard deviation) for the groups of clusters HCO$^+$-1
to 7 and CN-1 to 5.

  \begin{table*}
    \caption{Characteristic line intensities in each HCO$^+$ group: 
      median value, mean value and standard deviation. All
      values are in \Kkms{}.
    }
    \label{tab:clustersHCOp_intensities}
    \begin{center}
      \begin{tabular} {cc|ccccccc|c}
        \hline 
        & group & HCO$^+$-1 & HCO$^+$-2 & HCO$^+$-3 & HCO$^+$-4 & HCO$^+$-5  & HCO$^+$-6 & HCO$^+$-7 & global \\
\hline
 Nb. of pixels &  & 53049 & 21998 & 20541 & 25681 & 15885 & 2926 & 970 &141050 \\
\hline
 $^{12}$CO & median & 2.3 & 30.7 & 47.8 & 60.1 & 85.0  & 78.9 & 95.2 & 35.5 \\
 & mean & 5.7 & 29.4 & 47.2 & 60.2 & 84.2 & 81.6 & 98.7 & 36.4 \\
 & std & 8.7 & 8.2 & 8.3 & 11.2 & 17.2 & 16.2 & 15.0 & 30.3 \\
\hline
 $^{13}$CO & median & 0.3 & 2.5 & 6.1 & 11.7 & 17.7 & 27.1 & 38.0 & 3.6 \\
 & mean & 0.6 & 2.9 & 6.2 & 11.7 & 18.3 & 27.8 & 39.7 & 6.6 \\
 & std & 1.0 & 1.7 & 2.1 & 3.7 & 9.2 & 4.2 & 6.6 & 8.2 \\
\hline
C$^{18}$O & median & 0.02 & 0.07 & 0.2 & 0.6 & 1.1 & 3.3 & 5.8 & 0.2 \\
 & mean & 0.02 & 0.08 & 0.3 & 0.7 & 1.3 & 3.5 & 5.9 & 0.4 \\
 & std & 0.2 & 0.2 & 0.2 & 0.5 & 1.2 & 0.8 & 1.1 & 0.9 \\
 \hline
HCO$^+$ & median & 0.2 & 0.5 & 0.9 & 1.3 & 2.6 & 3.6 & 5.4 & 0.7 \\
 & mean & 0.3 & 0.5 & 0.9 & 1.4 & 3.0 & 4.0 & 6.3 & 1.1 \\
 & std & 0.6 & 0.3 & 0.4 & 0.5 & 1.6 & 1.3 & 2.5 & 1.3 \\
 \hline
CN & median & 0.07 & 0.09 & 0.3 & 0.4 & 1.1 & 1.2 & 1.9 & 0.2 \\
 & mean & 0.1 & 0.09 & 0.3 & 0.5 & 1.4 & 1.3 & 2.2 & 0.4 \\
 & std & 0.3 & 0.3 & 0.3 & 0.3 & 1.3 & 0.5 & 0.8 & 0.7 \\
        \hline 
      \end{tabular}
    \end{center}
  \end{table*}
{} %
  \begin{table*}
    \caption{Characteristic line intensities in each CN group:
      median value, mean value and standard deviation. All
      values are in \Kkms{}.
    }
    \label{tab:clustersCN_intensities}
    \begin{center}
      \begin{tabular} {cc|ccccc|c}
        \hline 
        & group & CN-1 & CN-2 & CN-3 & CN-4 & CN-5  & global\\
\hline
 Nb. of pixels &  & 84631 & 36638 & 6991 & 11820 & 970  & 141050 \\
\hline
 \twCO{} & median & 7.6 & 56.2 & 68.6 & 91.9 & 95.2  & 35.5 \\
 & mean & 17.0 & 55.4 & 70.7 & 91.5 & 98.7 & 36.4 \\
 & std & 17.8 & 12.9 & 12.0 & 14.8 & 15.0 & 30.3 \\
\hline
 \thCO{} & median & 0.6 & 9.9 & 20.6 & 17.2 & 38.0 & 3.6 \\
 & mean & 1.9 & 9.9 & 21.5 & 18.8 & 39.7 & 6.6 \\
 & std & 2.4 & 4.3 & 4.9 & 10.9 & 6.6 & 8.2 \\
\hline
 \CeiO{} & median & 0.05 & 0.5 & 2.0 & 0.8 & 5.8 & 0.2 \\
 & mean & 0.06 & 0.6 & 2.3 & 1.3 & 5.9 & 0.4 \\
 & std & 0.2 & 0.5 & 1.1 & 1.4 & 1.1 & 0.9 \\
\hline
 HCO$^+$ & median & 0.3 & 1.2 & 2.3 & 3.0 & 5.4 & 0.7 \\
 & mean & 0.4 & 1.3 & 2.7 & 3.4 & 6.3 & 1.1 \\
 & std & 0.5 & 0.5 & 1.3 & 1.7 & 2.5 & 1.3 \\
 \hline
 CN & median & 0.08 & 0.4 & 0.8 & 1.3 & 1.9 & 0.2 \\
 & mean & 0.1 & 0.5 & 0.9 & 1.7 & 2.2 & 0.4 \\
 & std & 0.3 & 0.3 & 0.4 & 1.4 & 0.8 & 0.7 \\
        \hline 
      \end{tabular}
    \end{center}
  \end{table*}
{} %

\section{Details of the CO LTE modeling}
\label{sec:CO-LTE:details}

Here, we describe the details of the modeling that allows us to derive the
physical and chemical conditions discussed in Sect.~\ref{sec:CO-LTE}.

\newcommand{\op}[1]{\emm{\tau_{#1}}}
\newcommand{\Tkin}[1]{\emm{T^\emr{kin}_\emr{#1}}}
\newcommand{\Tbrigth}[1]{\emm{T^\emr{B}_{#1}}}
\newcommand{\Intens}[1]{\emm{I_{#1}}}
\newcommand{\Tzero}[1]{\emm{T^{0}_{#1}}}
\newcommand{\Texc}[1]{\emm{T^\emr{exc}_{#1}}}
\newcommand{\width}[1]{\emm{w_{#1}}} \newcommand{\Dv}{\emm{\Delta v}}
\newcommand{\Tcmb}{\emm{T_\emr{cmb}}} \newcommand{\FWHM}{\emr{FWHM}}

\newcommand{\paren}[1] {\left( #1 \right) } \newcommand{\cbrace}[1]
{\left\{ #1 \right\}} \newcommand{\bracket}[1]{\left[ #1 \right] }

\subsection{Parametrization}

The intensity integrated over the line profile is defined as
\begin{equation}
  \label{eq:lte:gen}
  \Intens{ij} = \Tbrigth{ij} \, \width{ij},
\end{equation}
with $ij = 12$, 13, and 18 for the \twCO{}, \thCO{}, and \CeiO{} \Jone{}
lines, respectively. In this equation, \Tbrigth{ij} is the LTE intensity
that is defined as
\begin{equation}
  \Tbrigth{ij} 
  = \Tzero{ij} \bracket{1-\exp\paren{-\op{ij}}}
  \cbrace{\frac{1}{\exp\paren{\frac{\Tzero{ij}}{\Texc{ij}}}-1}
    -\frac{1}{\exp\paren{\frac{\Tzero{ij}}{\Tcmb}}-1}},
\end{equation}
where \op{ij} is the opacity at the line center, $\Tcmb = 2.73\K$ is the
cosmic microwave background temperature, and $\Tzero{ij} =
h\,\nu/k_\emr{B}$ ($= 5.53\K$ for \twCO{} \Jone, 5.29\K{} for \thCO, and
5.27\K{} for \CeiO).

Assuming LTE, we will parametrize the modeled curves with increasing
kinetic temperature (\Tkin{}) at constant \thCO{} opacity
\begin{equation}
  \Texc{13} = \Texc{18} = \Tkin{min}+(\Tkin{max}-\Tkin{min})\,\frac{i-1}{n-1}
  \quad \mbox{with} \quad
  1 \le i \le n.
\end{equation}
Moreover, we will use the additional freedom to have a higher excitation
temperature for \twCO{}, \ie{},
\begin{equation}
  \frac{\Texc{12}}{\Texc{13}} = \mbox{cste} \ge 1.
\end{equation}
The variation of the opacity with the temperature is computed using the
opacity at 20\K{} as reference, \ie,
\begin{equation}
  \op{ij} = \op{ij}^{20\K} \, \cbrace{\frac{20\K}{\Texc{ij}}} \, \cbrace{\frac{1-\exp{\paren{-\frac{\Tzero{ij}}{\Texc{ij}}}}}{1-\exp\paren{-\frac{\Tzero{ij}}{20\K}}}}
\end{equation}
\begin{equation}
  \mbox{with} \quad
  \frac{\op{12}}{\op{13}} = \frac{N_\emr{min}(\twCO)}{N(\thCO)},
\end{equation}
\begin{equation}
  \mbox{and} \quad
  \frac{\op{13}}{\op{18}} = \frac{N(\thCO)}{N(\CeiO)} = \frac{[\thCO]}{[\CeiO]},
\end{equation}
where $N$(CO) is the column density of each CO isotopologue (in fact, a
lower limit for \twCO{}), and [CO] is the abundance relative to H$_2$ of
each CO isotopologue. Only the $I(\thCO)/I(\CeiO)$ can be interpreted in
terms of relative abundances because the emission of these isotopologues is
probably co-spatial, while the \twCO{} line is mainly sensitive to
  the foreground part of the emission.

The second factor on the right-hand side of Eq.~\ref{eq:lte:gen} takes care
of the integration over a Gaussian line profile, including the opacity
broadening term
\begin{equation}
  \width{ij} = \FWHM \, \sqrt{\frac{1}{\log{2}}\,\log\cbrace{\frac{\op{ij}}{\log\bracket{\frac{2}{1+\exp(-\op{ij})}}}}},
\end{equation}
where \FWHM{} is the measured linewidth of the Gaussian profile for an
optically thin line. As we limited the range of velocity over which we
integrate the CO lines to $\Delta \mathrm{v}$, we saturate \width{ij} as
follows
\begin{equation}
  \width{ij} = \Delta \mathrm{v}
  \quad \mbox{where} \quad
  \width{ij} > \Delta \mathrm{v}.
\end{equation}

\subsection{Impact of each input parameter}

Here, we describe the specific influence that each input parameter has on
the curves in the histograms of Fig.~\ref{fig:CO:LTE}.  At constant \thCO{}
opacity, a change of the line FWHM has an homothetic effect on the modeled
curves in the intensity vs intensity histograms: the higher the FWHM, the
larger the amplitude of the curve. Limiting the interval of velocity over
which we integrate quickly leads to a saturation of the \thCO{} integrated
intensities. This is the reason why the upper edge of the \thCO{} vs
\twCO{} histogram is relatively sharp. This effect is less obvious for the
\twCO{} line because the opacity is so large that the line is already
saturated. Saturation of the \thCO{} emission also explains the range of
observed values of CO isotopologue ratios (bottom left panel).  When both
\twCO{} and \thCO{} get saturated the intensity ratio is mainly controlled
by the ratio of the excitation temperatures, with a modest influence of the
opacity broadening term.

The minimum kinetic temperature sets the lower edge of the histogram of
\thCO{} vs \CeiO{}. Using a lower temperature would result in curves that
go beyond the observed minimum \thCO{} intensity for each given \CeiO{}
intensity. The maximum kinetic temperature controls the regions of low
intensities at low column density for all three isotopologues. In other
words, we do not populate correctly the low intensity part of the histogram
when the maximum temperature is too low.

The $N(\twCO)/N(\thCO)$ column density ratio controls the observed lower
edge of the associated intensity vs intensity histogram. A too low value
underestimates the \thCO{} intensity at constant \twCO{} intensity and vice
versa. In a similar way, the $[\thCO]/[\CeiO]$ abundance ratio controls the
observed upper edge of the associated intensity vs intensity histogram.

Finally, the ratio of the $\Texc{\twCO}/\Texc{\thCO}$ excitation
temperatures controls the \thCO{} ``width'' of the curves.  A higher
$\Texc{\twCO}/\Texc{\thCO}$ implies both a lower and higher \thCO{}
intensity (at low and high \Texc{\thCO} value) for the same \twCO{}
intensity.  In other words, a higher $\Texc{\twCO}/\Texc{\thCO}$ would less
well describe both the lower and upper edge of the \thCO{} vs \twCO{}
histogram if all other parameters stay constant. A higher
$\Texc{\twCO}/\Texc{\thCO}$ also implies a higher slope of the
\thCO/\CeiO{} vs.\ \twCO/\thCO{} curves.

\section{Details of the CN, HCO$^+$, and \CeiO{} RADEX
  modeling}\label{Appendix-radex-models}

Here, we describe the detail of our modeling approach for deriving the
HCO$^+$ and CN abundances and abundance ratios discussed in
Sect.~\ref{CN-group-interpretation-Radex}.

\FigRadexConstraints{} %
\TabRadexMedians{} %

\subsection{RADEX non-LTE radiative transfer models}

The radiative transfer modeling is subtle for two reasons. First, the
critical densities of C$^{18}$O, HCO$^+$, and CN (for collisional
excitation with H$_2$) differ by two to three orders of magnitude ($\sim
2\times10^3$, $2\times10^5$, and $\sim2\times10^6\pccm$, respectively). The
emission of C$^{18}$O is mostly thermalized in all groups (except CN-1). In
contrast, the excitation of HCO$^+$, and CN is subthermal. We thus use a
non-LTE radiative transfer approach. Second, collisions with electrons can
dominate the excitation of high dipole moment species such as CN and
HCO$^+$ in regions where the electronic fraction is high
enough~\citep{Black1991,Liszt2012,Goldsmith2017}. This effect can be
important for the regions of low visual extinction, \eg{}, groups CN-1 and
to a lesser extent CN-2, and for the highly FUV-illuminated PDRs of group
CN-4.

We used the RADEX code \citep{VanDerTak2007} which uses the escape
probability approach of \citet{Sobolev1960} to compute the non-LTE level
populations and the emission from a region of given temperature and volume
density of collision partners within a given velocity interval. The
position-position-velocity data suggest to use a linewidth of 1\kms{}. The
cross-section coefficients for collisional excitation are obtained from the
LAMDA
database\footnote{\url{http://home.strw.leidenuniv.nl/~moldata/}}. The
coefficient data for collisional excitation of C$^{18}$O, HCO$^+$, and CN
with H$_2$ were computed by \citet{Yang2010}, \citet{Flower1999}, and
\citet{Lique2010}, respectively.  The data for the excitation by electrons
were computed by \citet{Faure2001,Fuente2008} for HCO$^+$, and
\citet{Allison1971} for CN.

For each CN group, we ran RADEX models for the median values of the volume
density and temperature corresponding to each group, and we adjusted the
column densities of C$^{18}$O, HCO$^+$ and CN to best reproduce the median
line intensities.  As electrons may be important for the collisional
excitation of high dipole moment molecules, we used two hypotheses for the
electron fraction: (1) The electron fraction $x_e$ is 0 and only H$_2$
contributes to the excitation; (2) The electron fraction is set by the
ionisation of all carbon atoms and $x_e = 1.4\times10^{-4}$.  Case (2) was
computed only for groups where this hypothesis can be relevant~: CN-1, 2
and 4. For the diffuse medium of group CN-1, case (2) is a good hypothesis
and we take only the corresponding column density values.  In group CN-2
and CN-4 however, the electronic fraction is more uncertain, and we give a
range of values corresponding to the extreme cases (1) and (2). We assumed
a negligible electronic fraction in the other groups.

\subsection{Uncertainties}

We estimated the uncertainties on the median line intensities by two
different methods. First, we used a perturbative Monte Carlo approach. We
produced 1000 perturbed datasets (where Gaussian noise is added to each
pixel's intensities according to the local noise rms). We then computed the
median intensities inside each group. And we finally took the standard
deviation of these results as the uncertainty on the median
intensity. Second, we used a bootstrapping method \citep{Feigelson2012}. We
produced 1000 bootstrapped datasets of the same size as the original one
(drawing with replacement from the initial dataset). And we took the
standard deviation of the medians of these datasets. Both approaches give
similar uncertainties.  The relative ($1\sigma$) uncertainties are below
2\% for all groups and for the three molecular lines considered here.

Due to the large number of pixels in each group, the median values are
highly statistically significant, even in group CN-1 where the three lines
stay undetected in most pixels. However, potential unknown biases in the
integrated intensities (baseline distortion, etc.) are not taken into
account. As a result, the median values for group CN-1 are less reliable
than for the other groups. Our estimates of the volume density and kinetic
temperature are likely to suffer from global biases. We thus did not try to
estimate a noise-associated uncertainty for these quantities.

\subsection{Column densities}

Figure~\ref{fig:RadexConstraints} illustrates how the column densities were
derived from the radiative transfer models. Table~\ref{tab:Radex_medians}
lists the median input data and the output results. For C$^{18}$O, the
curves of all groups are roughly superimposed (thermalized emission) and
the derived column densities are mostly proportional to the median
intensities of the groups. In contrast, the intensity vs column density
relations are strongly dependent on the median density of the group for
HCO$^+$ and CN.  For HCO$^+$ and CN, the excitation is subthermal, and
closer to the weak excitation regime of \cite{Liszt2016}. As a result, the
intensity-column density relations are strongly dependent on the median
volume density of the group (in the weak excitation regime, the intensity
is proportional to the product of column density and volume density).

\end{document}